%% file: Virgo-DetChar-Results.tex
\documentclass[12pt]{iopart}
\usepackage{xcolor}
\usepackage{xspace}
\usepackage{booktabs}

\newcommand{\pycbc}{\texttt{PyCBC}\xspace}

\input{definitions}

\let\Sectionmark\sectionmark
\def\sectionmark#1{\def\Sectionname{\uppercase{#1}}\Sectionmark{#1}}
\let\Subsectionmark\subsectionmark
\def\subsectionmark#1{\def\Subsectionname{#1}\Subsectionmark{#1}}

\input{acronyms}

\begin{document}
\leftline{Dated: \today}

\title{Virgo Detector Characterization and Data Quality: results from the O3 run}

\input{authors_list}

\begin{abstract}
\input{abstract}

\end{abstract}

\maketitle

\tableofcontents

\clearpage

\mainmatter

\setlength{\parindent}{0pt}
\setlength{\parskip}{\medskipamount}


\section{Introduction}
\markboth{\thesection. \Sectionname}{}

\input{introduction}


\section {The Advanced Virgo Detector}
\label{section:AdV}
\markboth{\thesection. \Sectionname}{}

\input{detector_intro}

\subsection{The path to the O3 run}
\input{detector_o3path}

\subsection{The O3 configuration}
\input{detector_o3config}

\subsection{Virgo data and DetChar products}
\label{section:data_detchar}
\input{detector_data_detchar}


\section{The O3 run}
\label{section:O3}
\markboth{\thesection. \Sectionname}{}
\input{run_virgo_intro}

\subsection{Organisation}

\subsubsection{Data taking}
\label{subsection:data_taking}
\input{run_virgo_organization}

\subsubsection{Detector steering}
\label{subsubsection:metatron}

\input{run_virgo_datataking}

\subsubsection{DetChar organization and tools}

\input{run_virgo_shifts}

\subsubsection{On call duty service and rapid response team meetings}
\label{section:oncall_RRT}

\input{run_virgo_oncall_RRT}

\subsection{Performance}

\subsubsection{Noise Budget}
\label{sec:tools:noise_budget}
\input{tools_noise_budget}

\subsubsection{Virgo O3 duty cycle}
\label{subsection:O3_perf}
\input{O3_perf}

\subsubsection{The Virgo O3 dataset}
\label{susbsection:O3_dataset}
\input{O3_dataset}


\section{Real-time data quality}
\label{sec:onlinedq}
\markboth{\thesection. \Sectionname}{}
\input{onlinedq_intro}

\subsection{The Virgo O3 online data quality framework}\label{sec:onlinedq:architecture}
\input{onlinedq_architecture}

\subsubsection{State vector}\label{sec:onlinedq:statevector}
\input{onlinedq_statevector}

\subsubsection{Online CAT1 vetoes}\label{sec:onlinedq:cat1}
\input{onlinedq_CAT1}

\subsubsection{SegOnline}\label{sec:onlinedq:segonline}
\input{onlinedq_segonline}

\subsection{Veto streams}\label{sec:onlinedq:vetostreams}
\input{onlinedq_vetostreams}


\section{Public alerts}
\label{section:public_alerts}
\markboth{\thesection. \Sectionname}{}
\input{alerts_intro}

\subsection{Performance of the Virgo O3 DQR framework}
\label{subsection:DQR}

\input{dqr}

\subsection{O3 public alerts}

\input{alerts_public}


\section{Global data quality studies}
\label{section:dq_studies}
\markboth{\thesection. \Sectionname}{}
\input{dq_intro}

\subsection{Glitches and pipeline triggers}

\input{dq_pipelines}

\subsection{Channel safety: channel (in)sensitivity to gravitational waves}

\input{dq_channel_safety}

\label{subsection:safety}

\subsection{Spectral noise}

\input{dq_spectral_noise}

\subsection{Offline data quality}

\subsubsection{Offline studies and checks}
\label{sec:dq_offline}

\input{dq_offline}

\subsubsection{Event validation}\label{sec:offlinedq:validation}
\input{validation}


\section{Preparation of the O4 run}
\label{section:outlook}
\markboth{\thesection. \Sectionname}{}
\input{outlook}


\section*{Acknowledgments}
\input{acknowledgments}

\markboth{\Sectionname}{}


\addcontentsline{toc}{section}{List of Abbreviations}
\printacronyms[name=List of Abbreviations]


\clearpage

\section*{References}
\markboth{REFERENCES}{}
\bibliographystyle{iopart-num-doi}
\bibliography{references}

\end{document}

%% file: acronyms.tex
\DeclareAcronym{adv}{
  short=AdV,
  long=Advanced Virgo,
}

\DeclareAcronym{asd}{
  short=ASD,
  long=amplitude spectral density,
}

\DeclareAcronym{bh}{
  short=BH,
  long=black hole,
}

\DeclareAcronym{bns}{
  short=BNS,
  long=binary neutron star,
}

\DeclareAcronym{bristol}{
  short=\texttt{BRiSTOL},
  long=Band-limited RMS Stationarity Test Tool,
}

\DeclareAcronym{brms}{
  short=BRMS,
  long=band-limited RMS,
}

\DeclareAcronym{bruco}{
  short=\texttt{BruCo},
  long=brute-force coherence tool,
}

\DeclareAcronym{bs}{
  short=BS,
  long=beam splitter,
}

\DeclareAcronym{carm}{
  short=CARM,
  long=common (i.e. average) length of the two arm cavities,
}

\DeclareAcronym{ceb}{
  short=CEB,
  long=central building,
}

\DeclareAcronym{cw}{
  short=CW,
  long=continuous gravitational waves,
}

\DeclareAcronym{daq}{
  short=DAQ,
  long=data acquisition system,
}

\DeclareAcronym{darm}{
  short=DARM,
  long=difference of the two arm cavity lengths,
}

\DeclareAcronym{dms}{
  short=\texttt{DMS},
  long=Detector Monitoring System,
}

\DeclareAcronym{dof}{
  short=DOF,
  long=degree of freedom,
}

\DeclareAcronym{dq}{
  short=DQ,
  long=data quality,
}

\DeclareAcronym{dqr}{
  short=\texttt{DQR},
  long=Data Quality Report,
}

\DeclareAcronym{dqsegdb}{
  short=\texttt{DQSEGDB},
  long=Data Quality SEGment Database,
}

\DeclareAcronym{eom}{
  short=EOM,
  long=electro-optical modulator,
}

\DeclareAcronym{fft}{
  short=FFT,
  long=fast Fourier transform,
}

\DeclareAcronym{gracedb}{
  short=\texttt{GraceDB},
  long=GRAvitational-wave Candidate Event DataBase,
}

\DeclareAcronym{gw}{
  short=GW,
  long=gravitational wave,
}

\DeclareAcronym{gwosc}{
  short=GWOSC,
  long=Gravitational Wave Open Science Center,
}

\DeclareAcronym{imc}{
  short=IMC,
  long=input mode-cleaner,
}

\DeclareAcronym{lvalert}{
  short=\texttt{LVAlert},
  long=LIGO-Virgo Alert System,
}

\DeclareAcronym{mich}{
  short=MICH,
  long=length difference between the Virgo Michelson interferometer short arms
}

\DeclareAcronym{monet}{
  short=\texttt{MONET},
  long=Modulated NoisE Tool
}

\DeclareAcronym{ne}{
  short=NE,
  long=north end,
}

\DeclareAcronym{neb}{
  short=NEB,
  long=north-end building,
}

\DeclareAcronym{ni}{
  short=NI,
  long=north input,
}

\DeclareAcronym{noemi}{
  short=\texttt{NoEMi},
  long=Noise Frequency Event Miner,
}

\DeclareAcronym{ns}{
  short=NS,
  long=neutron star,
}

\DeclareAcronym{omc}{
  short=OMC,
  long=output mode-cleaner,
}

\DeclareAcronym{pr}{
  short=PR,
  long=power recycling,
}

\DeclareAcronym{prcl}{
  short=PRCL,
  long=power recycling cavity length,
}

\DeclareAcronym{psd}{
  short=PSD,
  long=power spectral density,
}

\DeclareAcronym{rrt}{
  short=RRT,
  long=rapid-response team,
}

\DeclareAcronym{sgwb}{
  short=SGWB,
  long=stochastic gravitational-wave background,
}

\DeclareAcronym{sneb}{
  short=SNEB,
  long=suspended north-end bench,
}

\DeclareAcronym{snr}{
  short=SNR,
  long=signal-to-noise ratio,
}

\DeclareAcronym{sr}{
  short=SR,
  long=signal recycling,
}

\DeclareAcronym{ssfs}{
  short=SSFS,
  long=second-stage frequency stabilization system,
}

\DeclareAcronym{sweb}{
  short=SWEB,
  long=suspended west-end bench,
}

\DeclareAcronym{upv}{
  short=\texttt{UPV},
  long=Use-Percentage Veto,
}

\DeclareAcronym{vim}{
  short=\texttt{VIM},
  long=Virgo Interferometer Monitor,
}

\DeclareAcronym{vpm}{
  short=VPM,
  long=Virgo Process Monitoring
}

\DeclareAcronym{we}{
  short=WE,
  long=west end,
}

\DeclareAcronym{web}{
  short=WEB,
  long=west-end building,
}

\DeclareAcronym{wi}{
  short=WI,
  long=west input,
}

%% file: authors_list.tex
\author{%
F~Acernese$^{1,2}$, 
M~Agathos$^{3}$, 
A~Ain$^{4}$, 
S~Albanesi$^{5,6}$, 
A~Allocca\orcidlink{0000-0002-5288-1351}$^{7,2}$, 
A~Amato\orcidlink{0000-0001-9557-651X}$^{8}$, 
T~Andrade$^{9}$, 
N~Andres\orcidlink{0000-0002-5360-943X}$^{10}$, 
M~Andr\'es-Carcasona\orcidlink{0000-0002-8738-1672}$^{11}$, 
T~Andri\'c\orcidlink{0000-0002-9277-9773}$^{12}$, 
S~Ansoldi$^{13,14}$, 
S~Antier\orcidlink{0000-0002-7686-3334}$^{15,16}$, 
T~Apostolatos$^{17}$, 
E~Z~Appavuravther$^{18,19}$, %
M~Ar\`ene$^{20}$, %
N~Arnaud\orcidlink{0000-0001-6589-8673}$^{21,22}$, 
M~Assiduo$^{23,24}$, 
S~Assis~de~Souza~Melo$^{22}$, 
P~Astone\orcidlink{0000-0003-4981-4120}$^{25}$, 
F~Aubin\orcidlink{0000-0003-1613-3142}$^{24}$, 
S~Babak\orcidlink{0000-0001-7469-4250}$^{20}$, 
F~Badaracco\orcidlink{0000-0001-8553-7904}$^{26}$, 
M~K~M~Bader$^{27}$, %
S~Bagnasco\orcidlink{0000-0001-6062-6505}$^{6}$, 
J~Baird$^{20}$, %
T~Baka$^{28}$, 
G~Ballardin$^{22}$, 
G~Baltus\orcidlink{0000-0002-0304-8152}$^{29}$, 
B~Banerjee\orcidlink{0000-0002-8008-2485}$^{12}$, 
C~Barbieri$^{30,31,32}$, %
P~Barneo\orcidlink{0000-0002-8883-7280}$^{9}$, 
F~Barone\orcidlink{0000-0002-8069-8490}$^{33,2}$, 
M~Barsuglia\orcidlink{0000-0002-1180-4050}$^{20}$, 
D~Barta\orcidlink{0000-0001-6841-550X}$^{34}$, %
A~Basti$^{35,4}$, 
M~Bawaj\orcidlink{0000-0003-3611-3042}$^{18,36}$, 
M~Bazzan$^{37,38}$, 
F~Beirnaert\orcidlink{0000-0002-4003-7233}$^{39}$, 
M~Bejger\orcidlink{0000-0002-4991-8213}$^{40}$, 
I~Belahcene$^{21}$, %
V~Benedetto$^{41}$, %
M~Berbel\orcidlink{0000-0001-6345-1798}$^{42}$, 
S~Bernuzzi\orcidlink{0000-0002-2334-0935}$^{3}$, 
D~Bersanetti\orcidlink{0000-0002-7377-415X}$^{43}$, 
A~Bertolini$^{27}$, 
U~Bhardwaj\orcidlink{0000-0003-1233-4174}$^{16,27}$, 
A~Bianchi$^{27,44}$, 
S~Bini$^{45,46}$, 
M~Bischi$^{23,24}$, 
M~Bitossi$^{22,4}$, 
M-A~Bizouard\orcidlink{0000-0002-4618-1674}$^{15}$, 
F~Bobba$^{47,48}$, 
M~Bo\"{e}r$^{15}$, 
G~Bogaert$^{15}$, 
M~Boldrini$^{49,25}$, 
L~D~Bonavena$^{37}$, 
F~Bondu$^{50}$, 
R~Bonnand\orcidlink{0000-0001-5013-5913}$^{10}$, 
B~A~Boom$^{27}$, %
V~Boschi\orcidlink{0000-0001-8665-2293}$^{4}$, 
V~Boudart\orcidlink{0000-0001-9923-4154}$^{29}$, 
Y~Bouffanais$^{37,38}$, 
A~Bozzi$^{22}$, 
C~Bradaschia$^{4}$, 
M~Branchesi\orcidlink{0000-0003-1643-0526}$^{12,51}$, 
M~Breschi\orcidlink{0000-0002-3327-3676}$^{3}$, 
T~Briant\orcidlink{0000-0002-6013-1729}$^{52}$, 
A~Brillet$^{15}$, 
J~Brooks$^{22}$, %
G~Bruno$^{26}$, 
F~Bucci$^{24}$, 
T~Bulik$^{53}$, 
H~J~Bulten$^{27}$, 
D~Buskulic$^{10}$, 
C~Buy\orcidlink{0000-0003-2872-8186}$^{54}$, 
G~S~Cabourn~Davies\orcidlink{0000-0002-4289-3439}$^{55}$, %
G~Cabras\orcidlink{0000-0002-6852-6856}$^{13,14}$, 
R~Cabrita\orcidlink{0000-0003-0133-1306}$^{26}$, 
G~Cagnoli\orcidlink{0000-0002-7086-6550}$^{8}$, 
E~Calloni$^{7,2}$, 
M~Canepa$^{56,43}$, 
S~Canevarolo$^{28}$, 
M~Cannavacciuolo$^{47}$, %
E~Capocasa\orcidlink{0000-0003-3762-6958}$^{20}$, 
G~Carapella$^{47,48}$, 
F~Carbognani$^{22}$, 
M~Carpinelli$^{57,58,22}$, 
G~Carullo\orcidlink{0000-0001-9090-1862}$^{35,4}$, 
J~Casanueva~Diaz$^{22}$, 
C~Casentini$^{59,60}$, 
S~Caudill$^{27,28}$, 
F~Cavalier\orcidlink{0000-0002-3658-7240}$^{21}$, %
R~Cavalieri\orcidlink{0000-0001-6064-0569}$^{22}$, 
G~Cella\orcidlink{0000-0002-0752-0338}$^{4}$, 
P~Cerd\'a-Dur\'an$^{61}$, 
E~Cesarini\orcidlink{0000-0001-9127-3167}$^{60}$, 
W~Chaibi$^{15}$, 
P~Chanial\orcidlink{0000-0003-1753-524X}$^{22}$, %
E~Chassande-Mottin\orcidlink{0000-0003-3768-9908}$^{20}$, 
S~Chaty\orcidlink{0000-0002-5769-8601}$^{20}$, 
F~Chiadini\orcidlink{0000-0002-9339-8622}$^{62,48}$, 
G~Chiarini$^{38}$, 
R~Chierici$^{63}$, 
A~Chincarini\orcidlink{0000-0003-4094-9942}$^{43}$, 
M~L~Chiofalo$^{35,4}$, 
A~Chiummo\orcidlink{0000-0003-2165-2967}$^{22}$, 
S~Choudhary\orcidlink{0000-0003-0949-7298}$^{64}$, %
N~Christensen\orcidlink{0000-0002-6870-4202}$^{15}$, 
G~Ciani\orcidlink{0000-0003-4258-9338}$^{37,38}$, 
P~Ciecielag$^{40}$, 
M~Cie\'slar\orcidlink{0000-0001-8912-5587}$^{40}$, 
M~Cifaldi$^{59,60}$, 
R~Ciolfi\orcidlink{0000-0003-3140-8933}$^{65,38}$, 
F~Cipriano$^{15}$, %
S~Clesse$^{66}$, 
F~Cleva$^{15}$, 
E~Coccia$^{12,51}$, 
E~Codazzo\orcidlink{0000-0001-7170-8733}$^{12}$, 
P-F~Cohadon\orcidlink{0000-0003-3452-9415}$^{52}$, 
D~E~Cohen\orcidlink{0000-0002-0583-9919}$^{21}$, %
A~Colombo\orcidlink{0000-0002-7439-4773}$^{30,31}$, 
M~Colpi$^{30,31}$, 
L~Conti\orcidlink{0000-0003-2731-2656}$^{38}$, 
I~Cordero-Carri\'on\orcidlink{0000-0002-1985-1361}$^{67}$, 
S~Corezzi$^{36,18}$, 
D~Corre$^{21}$, %
S~Cortese\orcidlink{0000-0002-6504-0973}$^{22}$, 
J-P~Coulon$^{15}$, 
M~Croquette\orcidlink{0000-0002-8581-5393}$^{52}$, %
J~R~Cudell\orcidlink{0000-0002-2003-4238}$^{29}$, 
E~Cuoco$^{22,68,4}$, 
M~Cury{\l}o$^{53}$, 
P~Dabadie$^{8}$, %
T~Dal~Canton\orcidlink{0000-0001-5078-9044}$^{21}$, 
S~Dall'Osso\orcidlink{0000-0003-4366-8265}$^{12}$, %
G~D\'alya\orcidlink{0000-0003-3258-5763}$^{39}$, 
B~D'Angelo\orcidlink{0000-0001-9143-8427}$^{56,43}$, 
S~Danilishin\orcidlink{0000-0001-7758-7493}$^{69,27}$, 
S~D'Antonio$^{60}$, 
V~Dattilo$^{22}$, 
M~Davier$^{21}$, 
D~Davis\orcidlink{0000-0001-5620-6751}$^{70}$, %
J~Degallaix\orcidlink{0000-0002-1019-6911}$^{71}$, 
M~De~Laurentis$^{7,2}$, 
S~Del\'eglise\orcidlink{0000-0002-8680-5170}$^{52}$, 
F~De~Lillo\orcidlink{0000-0003-4977-0789}$^{26}$, 
D~Dell'Aquila\orcidlink{0000-0001-5895-0664}$^{57}$, 
W~Del~Pozzo$^{35,4}$, 
F~De~Matteis$^{59,60}$, %
A~Depasse\orcidlink{0000-0003-1014-8394}$^{26}$, 
R~De~Pietri\orcidlink{0000-0003-1556-8304}$^{72,73}$, 
R~De~Rosa\orcidlink{0000-0002-4004-947X}$^{7,2}$, 
C~De~Rossi$^{22}$, 
R~De~Simone$^{62}$, %
L~Di~Fiore$^{2}$, 
C~Di~Giorgio\orcidlink{0000-0003-2127-3991}$^{47,48}$, %
F~Di~Giovanni\orcidlink{0000-0001-8568-9334}$^{61}$, 
M~Di~Giovanni$^{12}$, 
T~Di~Girolamo\orcidlink{0000-0003-2339-4471}$^{7,2}$, 
A~Di~Lieto\orcidlink{0000-0002-4787-0754}$^{35,4}$, 
A~Di~Michele\orcidlink{0000-0002-0357-2608}$^{36}$, %
S~Di~Pace\orcidlink{0000-0001-6759-5676}$^{49,25}$, 
I~Di~Palma\orcidlink{0000-0003-1544-8943}$^{49,25}$, 
F~Di~Renzo\orcidlink{0000-0002-5447-3810}$^{35,4}$, 
L~D'Onofrio\orcidlink{0000-0001-9546-5959}$^{7,2}$, %
M~Drago\orcidlink{0000-0002-3738-2431}$^{49,25}$, 
J-G~Ducoin$^{21}$, 
U~Dupletsa$^{12}$, 
O~Durante$^{47,48}$, 
D~D'Urso\orcidlink{0000-0002-8215-4542}$^{57,58}$, 
P-A~Duverne$^{21}$, 
M~Eisenmann$^{10}$, %
L~Errico$^{7,2}$, 
D~Estevez\orcidlink{0000-0002-3021-5964}$^{74}$, 
F~Fabrizi\orcidlink{0000-0002-3809-065X}$^{23,24}$, 
F~Faedi$^{24}$, 
V~Fafone\orcidlink{0000-0003-1314-1622}$^{59,60,12}$, 
S~Farinon$^{43}$, %
G~Favaro\orcidlink{0000-0002-0351-6833}$^{37}$, 
M~Fays\orcidlink{0000-0002-4390-9746}$^{29}$, 
E~Fenyvesi\orcidlink{0000-0003-2777-3719}$^{34,75}$, %
I~Ferrante\orcidlink{0000-0002-0083-7228}$^{35,4}$, 
F~Fidecaro\orcidlink{0000-0002-6189-3311}$^{35,4}$, 
P~Figura\orcidlink{0000-0002-8925-0393}$^{53}$, 
A~Fiori\orcidlink{0000-0003-3174-0688}$^{4,35}$, 
I~Fiori\orcidlink{0000-0002-0210-516X}$^{22}$, 
R~Fittipaldi$^{76,48}$, %
V~Fiumara$^{77,48}$, %
R~Flaminio$^{10,78}$, 
J~A~Font\orcidlink{0000-0001-6650-2634}$^{61,79}$, 
S~Frasca$^{49,25}$, 
F~Frasconi\orcidlink{0000-0003-4204-6587}$^{4}$, 
A~Freise\orcidlink{0000-0001-6586-9901}$^{27,44}$, 
O~Freitas$^{80}$, 
G~G~Fronz\'e\orcidlink{0000-0003-0966-4279}$^{6}$, 
B~U~Gadre\orcidlink{0000-0002-1534-9761}$^{81,28}$, %
R~Gamba$^{3}$, 
B~Garaventa\orcidlink{0000-0003-2490-404X}$^{43,56}$, 
F~Garufi\orcidlink{0000-0003-1391-6168}$^{7,2}$, 
G~Gemme\orcidlink{0000-0002-1127-7406}$^{43}$, 
A~Gennai\orcidlink{0000-0003-0149-2089}$^{4}$, 
Archisman~Ghosh\orcidlink{0000-0003-0423-3533}$^{39}$, 
B~Giacomazzo\orcidlink{0000-0002-6947-4023}$^{30,31,32}$, 
L~Giacoppo$^{49,25}$, %
P~Giri\orcidlink{0000-0002-4628-2432}$^{4,35}$, 
F~Gissi$^{41}$, %
S~Gkaitatzis\orcidlink{0000-0001-9420-7499}$^{4,35}$, %
B~Goncharov\orcidlink{0000-0003-3189-5807}$^{12}$, 
M~Gosselin$^{22}$, 
R~Gouaty$^{10}$, 
A~Grado\orcidlink{0000-0002-0501-8256}$^{82,2}$, 
M~Granata\orcidlink{0000-0003-3275-1186}$^{71}$, 
V~Granata$^{47}$, 
G~Greco$^{18}$, 
G~Grignani$^{36,18}$, 
A~Grimaldi\orcidlink{0000-0002-6956-4301}$^{45,46}$, 
S~J~Grimm$^{12,51}$, %
P~Gruning$^{21}$, %
D~Guerra\orcidlink{0000-0003-0029-5390}$^{61}$, 
G~M~Guidi\orcidlink{0000-0002-3061-9870}$^{23,24}$, 
G~Guix\'e$^{9}$, 
Y~Guo$^{27}$, 
P~Gupta$^{27,28}$, 
L~Haegel\orcidlink{0000-0002-3680-5519}$^{20}$, 
O~Halim\orcidlink{0000-0003-1326-5481}$^{14}$, 
O~Hannuksela$^{28,27}$, 
T~Harder$^{15}$, 
K~Haris$^{27,28}$, 
J~Harms\orcidlink{0000-0002-7332-9806}$^{12,51}$, 
B~Haskell$^{40}$, 
A~Heidmann\orcidlink{0000-0002-0784-5175}$^{52}$, 
H~Heitmann\orcidlink{0000-0003-0625-5461}$^{15}$, 
P~Hello$^{21}$, 
G~Hemming\orcidlink{0000-0001-5268-4465}$^{22}$, 
E~Hennes\orcidlink{0000-0002-2246-5496}$^{27}$, 
S~Hild$^{69,27}$, 
D~Hofman$^{71}$, %
V~Hui\orcidlink{0000-0002-0233-2346}$^{10}$, 
B~Idzkowski\orcidlink{0000-0001-5869-2714}$^{53}$, 
A~Iess$^{59,60}$, %
P~Iosif\orcidlink{0000-0003-1621-7709}$^{83}$, 
T~Jacqmin\orcidlink{0000-0002-0693-4838}$^{52}$, 
P-E~Jacquet\orcidlink{0000-0001-9552-0057}$^{52}$, %
S~P~Jadhav$^{64}$, %
J~Janquart$^{28,27}$, 
K~Janssens\orcidlink{0000-0001-8760-4429}$^{84,15}$, 
P~Jaranowski\orcidlink{0000-0001-8085-3414}$^{85}$, 
V~Juste$^{74}$, 
C~Kalaghatgi$^{28,27,86}$, 
C~Karathanasis\orcidlink{0000-0002-0642-5507}$^{11}$, 
S~Katsanevas\orcidlink{0000-0003-0324-0758}$^{22}$\footnote{Deceased, November 2022.}, 
F~K\'ef\'elian$^{15}$, 
N~Khetan$^{12,51}$, %
G~Koekoek$^{27,69}$, 
S~Koley\orcidlink{0000-0002-5793-6665}$^{12}$, 
M~Kolstein\orcidlink{0000-0002-5482-6743}$^{11}$, 
A~Kr\'olak\orcidlink{0000-0003-4514-7690}$^{87,88}$, 
P~Kuijer\orcidlink{0000-0002-6987-2048}$^{27}$, 
P~Lagabbe$^{10}$, 
D~Laghi\orcidlink{0000-0001-7462-3794}$^{54}$, 
M~Lalleman$^{84}$, 
A~Lamberts$^{15,89}$, 
I~La~Rosa$^{10}$, 
A~Lartaux-Vollard$^{21}$, %
C~Lazzaro$^{37,38}$, 
P~Leaci\orcidlink{0000-0002-3997-5046}$^{49,25}$, 
A~Lema{\^i}tre$^{90}$, 
M~Lenti\orcidlink{0000-0002-2765-3955}$^{24,91}$, 
E~Leonova$^{16}$, %
N~Leroy\orcidlink{0000-0002-2321-1017}$^{21}$, 
N~Letendre$^{10}$, 
K~Leyde$^{20}$, 
F~Linde$^{86,27}$, 
L~London$^{16}$, 
A~Longo\orcidlink{0000-0003-4254-8579}$^{92}$, 
M~Lopez~Portilla$^{28}$, %
M~Lorenzini\orcidlink{0000-0002-2765-7905}$^{59,60}$, 
V~Loriette$^{93}$, 
G~Losurdo\orcidlink{0000-0003-0452-746X}$^{4}$, 
D~Lumaca\orcidlink{0000-0002-3628-1591}$^{59,60}$, 
A~Macquet$^{15}$, 
C~Magazz\`u\orcidlink{0000-0002-9913-381X}$^{4}$, %
M~Magnozzi\orcidlink{0000-0003-4512-8430}$^{43,56}$, 
E~Majorana$^{49,25}$, 
I~Maksimovic$^{93}$, %
N~Man$^{15}$, 
V~Mangano\orcidlink{0000-0001-7902-8505}$^{49,25}$, 
M~Mantovani\orcidlink{0000-0002-4424-5726}$^{22}$, 
M~Mapelli\orcidlink{0000-0001-8799-2548}$^{37,38}$, 
F~Marchesoni$^{19,18,94}$, 
D~Mar\'{\i}n~Pina\orcidlink{0000-0001-6482-1842}$^{9}$, 
F~Marion$^{10}$, 
A~Marquina$^{67}$, 
S~Marsat\orcidlink{0000-0001-9449-1071}$^{20}$, 
F~Martelli$^{23,24}$, 
M~Martinez$^{11}$, 
V~Martinez$^{8}$, 
A~Masserot$^{10}$, 
S~Mastrogiovanni\orcidlink{0000-0003-1606-4183}$^{20}$, 
Q~Meijer$^{28}$, 
A~Menendez-Vazquez$^{11}$, %
L~Mereni$^{71}$, 
M~Merzougui$^{15}$, %
A~Miani\orcidlink{0000-0001-7737-3129}$^{45,46}$, 
C~Michel\orcidlink{0000-0003-0606-725X}$^{71}$, 
L~Milano$^{7}$\footnote{Deceased, April 2021.}, %
A~Miller$^{26}$, 
B~Miller$^{16,27}$, %
E~Milotti$^{95,14}$, 
Y~Minenkov$^{60}$, 
Ll~M~Mir$^{11}$, 
M~Miravet-Ten\'es\orcidlink{0000-0002-8766-1156}$^{61}$, 
M~Montani$^{23,24}$, 
F~Morawski$^{40}$, 
B~Mours\orcidlink{0000-0002-6444-6402}$^{74}$, 
C~M~Mow-Lowry\orcidlink{0000-0002-0351-4555}$^{27,44}$
% ...
}%
\address{$^{1}$Dipartimento di Farmacia, Universit\`a di Salerno, I-84084 Fisciano, Salerno, Italy}
\address{$^{2}$INFN, Sezione di Napoli, Complesso Universitario di Monte S. Angelo, I-80126 Napoli, Italy}
\address{$^{3}$Theoretisch-Physikalisches Institut, Friedrich-Schiller-Universit\"at Jena, D-07743 Jena, Germany}
\address{$^{4}$INFN, Sezione di Pisa, I-56127 Pisa, Italy}
\address{$^{5}$Dipartimento di Fisica, Universit\`a degli Studi di Torino, I-10125 Torino, Italy}
\address{$^{6}$INFN Sezione di Torino, I-10125 Torino, Italy}
\address{$^{7}$Universit\`a di Napoli ``Federico II'', Complesso Universitario di Monte S. Angelo, I-80126 Napoli, Italy}
\address{$^{8}$Universit\'e de Lyon, Universit\'e Claude Bernard Lyon 1, CNRS, Institut Lumi\`ere Mati\`ere, F-69622 Villeurbanne, France}
\address{$^{9}$Institut de Ci\`encies del Cosmos (ICCUB), Universitat de Barcelona, C/ Mart\'{\i} i Franqu\`es 1, Barcelona, 08028, Spain}
\address{$^{10}$Univ. Savoie Mont Blanc, CNRS, Laboratoire d'Annecy de Physique des Particules - IN2P3, F-74000 Annecy, France}
\address{$^{11}$Institut de F\'{\i}sica d'Altes Energies (IFAE), Barcelona Institute of Science and Technology, and  ICREA, E-08193 Barcelona, Spain}
\address{$^{12}$Gran Sasso Science Institute (GSSI), I-67100 L'Aquila, Italy}
\address{$^{13}$Dipartimento di Scienze Matematiche, Informatiche e Fisiche, Universit\`a di Udine, I-33100 Udine, Italy}
\address{$^{14}$INFN, Sezione di Trieste, I-34127 Trieste, Italy}
\address{$^{15}$Artemis, Universit\'e C\^ote d'Azur, Observatoire de la C\^ote d'Azur, CNRS, F-06304 Nice, France}
\address{$^{16}$GRAPPA, Anton Pannekoek Institute for Astronomy and Institute for High-Energy Physics, University of Amsterdam, Science Park 904, 1098 XH Amsterdam, Netherlands}
\address{$^{17}$Department of Physics, National and Kapodistrian University of Athens, School of Science Building, 2nd floor, Panepistimiopolis, 15771 Ilissia, Greece}
\address{$^{18}$INFN, Sezione di Perugia, I-06123 Perugia, Italy}
\address{$^{19}$Universit\`a di Camerino, I-62032 Camerino, Italy}
\address{$^{20}$Universit\'e de Paris, CNRS, Astroparticule et Cosmologie, F-75006 Paris, France}
\address{$^{21}$Universit\'e Paris-Saclay, CNRS/IN2P3, IJCLab, 91405 Orsay, France}
\address{$^{22}$European Gravitational Observatory (EGO), I-56021 Cascina, Pisa, Italy}
\address{$^{23}$Universit\`a degli Studi di Urbino ``Carlo Bo'', I-61029 Urbino, Italy}
\address{$^{24}$INFN, Sezione di Firenze, I-50019 Sesto Fiorentino, Firenze, Italy}
\address{$^{25}$INFN, Sezione di Roma, I-00185 Roma, Italy}
\address{$^{26}$Universit\'e catholique de Louvain, B-1348 Louvain-la-Neuve, Belgium}
\address{$^{27}$Nikhef, Science Park 105, 1098 XG Amsterdam, Netherlands}
\address{$^{28}$Institute for Gravitational and Subatomic Physics (GRASP), Utrecht University, Princetonplein 1, 3584 CC Utrecht, Netherlands}
\address{$^{29}$Universit\'e de Li\`ege, B-4000 Li\`ege, Belgium}
\address{$^{30}$Universit\`a degli Studi di Milano-Bicocca, I-20126 Milano, Italy}
\address{$^{31}$INFN, Sezione di Milano-Bicocca, I-20126 Milano, Italy}
\address{$^{32}$INAF, Osservatorio Astronomico di Brera sede di Merate, I-23807 Merate, Lecco, Italy}
\address{$^{33}$Dipartimento di Medicina, Chirurgia e Odontoiatria ``Scuola Medica Salernitana'', Universit\`a di Salerno, I-84081 Baronissi, Salerno, Italy}
\address{$^{34}$Wigner RCP, RMKI, H-1121 Budapest, Konkoly Thege Mikl\'os \'ut 29-33, Hungary}
\address{$^{35}$Universit\`a di Pisa, I-56127 Pisa, Italy}
\address{$^{36}$Universit\`a di Perugia, I-06123 Perugia, Italy}
\address{$^{37}$Universit\`a di Padova, Dipartimento di Fisica e Astronomia, I-35131 Padova, Italy}
\address{$^{38}$INFN, Sezione di Padova, I-35131 Padova, Italy}
\address{$^{39}$Universiteit Gent, B-9000 Gent, Belgium}
\address{$^{40}$Nicolaus Copernicus Astronomical Center, Polish Academy of Sciences, 00-716, Warsaw, Poland}
\address{$^{41}$Dipartimento di Ingegneria, Universit\`a del Sannio, I-82100 Benevento, Italy}
\address{$^{42}$Departamento de Matem\'aticas, Universitat Aut\`onoma de Barcelona, Edificio C Facultad de Ciencias 08193 Bellaterra (Barcelona), Spain}
\address{$^{43}$INFN, Sezione di Genova, I-16146 Genova, Italy}
\address{$^{44}$Vrije Universiteit Amsterdam, 1081 HV Amsterdam, Netherlands}
\address{$^{45}$Universit\`a di Trento, Dipartimento di Fisica, I-38123 Povo, Trento, Italy}
\address{$^{46}$INFN, Trento Institute for Fundamental Physics and Applications, I-38123 Povo, Trento, Italy}
\address{$^{47}$Dipartimento di Fisica ``E.R. Caianiello'', Universit\`a di Salerno, I-84084 Fisciano, Salerno, Italy}
\address{$^{48}$INFN, Sezione di Napoli, Gruppo Collegato di Salerno, Complesso Universitario di Monte S. Angelo, I-80126 Napoli, Italy}
\address{$^{49}$Universit\`a di Roma ``La Sapienza'', I-00185 Roma, Italy}
\address{$^{50}$Univ Rennes, CNRS, Institut FOTON - UMR6082, F-3500 Rennes, France}
\address{$^{51}$INFN, Laboratori Nazionali del Gran Sasso, I-67100 Assergi, Italy}
\address{$^{52}$Laboratoire Kastler Brossel, Sorbonne Universit\'e, CNRS, ENS-Universit\'e PSL, Coll\`ege de France, F-75005 Paris, France}
\address{$^{53}$Astronomical Observatory Warsaw University, 00-478 Warsaw, Poland}
\address{$^{54}$L2IT, Laboratoire des 2 Infinis - Toulouse, Universit\'e de Toulouse, CNRS/IN2P3, UPS, F-31062 Toulouse Cedex 9, France}
\address{$^{55}$University of Portsmouth, Portsmouth, PO1 3FX, United Kingdom}
\address{$^{56}$Dipartimento di Fisica, Universit\`a degli Studi di Genova, I-16146 Genova, Italy}
\address{$^{57}$Universit\`a degli Studi di Sassari, I-07100 Sassari, Italy}
\address{$^{58}$INFN, Laboratori Nazionali del Sud, I-95125 Catania, Italy}
\address{$^{59}$Universit\`a di Roma Tor Vergata, I-00133 Roma, Italy}
\address{$^{60}$INFN, Sezione di Roma Tor Vergata, I-00133 Roma, Italy}
\address{$^{61}$Departamento de Astronom\'{\i}a y Astrof\'{\i}sica, Universitat de Val\`encia, E-46100 Burjassot, Val\`encia, Spain}
\address{$^{62}$Dipartimento di Ingegneria Industriale (DIIN), Universit\`a di Salerno, I-84084 Fisciano, Salerno, Italy}
\address{$^{63}$Universit\'e Lyon, Universit\'e Claude Bernard Lyon 1, CNRS, IP2I Lyon / IN2P3, UMR 5822, F-69622 Villeurbanne, France}
\address{$^{64}$Inter-University Centre for Astronomy and Astrophysics, Post Bag 4, Ganeshkhind, Pune 411 007, India}
\address{$^{65}$INAF, Osservatorio Astronomico di Padova, I-35122 Padova, Italy}
\address{$^{66}$Universit\'e libre de Bruxelles, Avenue Franklin Roosevelt 50 - 1050 Bruxelles, Belgium}
\address{$^{67}$Departamento de Matem\'aticas, Universitat de Val\`encia, E-46100 Burjassot, Val\`encia, Spain}
\address{$^{68}$Scuola Normale Superiore, Piazza dei Cavalieri, 7 - 56126 Pisa, Italy}
\address{$^{69}$Maastricht University, P.O. Box 616, 6200 MD Maastricht, Netherlands}
\address{$^{70}$LIGO Laboratory, California Institute of Technology, Pasadena, CA 91125, USA}
\address{$^{71}$Universit\'e Lyon, Universit\'e Claude Bernard Lyon 1, CNRS, Laboratoire des Mat\'eriaux Avanc\'es (LMA), IP2I Lyon / IN2P3, UMR 5822, F-69622 Villeurbanne, France}
\address{$^{72}$Dipartimento di Scienze Matematiche, Fisiche e Informatiche, Universit\`a di Parma, I-43124 Parma, Italy}
\address{$^{73}$INFN, Sezione di Milano Bicocca, Gruppo Collegato di Parma, I-43124 Parma, Italy}
\address{$^{74}$Universit\'e de Strasbourg, CNRS, IPHC UMR 7178, F-67000 Strasbourg, France}
\address{$^{75}$Institute for Nuclear Research, Bem t'er 18/c, H-4026 Debrecen, Hungary}
\address{$^{76}$CNR-SPIN, c/o Universit\`a di Salerno, I-84084 Fisciano, Salerno, Italy}
\address{$^{77}$Scuola di Ingegneria, Universit\`a della Basilicata, I-85100 Potenza, Italy}
\address{$^{78}$Gravitational Wave Science Project, National Astronomical Observatory of Japan (NAOJ), Mitaka City, Tokyo 181-8588, Japan}
\address{$^{79}$Observatori Astron\`omic, Universitat de Val\`encia, E-46980 Paterna, Val\`encia, Spain}
\address{$^{80}$Centro de F\'{\i}sica das Universidades do Minho e do Porto, Universidade do Minho, Campus de Gualtar, PT-4710 - 057 Braga, Portugal}
\address{$^{81}$Max Planck Institute for Gravitational Physics (Albert Einstein Institute), D-14476 Potsdam, Germany}
\address{$^{82}$INAF, Osservatorio Astronomico di Capodimonte, I-80131 Napoli, Italy}
\address{$^{83}$Department of Physics, Aristotle University of Thessaloniki, University Campus, 54124 Thessaloniki, Greece}
\address{$^{84}$Universiteit Antwerpen, Prinsstraat 13, 2000 Antwerpen, Belgium}
\address{$^{85}$University of Bia{\l}ystok, 15-424 Bia{\l}ystok, Poland}
\address{$^{86}$Institute for High-Energy Physics, University of Amsterdam, Science Park 904, 1098 XH Amsterdam, Netherlands}
\address{$^{87}$Institute of Mathematics, Polish Academy of Sciences, 00656 Warsaw, Poland}
\address{$^{88}$National Center for Nuclear Research, 05-400 {\' S}wierk-Otwock, Poland}
\address{$^{89}$Laboratoire Lagrange, Universit\'e C\^ote d'Azur, Observatoire C\^ote d'Azur, CNRS, F-06304 Nice, France}
\address{$^{90}$NAVIER, \'{E}cole des Ponts, Univ Gustave Eiffel, CNRS, Marne-la-Vall\'{e}e, France}
\address{$^{91}$Universit\`a di Firenze, Sesto Fiorentino I-50019, Italy}
\address{$^{92}$INFN, Sezione di Roma Tre, I-00146 Roma, Italy}
\address{$^{93}$ESPCI, CNRS, F-75005 Paris, France}
\address{$^{94}$School of Physics Science and Engineering, Tongji University, Shanghai 200092, China}
\address{$^{95}$Dipartimento di Fisica, Universit\`a di Trieste, I-34127 Trieste, Italy}
% ...

%% file: abstract.tex
The Advanced Virgo detector has contributed with its data to the rapid growth
of the number of detected gravitational-wave (GW) signals in the past few years,
alongside the two Advanced LIGO instruments.
First during the last month of the Observation Run 2 (O2) in August 2017 (with,
most notably, the compact binary mergers GW170814 and GW170817), and then during
the full Observation Run 3 (O3): an 11-months data taking period, between April
2019 and March 2020, that led to the addition of 79 events to the catalog
of transient GW sources maintained by LIGO, Virgo and now KAGRA.
These discoveries and the manifold exploitation of the detected waveforms
benefit from an accurate characterization of the quality of the data, such as
continuous study and monitoring of the detector noise sources.
These activities, collectively named {\em detector characterization and data quality} or
{\em DetChar}, span the whole workflow of the Virgo data, from the instrument
front-end hardware to the final analyses.
They are described in detail in the following article, with a focus on the
results achieved by the Virgo DetChar group during the O3 run.
Concurrently, a companion article describes the tools that have been used by the Virgo
DetChar group to perform this work.

%% file: introduction.tex
A century after being predicted by Albert Einstein in the framework of general relativity, \acp{gw} have been detected by a global network of ground-based interferometric detectors~\cite{Abbott:2016blz}.
The LIGO~\cite{TheLIGOScientific:2014jea} and Virgo~\cite{TheVirgo:2014hva} collaborations, now joined by the KAGRA~\cite{10.1093/ptep/ptab018} collaboration, have observed in the past seven years dozens of \ac{gw} signals coming from merging compact binary systems.
Compact binaries composed of two \acp{bh}, two \acp{ns}, or both kinds of compact object have now all been observed.
GW150914~\cite{Abbott:2016blz}, the first \ac{gw} signal ever detected (at that time by the two Advanced LIGO detectors only) was a binary \ac{bh} merger.
Two years later, shortly after the \ac{adv} detector had started operating, the LIGO-Virgo 3-interferometer network detected the signal GW170817~\cite{TheLIGOScientific:2017qsa}, emitted by the fusion of two \acp{ns} and associated with counterparts in the entire electromagnetic spectrum, leading to the birth of multi-messenger astronomy with \acp{gw}.
More recently, LIGO, Virgo and KAGRA (in short ``LVK'') have announced the first detections of \ac{ns}-\ac{bh} mergers in data taken in January 2020~\cite{NSBHDiscovery}.

All these events add up in a \ac{gw} Transient Catalog whose first four versions ---  GWTC-1~\cite{LIGOScientific:2018mvr}, GWTC-2~\cite{Abbott:2020niy}, GWTC-2.1~\cite{GWTC2p1} and GWTC-3~\cite{GWTC3} ---  have been successively released.
Such catalogs allow scientists to analyze all detections globally: to probe the populations of compact stars, estimate the merger rate of binary systems, test general relativity in the strong-field regime, and perform searches for counterparts using archival data from other observatories.
The reconstructed \ac{gw} strain data --- the so-called $h(t)$ streams --- are regularly released in chunks of several months on the \ac{gwosc} website~\cite{GWOSC}.

Producing these results requires a thorough characterization of the data quality, a large part of which involves studying and monitoring the noise of \ac{gw} detectors.
This activity, some aspects of which are often referred to as {\em detector characterization} or {\em DetChar}, is an expertise which has been constructed over many years, starting more than two decades ago, first on simulated~\cite{PhysRevD.60.022001} or prototype~\cite{PhysRevD.60.021101} data, then with the initial detectors~\cite{Aasi:2012wd,LIGOScientific:2014qfs}.
Analysis methods and tools have been developed and implemented to characterize the Virgo data both with low latency\footnote{Throughout this article, the word {\em latency} is used to indicate the period between when an information becomes available (for instance some data quality assessment) and the time when the corresponding data have been acquired by the Virgo detector.} and offline.
The various analyses cover a physics-driven chain that starts from the raw data recorded by the instrument and extends all the way to the final set of \ac{gw} events and the related analyses.
The results of DetChar studies are used to improve the detector performances during the commissioning periods and to maximize the sensitivity to \ac{gw} signals during an {\em Observation Run}, when good-quality data are recorded.
They are also used to define the final Virgo dataset --- used by the LVK analyses and later published on the GWOSC website ---, and to vet all \ac{gw} candidates, found either in low-latency or offline.

This article reports the work of the Virgo DetChar group over the past few years. It will mainly focus on the activities carried out for the preparation of the third LIGO-Virgo observation run O3, from April 2019 to March 2020 as well as on the final results achieved after the run. The experience accumulated in view of the future runs of the LVK network will also be presented.
These achievements stem from the developments made before and during the O2 run (for Virgo: 25 days of data taking in August 2017) and those will also be described here when appropriate. 

Concurrently, a companion article~\cite{O3DetChar_tools} describes in detail all the tools the Virgo DetChar group has been relying on, in order to obtain the results presented here.

This article is organized as follows.
Section~\ref{section:AdV} provides an overview of the \ac{adv} detector configuration during the O3 run, preceded by a short summary of the path that led to this data taking period.
The same section also introduces notions and concepts that will be extensively used in the rest of the article, and defines a few related abbreviations.
Then, section~\ref{section:O3} summarizes the O3 run from a Virgo perspective: how the data taking was organized, what the performance of the detector and the final O3 dataset were.
Section~\ref{sec:onlinedq} presents the Virgo online data quality framework built for the O3 run.
Section~\ref{section:public_alerts} deals with the software developed to vet signal candidates to be released as public alerts to the astronomical community.
Section~\ref{section:dq_studies} presents the main DetChar analyses done on the O3 dataset to study noise transients, their impact on \ac{gw} searches, the noise spectrum, and the final validation of events.
Finally, section~\ref{section:outlook} provides some information about the ongoing preparation of the future O4 run, that is currently scheduled to begin in Spring 2023. A list of the main abbreviations used throughout the article is provided as well, for reference.

\begin{figure}[t!]
    \centering
    \includegraphics[trim=0 80 10 0, clip, width=\textwidth]{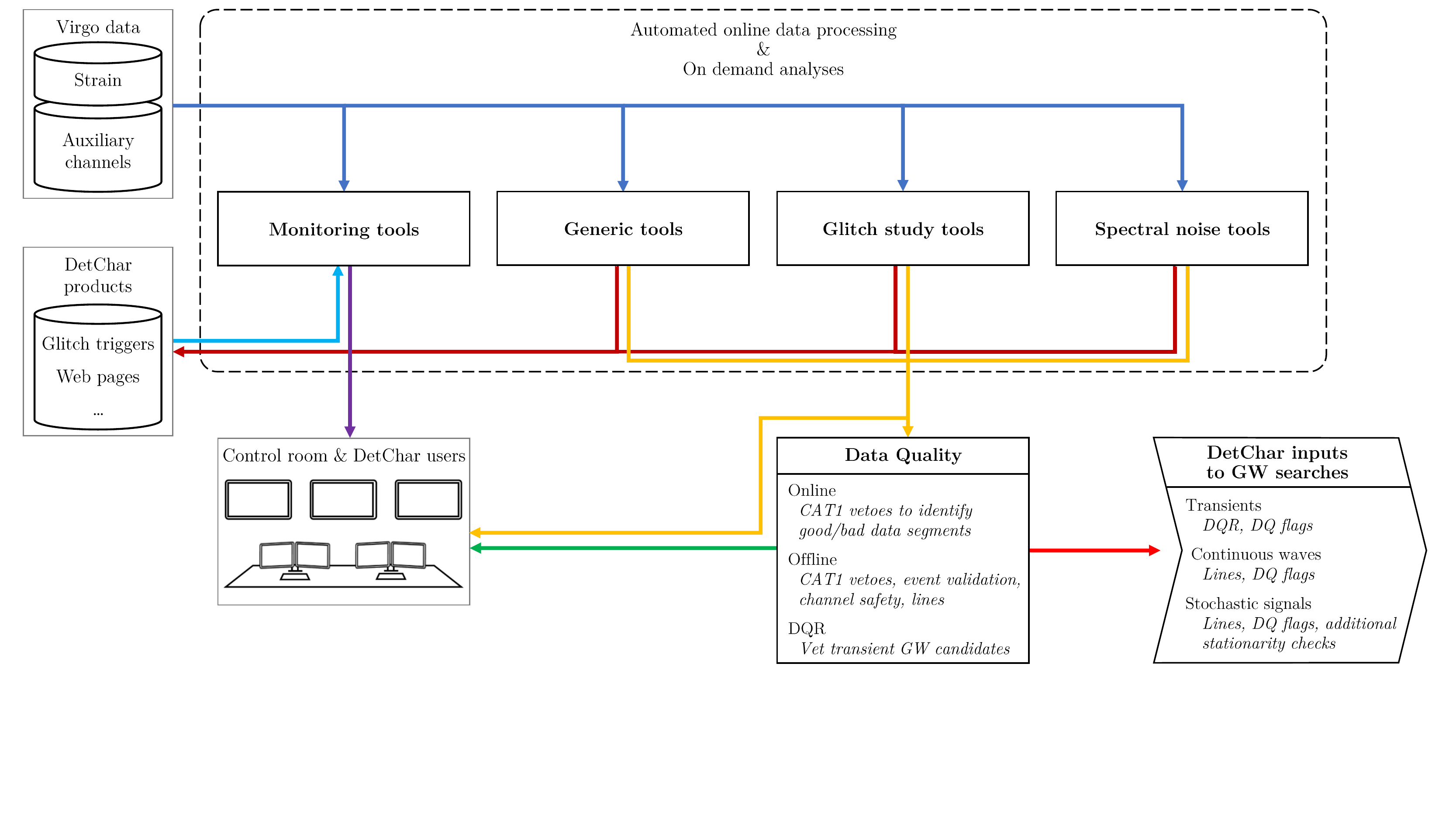}
    \caption{Flowchart of the Virgo DetChar tools and monitors described in~\cite{O3DetChar_tools}. The corresponding data quality (in short ``DQ'') products are used to monitor the detector and as inputs to \ac{gw} searches.}
    \label{fig:results_flowchart}
  \end{figure}

Finally, figure~\ref{fig:results_flowchart} (a simplified version of the flowchart used in Reference~\cite{O3DetChar_tools} to show an overall view of the DetChar tools) summarizes the Virgo data flow and describes how the Virgo DetChar products are combined to provide inputs to the \ac{gw} searches.

%% file: detector_intro.tex
This section focuses on the Virgo detector during the O3 run. First, we briefly review the main steps of the \ac{adv} project up to the beginning of O3. In particular, we emphasize its participation to the last four weeks of the O2 run in August 2017 that were rich of discoveries. Then, we summarize the activities during the 1.5 year-long shutdown between O2 and O3 that allowed the Virgo Collaboration to improve the instrument significantly. Finally, we describe the detector configuration during O3 and  present the main features of the data it has collected.

%% file: detector_o3path.tex
Virgo~\cite{2012JInst...7.3012A} is an interferometric detector of \acp{gw} located at the European Gravitational Observatory (EGO) in Cascina, Italy. The \ac{adv} project~\cite{TheVirgo:2014hva} allowed to upgrade the original instrument to a second-generation detector, similarly to what LIGO has done with its two interferometers~\cite{TheLIGOScientific:2014jea}, located in Hanford (WA, USA) and Livingston (LA, USA). The funding of \ac{adv} was approved in December 2009 by CNRS (France) and INFN (Italy), with an in-kind contribution from Nikhef (The Netherlands). The decommissioning of the first-generation Virgo detector started in Fall 2011, after the completion of the science run VSR4~\cite{PhysRevD.91.022004}, pursued together with the GEO600 detector~\cite{0264-9381-33-7-075009} (the Advanced LIGO upgrade project had already started).
The installation of the Advanced Virgo equipment started mid-2012 and was completed in 2016. The upgraded interferometer was robustly controlled in March 2017 and the next few months were dedicated to commissioning activities: noise hunting and sensitivity improvement. At the end of July, the detector was very stable and had a sensitivity corresponding to a \ac{bns} range\footnote{The \ac{bns} range is the average distance up to which the merger of a \ac{bns} system can be detected. The average is taken over the source location in the sky and the \ac{bns} system orientation, while a detection is defined as a signal-to-noise ratio of 8 or above} of ${\sim} 30$~Mpc, that is more than a factor two above the performance of the Virgo+ detector during the VSR4 run.

Therefore, \ac{adv} started taking data on August 1$^{st}$, 2017, joining the second Observing Run O2, which had started on November 30$^{th}$, 2016 for the two LIGO interferometers~\cite{0264-9381-32-7-074001}. 
On August 14$^{th}$, 2017, the \ac{adv} detector made its first detection of a \ac{gw}. That event, labeled GW170814~\cite{2017PhRvL.119n1101A}, was also recorded by the two LIGO interferometers. It was the first ever triple detection of a binary black hole coalescence, allowing an unprecedented accuracy in the localization of the source in the sky. A few days later, on August 17, the three interferometers jointly detected, for the first time, a \ac{gw} signal emitted by the coalescence of two neutron stars~\cite{TheLIGOScientific:2017qsa}. This event, known as GW170817, was accompanied by the almost simultaneous detection of a gamma-ray burst by the Fermi Gamma-ray and Integral space telescopes~\cite{GW-GRB}. The accuracy in the localization of the \ac{gw} source (approx. 30$\deg^2$) allowed to identify the optical counterpart in the galaxy NGC4993~\cite{MMA}. The O2 run ended on August 25$^{th}$, 2017.

The LIGO-Virgo shutdown between O2 and the third Observation Run O3, lasted 19 months.
On the Virgo side, it was divided into four periods:
\begin{itemize}
\item A post-O2 commissioning phase, until early December 2017. \\
The goal was twofold: to make a series of measurements on the O2 detector configuration that would have been too invasive during the run, and to perform some tests to try to further improve the instrument.
\item Hardware upgrades until mid-March 2018. Four main projects were pursued:
\begin{itemize}
\item {\bf The re-installation of the mirror suspensions and various vacuum upgrades.} \\
The steel wires, with which the \ac{adv} arm cavity mirrors were suspended for the O2 run, were replaced with quartz fibers in order to reduce the friction at the mirror-wire contact points --- a source of thermal noise. Fused silica ``monolithic''\footnote{The wires are welded to the mirrors and both are composed of the same fused silica.} suspensions, successfully tested in the Virgo+ configuration~\cite{doi:10.1063/1.1392338,art4-MSProc}, were foreseen in the \ac{adv} Technical Design Report~\cite{TDR}. Yet, multiple breakages of fused silica fibers when installed in vacuum were observed during Fall 2016, forcing the recourse of steel wires to preserve the participation of Virgo to the O2 run. The fiber breaking issue was eventually demonstrated to be caused by a spurious dust contamination generated by some vacuum pumps~\cite{Travasso_2018}. Therefore, the Virgo vacuum system was improved in order to avoid dust contamination, while the suspension fibers were shielded to prevent them from being hit by dust particles set in motion by ``air flows'' when acting on the vacuum system.
\item {\bf A higher laser power.} \\
The power of the laser injected into the interferometer was increased, reducing the photon shot noise that is limiting the high-frequency sensitivity: 10~W (19~W) were injected in Virgo during the O2 run (at the beginning of the O3 run).
\item {\bf The installation of a squeezed light source.} \\
This allows to further reduce the shot noise limit at high frequencies by modifying the quantum properties of the light coming out of the interferometer~\cite{PhysRevLett.123.231108}.
\item {\bf The test installation of an array of seismic sensors.} \\
An in-depth characterization of the seismic noise field at the test mass locations was performed in order to prepare for the subtraction of the Newtonian noise contribution that may limit the low-frequency sensitivity in the future~\cite{2015LRR....18....3H}.
\end{itemize}
\item A commissioning period, until Fall 2018, to improve the sensitivity and the duty cycle of the detector.
\item Finally, the transition phase to the O3 run, that officially started on April 1$^{st}$, 2019 at 15:00 UTC.
\end{itemize}

%% file: detector_o3config.tex
The \ac{adv} detector~\cite{TDR,AdVPlus} has been designed to achieve a sensitivity about one order of magnitude better compared to the initial Virgo detector, corresponding to an increase in the detection rate by about three orders of magnitude.
The \ac{adv} design choices were made on the basis of the outcome of the different research and development activities carried out within the \ac{gw} community and the experience gained with initial Virgo, while also taking into account budget and schedule constraints.

\begin{figure}
  \centering
  \includegraphics[width=0.95\textwidth]{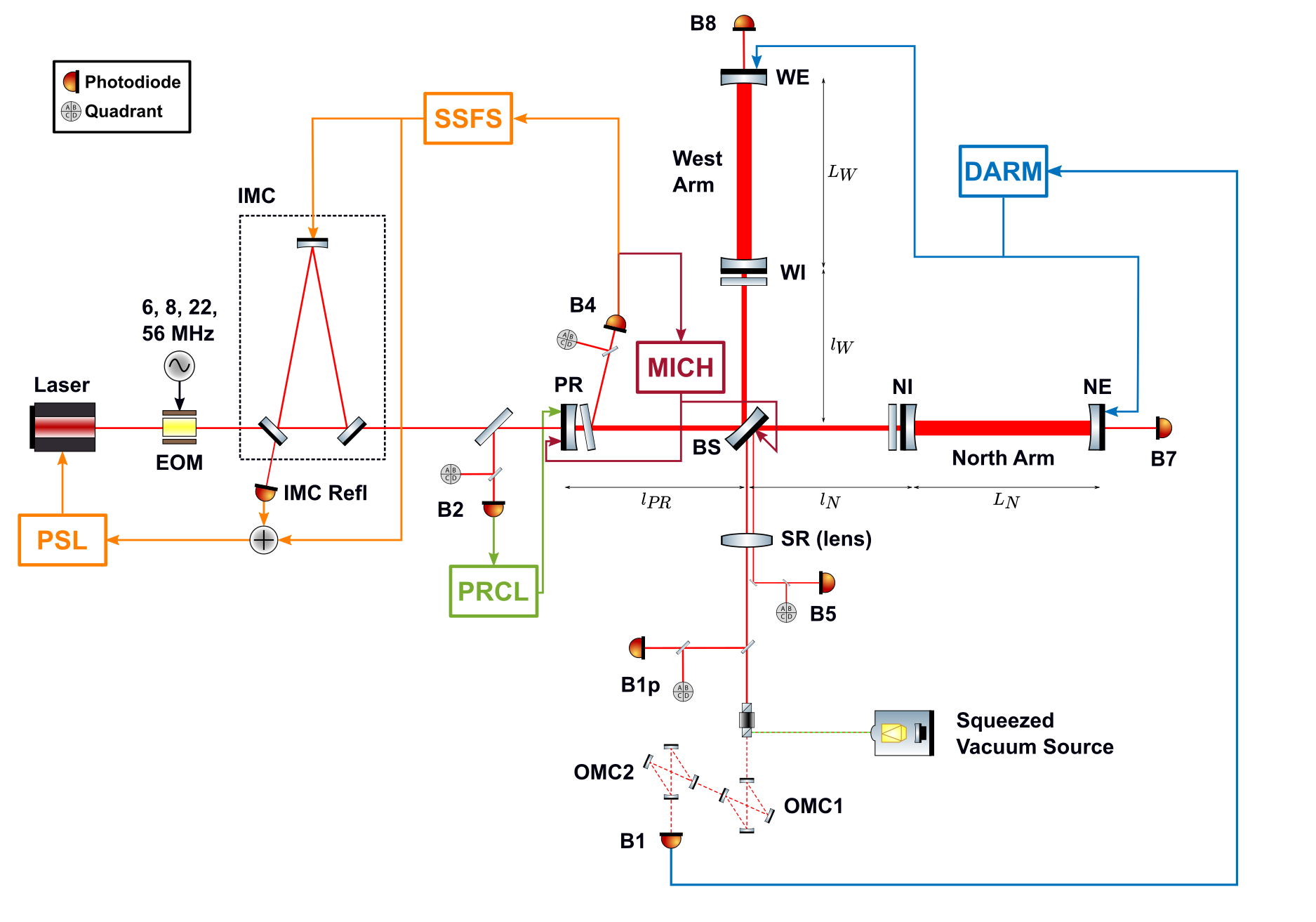}
  \caption{Schematics of the \ac{adv} configuration during the O3 run (not to scale), showing optics, photodiodes and quadrant photodiodes, such as the main components of the global feedback system used to steer the detector. The suspended optical benches introduced in the text are not represented here. This figure is taken from~\cite{galaxies8040085}.}
  \label{fig:AdVsetup}
\end{figure}

The simplified optical schematic of \ac{adv} during the O2 and O3 runs is shown in figure~\ref{fig:AdVsetup}. In the following, we briefly outline the different parts of the detector layout and define the main abbreviations that are labeled on the schematic or used later in the article. Further information about the O3 configuration and control system of the Virgo detector can be found in~\cite{galaxies8040085}.

The Virgo power-stabilized infrared laser beam (PSL, wavelength: 1.064~$\mu$m) is filtered at the interferometer input by a 144~m triangular cavity called the \ac{imc}. The two flat mirrors of the \ac{imc} are located on the first suspended injection bench (SIB1), that also hosts various optics for beam matching\footnote{Making the laser beam wavefront matching the one resonating in the arm cavities.}. Then, the beam goes through the partially reflective \ac{pr} mirror  before being split into two perpendicular beams at the \ac{bs} mirror. The two 3~km-long arms hosting Fabry-Perot cavities are called ``North'' and ``West'' as they are roughly oriented along these geographical directions. The cavity mirrors closest (furthest away) from the \ac{bs} are called ``input'' (``end'') mirrors. So, following these conventions, the test masses (the four mirrors forming the two 3-km long Fabry-Perot cavities) are labeled \ac{ni}, \ac{ne}, \ac{wi} and \ac{we}, respectively. Both arms end with a suspended terminal bench --- called \ac{sneb} or \ac{sweb} --- hosting a photodiode (B7 or B8) receiving the cavity transmitted beam. More generally, most optical components and sensors are located on suspended benches (not displayed on the schematic) to reduce the impact of the residual seismic motion. After propagation and storage in the kilometric cavities, the arm beams recombine on the \ac{bs} and the beam resulting from this interference goes to the interferometer output port.
As indicated in figure~\ref{fig:AdVsetup}, the location of the foreseen \ac{sr} mirror was occupied by the first lens of the detection system during the O3 run (and during O2 as well)\footnote{The installation of that additional mirror only took place during the shutdown period that followed the end of O3.}.
The beam from the frequency-independent squeezed light source~\cite{PhysRevLett.123.231108} enters the detector between the \ac{sr} lens and the interferometer output. Finally, prior to being detected on the B1 photodiode located on the suspended detection bench 2 (SDB2), the output port beam is filtered in sequence by two \ac{omc} cavities, \ac{omc}1 and \ac{omc}2, located on the suspended detection bench 1 (SDB1).

A complex active feedback system, made of several automated control feedback loops, is necessary to bring and maintain the detector at its global working point. In particular, it aims at controlling the four main longitudinal \acp{dof} of the \ac{adv} detector which, in its O2-O3 configuration (see figure~\ref{fig:AdVsetup} for the definition of the different lengths used below), are:

\begin{itemize}
\item The \ac{mich}, $l_N - l_W$, sets the destructive interference (``dark fringe'') optimal condition.
\item The \ac{prcl}, $l_{PR} + (l_N + l_W) / 2$, must be resonant.
\item The lengths of the kilometric Fabry-Perot cavities, $L_N$ and $L_W$, must be resonant as well, or rather their average and difference that are more physical:
    \begin{itemize}
    \item The \ac{carm}, $(L_N + L_W) / 2$, used as a length etalon by the \ac{ssfs} to stabilize further the frequency of the input laser.
    \item The \ac{darm}, $L_N - L_W$, the quantity sensitive to a passing \ac{gw}.
    \end{itemize}
\end{itemize}

This global control relies on radio-frequency sidebands for the carrier beam that are generated by the \ac{eom} located in between the laser source and the \ac{imc} on Figure~\ref{fig:AdVsetup}. The 6, 8 and 56~MHz sidebands\footnote{The exact frequencies of the different sidebands are provided for instance in reference~\cite{TDR}; they have been chosen to make these sidebands resonant / anti-resonant / non-resonant in the different optical cavities, depending on their usage by the Virgo global control system.} are used to control the interferometer, while the 22~MHz one is used to control the injection system.

%% file: detector_data_detchar.tex
The \ac{gw} strain data stream $h(t)$ reconstructed at the Virgo detector is dominated by noise with, up-to-now, rare and weak \ac{gw} signals. The noise contributions can be roughly classified into two main categories:

\begin{itemize}
\item Fundamental noises, that are inherent to the instrument and represent the ultimate limit of its sensitivity.
Their combined contribution is overall stationary and Gaussian~\cite{Aasi:2012wd}, two properties that are thoroughly verified as part of the DetChar activities, and in particular in correspondence of candidate \ac{gw} events~\cite{O3DetChar_tools}.
\item Various noise artifacts, whose origins are manifold (hardware components of the detector, feedback control loops, interaction with the external environment, etc.) and that represent potential issues, not only because they may impact the running of the instrument but also --- and above all --- because they show up in the background of searches for \ac{gw} signals, thus limiting their sensitivity. Noise transients, called {\em glitches}, can either look like real signals or overlap in time with one, either impairing its detection or confusing the inference of its source parameters. These glitches are monitored and studied with time-frequency representations that are used to classify their numerous signatures into families and separate them from real \ac{gw} events. In addition, long-lasting noise excesses, also called {\it spectral noises}, are also seen around particular frequencies (power main frequency and its harmonics, suspension resonating modes, etc.): the narrow ones, (nearly) monochromatic, are called {\em lines} and the wider ones {\em bumps}.
Both can manifest themselves in several ``flavours''. For instance, lines can exist individually, but sometimes appear as {\it combs}, that is families of lines separated by a constant frequency interval. They are typically due to processes with a strict time periodicity, like electronic clock signals. Bumps may have some specific structure, depending on the source. Both lines and bumps can exhibit structures symmetric around their main frequency, called {\it sidebands}, that are due to non-linear interactions among different disturbances. Moreover, spectral noises can be persistent across a full run, or only be present in a portion of it.

 Both the glitch rate in a particular frequency band and the properties (amplitude, peak frequency and bandwidth) of spectral noises can vary in time, to reflect changes occurring at the level of the detector or its environment.
\end{itemize}

To allow investigating these variations, hundreds of {\em auxiliary channels} are acquired by the Virgo \ac{daq}, providing both a detailed status of the detector control systems and a complete monitoring of the local environment~\cite{EnvHuntVirgoO3,o3virgoenv}.
Integer GPS ranges used to flag data with common properties (data quality level, particular detector condition, etc.) are called {\em segments} in the following.

\input{tools_intro}

%% file: tools_intro.tex
\begin{figure}
  \centering
  \includegraphics[width=\textwidth]{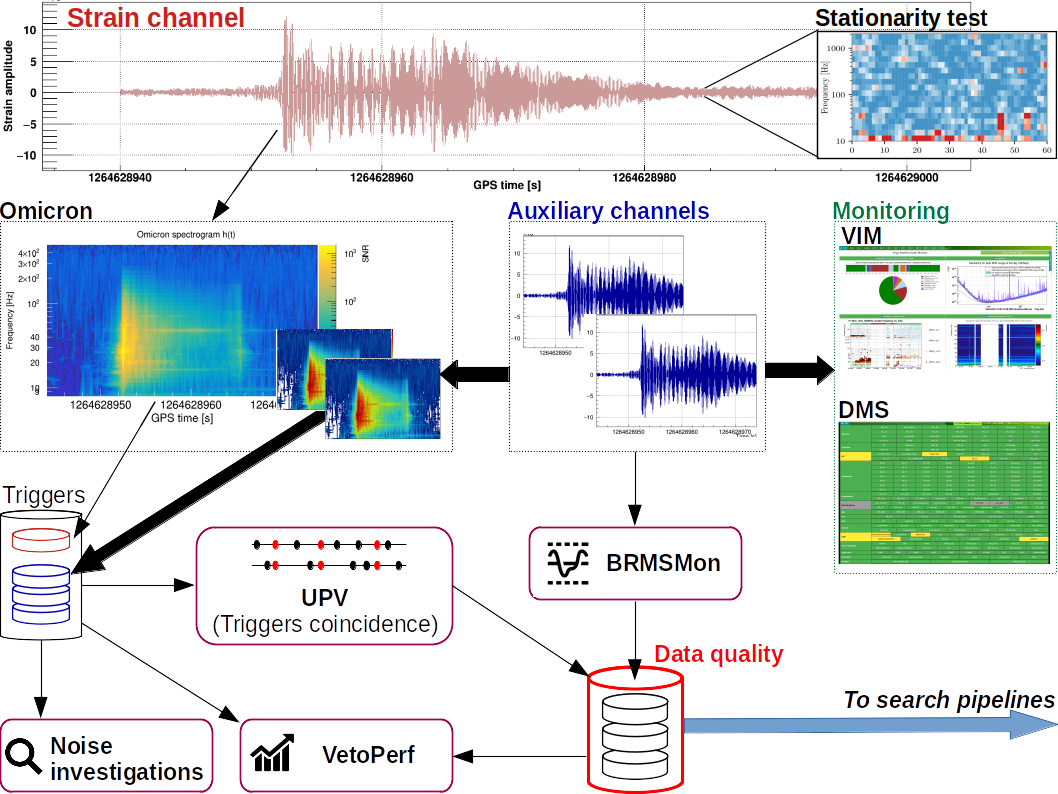}
  \caption{Generic workflow of transient noise studies in Virgo, using the full set of DetChar and monitoring tools described in Reference~\cite{O3DetChar_tools}.}
  \label{fig:glitch_flowchart}  
\end{figure}

The Virgo \ac{gw} strain data $h(t)$ and the many associated auxiliary channels are analysed by a wide set of DetChar tools described in detail in Reference~\cite{O3DetChar_tools}. As an example, figure~\ref{fig:glitch_flowchart} describes the joint application of various DetChar analysis and monitoring tools to the study of transient noise. This flowchart focuses specifically on how these tools are used and complement each other to investigate transient noise. \texttt{Omicron} is the main tool to identify glitches in all the relevant \ac{daq} channels. Those glitch triggers are stored on disk and mined by the \ac{upv} tool, to look for coincidences between them, allowing to assign confidently a terrestrial origin to a fraction of these triggers. \texttt{VetoPerf} monitors the performance of these tools to find an optimum balance between the fraction of glitches flagged as non-astrophysical and the amount of data removed by these associations. Other tools like \texttt{BRMSMon} look for patterns in the data that are known to be due to noise. All these inputs are used to trigger further noise investigations and are gathered to allow a global assessment of the quality of the data. The hypothesis of stationary (and Gaussian) data are also tested around all \ac{gw} candidates as they are basic assumptions for the algorithms searching for \acp{gw} in the data, whether they are run in real-time or offline. Parallel to this dataflow, dedicated monitoring tools like the \ac{dms} and the \ac{vim} continuously provide information about the status of all the Virgo components, from the hardware blocks to the online software processes.

%% file: run_virgo_intro.tex
The joint LIGO-Virgo Observing Run 3 O3,  has been divided into two consecutive sub-data-taking periods, separated by a one-month commissioning break in October 2019:
\begin{itemize}
\item O3a: from April 1$^{st}$, 2019 at 15:00 UTC (GPS: 1238166018), to October 1$^{st}$, 2019 at 15:00 UTC (GPS: 1253977218).
\item O3b: from November 1$^{st}$, 2019 at 15:00 UTC (GPS: 1256655618), to March 27$^{th}$, 2020 at 17:00 UTC (GPS: 1269363618).
\end{itemize}
All three detectors have participated to the whole run. The O3b end date has been moved forward by more than a month, due to the worldwide Covid-19 pandemic.

This section presents the LIGO-Virgo O3 run, seen from a Virgo perspective. First, we describe the main activities into which the data taking was divided, before summarizing how the detector was steered from the EGO control room. Then, we focus on actions taken to maximize the amount of data collected and to ensure their good quality. In particular, we highlight the main DetChar activities during O3, explaining how they fit and complement each other, following the flow of data from the detector to the final analyses. Key to achieve this level of performance and to maintain it over almost a year, were the 24/7 on-call duty service and the \ac{rrt}: both will be briefly described as well.

Then, we review the performance of the Virgo detector during O3, mainly from the point of view of the duty cycle. A high duty cycle requires not only a stable and robust detector against external disturbances (see~\cite{o3virgoenv} for a comprehensive study of that topic) but also a quick and reliable procedure to bring the instrument to its working point, starting from an uncontrolled global state. The main statistics of the Virgo O3 global control acquisition are thus provided, before studying the actual duty cycle. We also present the evolution of the \ac{adv} detector sensitivity, from the O2 run to the end of O3.

This section ends with a brief overview of the final Virgo O3 dataset, describing how it was constructed offline, building upon the preliminary dataset established by the live monitoring and data quality checks.

%% file: run_virgo_organization.tex
While data acquisition was the highest priority during the O3 run, a limited fraction of the time had to be dedicated to other activities. The two main recurring ones were:
\begin{itemize}
\item The maintenance periods, held every Tuesday morning, staggered with respect to the similar times in LIGO, in order to maximize the two-detector network coverage. Maintenance, limited to about 4 hours per week, was used to look after the detector components, to perform various cleaning actions, and to host noisy activities incompatible with data taking ---  for instance the refilling of liquid nitrogen tanks located nearby the \ac{ceb}, \ac{neb} and \ac{web}, delivered by heavy trucks.
\item The calibration shifts, held almost every week on Wednesday afternoons or evenings. These campaigns allowed to check the accuracy of the reconstruction of the $h(t)$ stream~\cite{VIRGO:2021umk}, to monitor its stability over time and to test new, complimentary calibration methods, like the use of a Newtonian calibration system~\cite{Estevez_2021_2} in addition to the usual photon calibrators~\cite{Estevez_2021_1}.
\end{itemize}

In addition, commissioning time was allocated irregularly to tune or optimize some aspects of the detector, depending on the needs and opportunities. Finally, some time was spent studying and fixing problems impacting the data taking.

%% file: run_virgo_datataking.tex
The Virgo data taking is largely automated and usually only requires a single operator on duty in the control room. Operators are present 24/7 during a run and take shifts every 8 hr.

The \ac{adv} detector automation, called \texttt{Metatron}, relies on the Guardian~\cite{Rollins:2016hlk,guardian} framework, developed by LIGO and based on hierarchical finite state machines. The Virgo implementation links this framework to the \ac{daq}: automation nodes become \ac{daq} nodes that get data directly from shared memories and are synchronized with a 1~s data availability period. A generic mechanism to read and write \ac{daq} channels has been introduced and can be used within user codes via dedicated functions.

\begin{figure}[htb!]
  \centering
  \includegraphics[width=\textwidth]{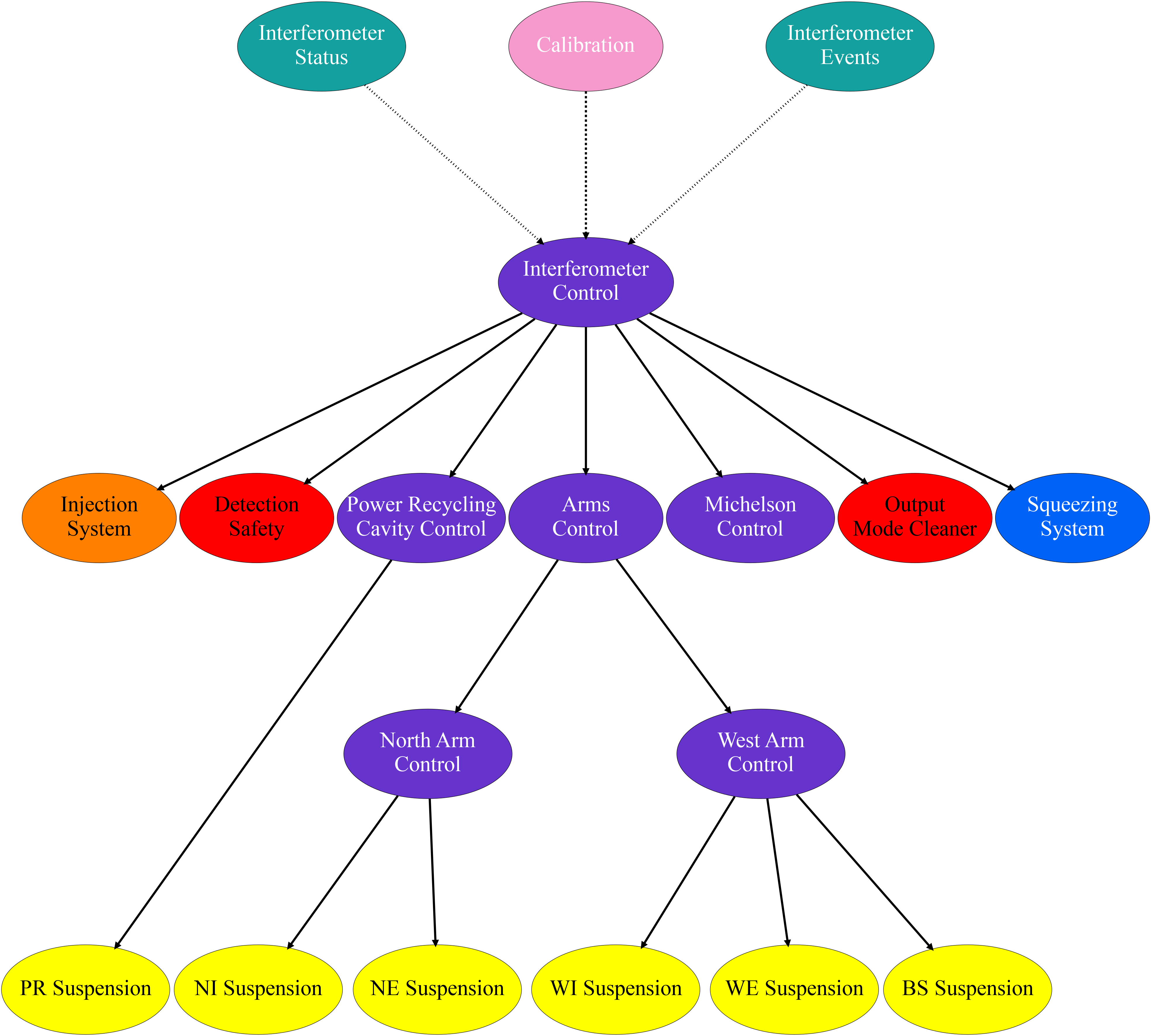}
  \caption{\texttt{Metatron} nodes hierarchy used during the O3 run.}
  \label{fig:metatron_hierarchy}
\end{figure}

The full Virgo control acquisition procedure has been implemented in \texttt{Metatron}, initially prior to the O2 run and then updated for the O3 configuration, the main difference being the addition of the frequency-independent squeezing~\cite{PhysRevLett.123.231108}. The scheme adopted, depicted in figure~\ref{fig:metatron_hierarchy}, strictly follows a top-down approach, with the lower-level nodes being automatically managed by higher-level ones.

The suspension nodes (yellow background in the graph) are tasked to align/misalign the Virgo optics, each of them is managed by the most appropriate control node (purple background), divided on the basis of the degrees of freedom to be controlled. The main node --- \texttt{Interferometer Control} ---  is usually the only one operated manually to steer the detector. It defines the control paths, such as for instance the main global control procedure that allows reaching the Science mode (the nominal data taking state), plus other procedures to control various configurations of the optics, or to perform automated calibrations. It relies on the underlying managed nodes to perform these actions on the instrument. During the final steps of the control procedure, each single part of the interferometer is ultimately entangled with the others, and the interferometer is naturally treated as a single system. For these reasons, the last part of the procedure is directly managed by the upper level node, which sets the control parameters to the whole system, while the lower level nodes are only used as watchdogs for the correct functioning of their own sub-systems.

Additionally, the \texttt{Metatron} main node manages:
\begin{itemize}
\item The injection system, from the laser source to the \ac{imc} (\texttt{Metatron} node \texttt{Injection System}, orange background);
\item The two \acp{omc}, which are controlled in sequence in the final steps of the nominal control acquisition procedure (\texttt{Metatron} node \texttt{Output Mode Cleaner}, red background);
\item The detection system at the interferometer output port (\texttt{Metatron} node \texttt{Detection Safety}, red background); 
\item The frequency-independent squeezing system (\texttt{Metatron} node \texttt{Squeezing System}, blue background), whose control proceeds in parallel to the one followed for the main detector. As Virgo can take valid Science data with or without this system being in its nominal state, the corresponding \texttt{Metatron} node is a bit apart from the others logic-wise.
\end{itemize}

Only during the calibration measurements, the \texttt{Interferometer Control} node is automatically managed by the \texttt{Calibration} node (pink background). 

The \texttt{Metatron} framework also takes care of generating high-level flags that provide the overall status of the interferometer: this is done within the \texttt{Interferometer Status} node (green background). Finally, the \texttt{Interferometer Events} node (green background) records all state transitions of the detector. Information from these last two nodes is passed onto the Virgo live monitoring system, documented in Reference~\cite{O3DetChar_tools}.

%% file: run_virgo_shifts.tex
\begin{figure}
  \center
  \includegraphics[width=\textwidth]{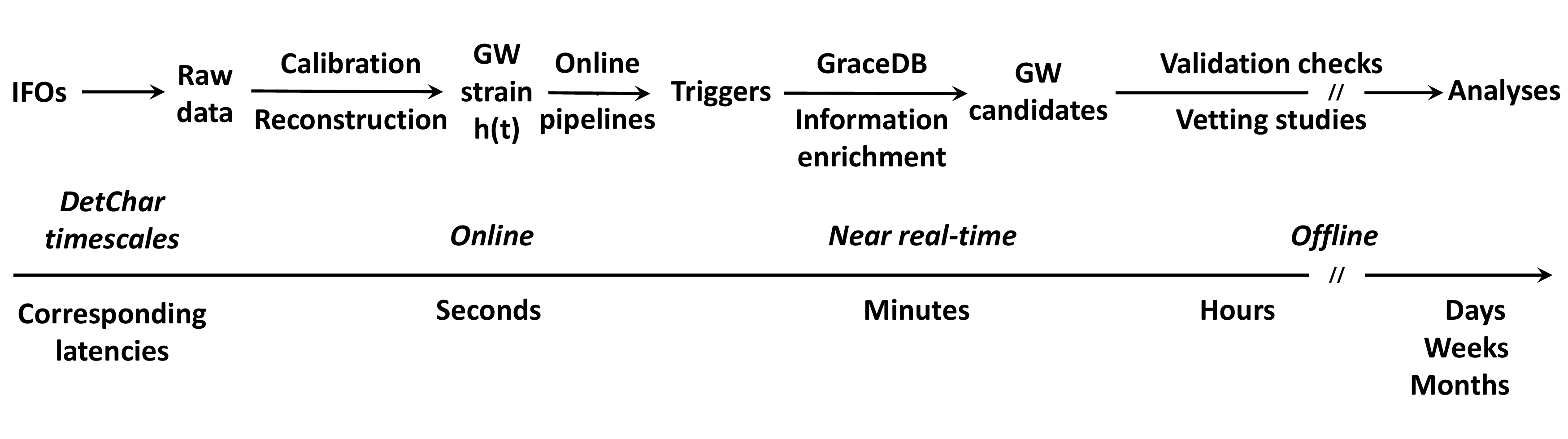}
  \caption{Dataflow from the interferometers (labeled "IFOs" on the left) to the offline validation of \ac{gw} candidates and the completion of the final dataset (right). It focuses on the generation and the vetting of the public alerts that are a key product of the LIGO-Virgo observing runs. It shows the three main timescales at which the Virgo DetChar group operates: online, near real-time and offline (see text for details).}
  \label{fig:DataflowDetChar}  
\end{figure}

Figure~\ref{fig:DataflowDetChar} shows the flow of data, from the interferometers (IFOs, on the left), to the physics analyses (on the right). While focusing on the \ac{gw} candidates, this schematic highlights the three main pillars of DetChar activities during a run:

\begin{itemize}
\item The first timescale on which DetChar activities take place is online (latency: $\mathcal{O}(\textrm{s})$). Quick automated checks are run on live data to mark out (quality: good or bad) the data stream used as input by the ``pipelines'' --- that is the algorithms that scan the network data in real time, as soon as they become available. Initial data quality information is indeed shipped alongside the reconstructed \ac{gw} stream, as explained in section~\ref{sec:onlinedq}.
\item The second timescale is near real-time (latency: $\mathcal{O}(\textrm{min})$), crucial to assess the quality of the \ac{gw} candidate public alerts. Thanks to a dedicated framework that is described in section~\ref{section:public_alerts}, the data around a significant candidate are vet for each detector and a global decision is then taken: either to confirm the public alert sent to the telescopes or to retract it (see section~\ref{section:oncall_RRT} below for a description of the procedure).
\item Finally, the last timescale is offline (much higher latency: up to months after the data taking). The goals of these studies are twofold: first, to finalize the dataset that all offline analyses will use, regardless of whether they look for transient or continuous signals; then, to validate the events that will be included in the final publications and whose parameters will be used to extract astrophysical information.
\end{itemize}

To ensure a continuous monitoring of the data quality, DetChar shifts were organized during the entire O3 run on a weekly basis, with two people (working onsite or remotely) on duty. The shifter crew changed every Tuesday morning, during the weekly maintenance of the Virgo detector. In addition to attending all relevant meetings, DetChar shifters usually reported their findings at the weekly DetChar meeting on Fridays and at the weekly detector meeting on Tuesdays (thus at the end of their weekly shift).

%% file: run_virgo_oncall_RRT.tex
An on-call service was organized during the O3 run to ensure a 24/7 expert coverage for all the Virgo detector components, from hardware systems to online computing and DetChar. In case of a problem, the operator on duty would contact the relevant experts from the control room, as well as the data taking coordinators if needed.

In addition, a joint LIGO-Virgo low-latency automated alert system was setup to contact the \ac{rrt} experts --- specialists of data taking, data quality or \ac{gw} transient searches --- who would meet remotely on short notice each time a public alert candidate was identified in real time. They would vet that candidate, using all raw information available, plus the output of several data quality checks, triggered automatically by the generation of the signal candidate: the \ac{dqr} (see section~\ref{subsection:DQR} for details). The outcome of an \ac{rrt} meeting could be twofold: either to confirm the public alert, or to retract it when the astrophysical origin of the candidate was questionable.

%% file: tools_noise_budget.tex
The noise budget compares the measured detector sensitivity with the incoherent sum of all known noise contributions. Each noise projection depends on the noise level, as measured by external probes, and of its coupling to the strain channel $h(t)$, that is estimated by dedicated measurements called noise injections~\cite{EnvHuntVirgoO3}.

The \ac{adv} noise budget is based on the \texttt{SimulinkNb}
\cite{SimulinkNb_git} software package. It includes a complete model
of the four main longitudinal \acp{dof} of the interferometer
(\ac{darm}, \ac{carm}, \ac{mich}, \ac{prcl}), with the interferometer optical response
simulated using \texttt{Optickle}~\cite{optickle_git}. The mirror
suspensions are approximated by a double pendulum state space model of
the mirror and marionette
(the steel body to which the mirror is suspended, a component of the Virgo suspension's last stage, called payload~\cite{art2-PayloadProceeding}).
 It also includes the feedback response measured from the
transfer function between the photodiode signal and the mirror and
marionette corrections. This approach allows to simply add different
noise sources at their physical entry into the interferometer control
loop, and also includes the expected cross couplings between the
longitudinal \acp{dof}.

\begin{figure}
  \center
\includegraphics[width=\textwidth]{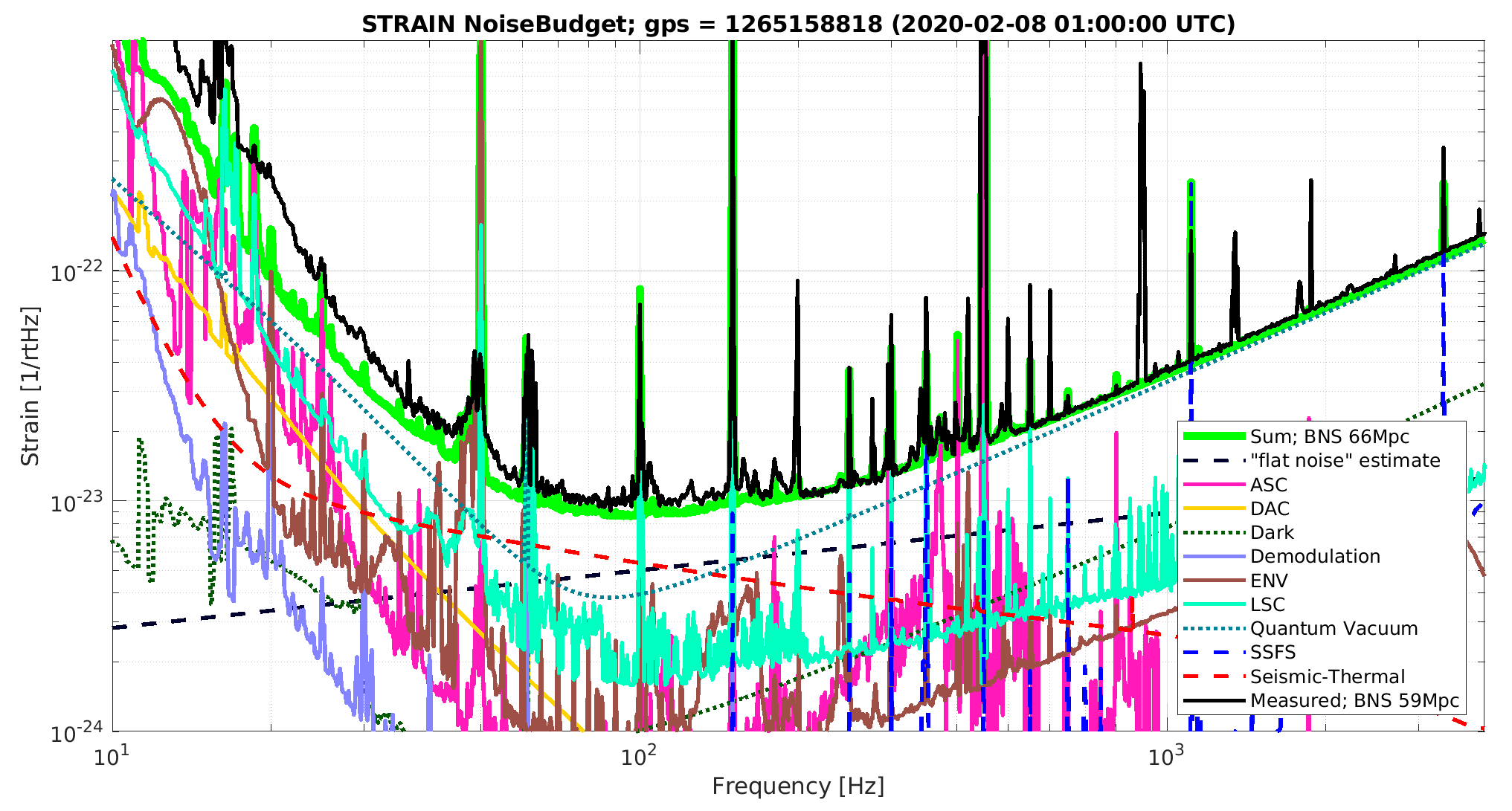}
\caption{Snapshot of the \ac{adv} O3 noise budget generated at a time
    of near best sensitivity of the detector (February 8$^{th}$, 2020). The different noise
    sources shown are described in the text. The green line (\ac{bns} range: 66~Mpc) represents the
    sum of these noises: it can be compared to the measured total
    noise shown in black (\ac{bns} range: 59~Mpc).}
 \label{fig:noise_budget_O3b}  
\end{figure}

This model has been verified to match the measured open loop transfer
functions of the four modeled \acp{dof}, and to reproduce the
interferometer strain data calibration with errors smaller than
10\%. In total, more than 100 noise sources are taken into
account, and the total of those noises is summarized in
figure~\ref{fig:noise_budget_O3b}. 

The noises are summed in log-spaced
frequency bins, which allows resolving narrow lines at low
frequencies and a low statistical error on the broadband noise
estimation at high frequencies. The noises taken into account are the following:

\begin{description}
\item[ASC] -- Angular Sensing and Control. This represents the control
  noise of 12 angular \acp{dof} of the interferometer (two per
  mirror) and four \acp{dof} of the beam injected into the
  interferometer. The coupling of these noises has been measured by
  injecting broadband noise into each \ac{dof}~\cite{thesisBader}.
\item[DAC] -- Digital Analog Converter. This is the electronic noise
  of the digital to analog converters used to drive the six main mirrors
  and marionettes of the interferometer. This electronic noise has
  been measured in the laboratory before installation, and the
  noise coupling is modeled using \texttt{SimulinkNb}.
\item[Dark.] This is the electronic and dark noise of the photodiodes used in the four
  longitudinal \acp{dof} control. The dark noise of each photodiode is measured by closing
  the mechanical shutter in front of it. The noise coupling is part of the model
  of the longitudinal control loops in \texttt{SimulinkNb}.
\item[Demodulation.] This is the phase noise of the demodulation of
  radio-frequency signals from photodiodes to control \ac{carm}, \ac{mich} and
  \ac{prcl}. That phase noise mixes the two demodulation quadratures. This
  bi-linear noise source is measured, and the noise coupling is
  modeled using \texttt{SimulinkNb}.
\item[ENV] -- Environment. This is the sum of three contributions:
  acoustic, magnetic and scattered light. The acoustic and magnetic
  noises are measured with four microphones and three 3-axis
  magnetometers, located in the experimental buildings near the interferometer
  components (see~\cite{EnvHuntVirgoO3,o3virgoenv} for details).
  Their couplings are measured by broadband and sweeping
  sine noise injections. Scattered light noise is projected in two ways:
  i) using the measured relative intensity noise on auxiliary photodiodes
  and measuring a linear coupling by shaking the bench hosting that photodiode
  to elevate the noise in the detector sensitivity; ii) using position
  sensors of suspended benches that couple in a non-linear way, with a modeled
  coupling that is scaled based on measurement obtained by displacing
  intentionally the bench by tens of microns per second to elevate the noise
  in the detector sensitivity~\cite{Was2021}.
\item[LSC] -- Length Sensing and Control. This represents the control
  noise of four \acp{dof}: \ac{mich}, \ac{prcl}, \ac{omc} length, and residual intensity
  noise. The noise is measured in all cases, the coupling is measured
  for all except for the \ac{omc} length where it is modeled. Note that this
  results in double counting the dark and quantum noise of the sensors
  used for \ac{mich} and \ac{prcl} control, however these double counted
  contributions are negligible.
\item[Quantum.] Quantum noise of the detector and shot noise of the
  sensors used for \ac{mich}, \ac{prcl} and \ac{carm} control. The noise and the
  coupling are modeled using \texttt{SimulinkNb}.
\item[\ac{ssfs}.] This represents the
  control noise of the relative error between \ac{carm} and the laser
  wavelength. The noise is measured, the  frequency dependent coupling
  is modeled using \texttt{SimulinkNb} and a time dependent scaling factor is measured.  
\item[Seismic-Thermal.] This is the sum of the negligible seismic noise
  and three thermal noise contributions: suspension, mirror coatings and
  residual gas pressure in the arm vacuum tubes. The noise sources and
  the couplings are modeled using analytical functions in separate dedicated codes.
\item[``flat noise''.] It is a noise source of not yet understood physical origin. Its level has been measured proportional to the
  square root of the \ac{darm} offset used to obtain the interferometer DC
  readout~\cite{Hild2009, Fricke2011}. 
\end{description}

The sum of the noises described above correspond to a \ac{bns} range of
66~Mpc, while the actual \ac{bns} range in the corresponding
data was measured at 59~Mpc. Hence, about 10\% of the noise limiting \ac{bns} detections
is unaccounted for, not understood and not described in this section.

More in detail, at frequencies above 1~kHz the sensitivity is mostly
limited by quantum shot noise. The measured level is about 5\% higher
than expected. This is due to a slow degradation of the frequency-independent light squeezing during O3,
from 3~dB at the beginning of the run to about 2.5~dB at the end of it.

In the most sensitive frequency range, between 80~Hz and 200~Hz, there
are significant contributions from three sources: quantum shot noise,
mirror coating thermal noise and the ``flat noise'' of unknown physical
origin. Assuming that the ``flat noise'' estimate is correct,
removing completely this unknown noise source would have resulted in
10~Mpc improvement in the \ac{bns} range.

At low frequencies between 20~Hz and 50~Hz, the dominant noise
sources are quantum radiation pressure noise that is increased by the
frequency independent light squeezing and the laser intensity noise. However,
30\% of the noise remains not understood in that frequency range, so
other significant noise sources are yet to be identified.

%% file: O3_perf.tex
Table~\ref{table:O3_locking} summarizes the performance of the global control acquisition procedure for the Virgo detector during O3. This performance has been stable over the whole run, showing the robustness of that procedure. As not all control acquisition {\em attempts} are successful, a global control acquisition {\em sequence} is defined as a set of successive control attempts that leads to the global control of the instrument.

The dataset analysed here spans the whole O3 run and includes more than 700 successful global control acquisition sequences. Only the periods during which detector activities incompatible with data taking (maintenance, commissioning, calibration and known hardware problems) were ongoing have been excluded. In order to be less sensitive to the tails of the statistical distributions --- which do impact the duty cycle (see the 'Locking' contributions to the pie charts below) but can have multiple origins (human errors, hardware or software failures possibly hard to diagnose quickly, or external conditions like bad weather) which are not directly related to the global control acquisition procedure -- we have decided to report median durations in the following. 

The median duration of a successful global control  acquisition attempt is 18~min: 30\% of this time is spent reaching the detector working point (Michelson interferometer at the dark fringe, power recycling cavity and arm cavities resonant, \ac{ssfs} enabled); 50\% is spent to control the two \acp{omc} at the Virgo output port; the final 20\% are used to reach the lowest noise configuration at the level of the suspension actuation. The median number of attempts needed to complete a global control sequence is 2 and the median duration of a successful global control acquisition sequence is 25~min. During O3 the quickest sequence took however 13~min.

\begin{table}[htbp!]
\centering
\caption{\label{table:O3_locking}Summary of the Virgo global control acquisition performance during O3: the control is acquired after a successful control acquisition {\it sequence} that counts one or more control acquisition {\it attempts}.}
\begin{tabular}{cr|c}
\toprule
\multicolumn{3}{c}{\textbf{Global control acquisition attempt}} \\\midrule
~ & Median duration & 18 minutes \\
\hline ~ & Distribution of this time & ~ \\
~ & Reaching the detector working point & ${\sim}$30\% \\
~ & Controlling the two \acp{omc} & ${\sim}50$\% \\
~ & Acquiring the lowest noise configuration & ${\sim}20$\% \\
\bottomrule
\multicolumn{3}{c}{\textbf{Global control acquisition sequence}} \\\midrule
~ & Median number of attempts & 2 \\
~ & Median duration & 25~minutes \\
\bottomrule
\end{tabular}

\end{table}

\begin{table}[htbp!]
\caption{\label{table:O3_perf}Summary of the O3 data taking performance of the Virgo detector. The last three rows of the table provide duty cycles for different configurations of the 3-detector LIGO-Virgo global network: the fraction of the time during which at least one the three instruments is taking data, at least two are and finally all three are.}
\centering
\begin{tabular}{lr|ccc}
\toprule
~ & ~ & O3a & O3b & O3 \\ \midrule
\multirow{2}{*}{Virgo global control segments} & Mean [hr]   & 6.1 & 6.4 & 6.3 \\
 & Median [hr] & 2.7 & 1.8 & 2.2 \\ \midrule
\multirow{2}{*}{Virgo Science segments} & Mean [hr] & 5.0 & 4.0 & 4.5 \\
 & Median [hr] & 2.6 & 1.4 & 1.9 \\ \midrule
\multirow{4}{*}{Duty cycles} & Virgo [\%] & 76.3 & 75.6 & 76.0 \\
 & Network --- at least 1/3 [\%] & 96.8 & 96.6 & 96.7 \\
 & Network --- at least 2/3 [\%] & 81.9 & 85.4 & 83.4 \\
 & Network --- 3/3 [\%]          & 44.5 & 51.0 & 47.4 \\
 \bottomrule
\end{tabular}

\end{table}

Table~\ref{table:O3_perf} details the control stability of the Virgo detector, separately for the sub-runs O3a and O3b, and averaged over the whole O3 run. The ``global control segments'' are stretches of data during which Virgo is controlled in its nominal low-noise configuration, while, as already defined, the ``Science segments'' are the subset of the global control segments during which Virgo is taking data of good quality, to be used by analyses. The difference of duration between the global control and Science segments is dominated by limited disruptions of the data taking, that usually stop the Science mode for a short time. The dominant source of these breaks is the frequency-independent squeezer that lost its nominal configuration about 240 times during the O3 run; the median time to restore it and switch back to Science data taking was about 140~s.

We note that the Virgo segment duration summary numbers listed here are lower than those reported by LIGO~\cite{Buikema:2020dlj,LIGO:2021ppb}. Yet, this difference has no significant impact on the duty cycle that is very similar for the three detectors of the global LIGO-Virgo network. The comparison between the O3a and O3b sub-runs shows that the impact of the winter season (larger sea seismic activity, wind, and more generally bad weather), although real, has been limited. Overall, the global network duty cycle has improved during O3, mainly due to the increase of the LIGO detectors duty cycle, while the Virgo one has been very stable. With an average of 76\%, the Virgo O3 duty cycle is lower than that measured during August 2017, the final weeks of the O2 run Virgo took part of: ${\sim}$85\%. Yet, the O3 performance has been achieved over 11 months spanning a whole calendar year and cannot be directly compared to the duty cycle of a short (only 25~days) run in Summer time, the most favorable period to operate an instrument like Virgo. Running one full year instead of one month is also more complex person-power wise, and the Virgo organization implemented during O3, although perfectible, held on during the whole run. This experience represents a good base on which to build upon in order to improve the Virgo performance for the O4 run and beyond.

\begin{figure}
  \center
  \includegraphics[width=\textwidth]{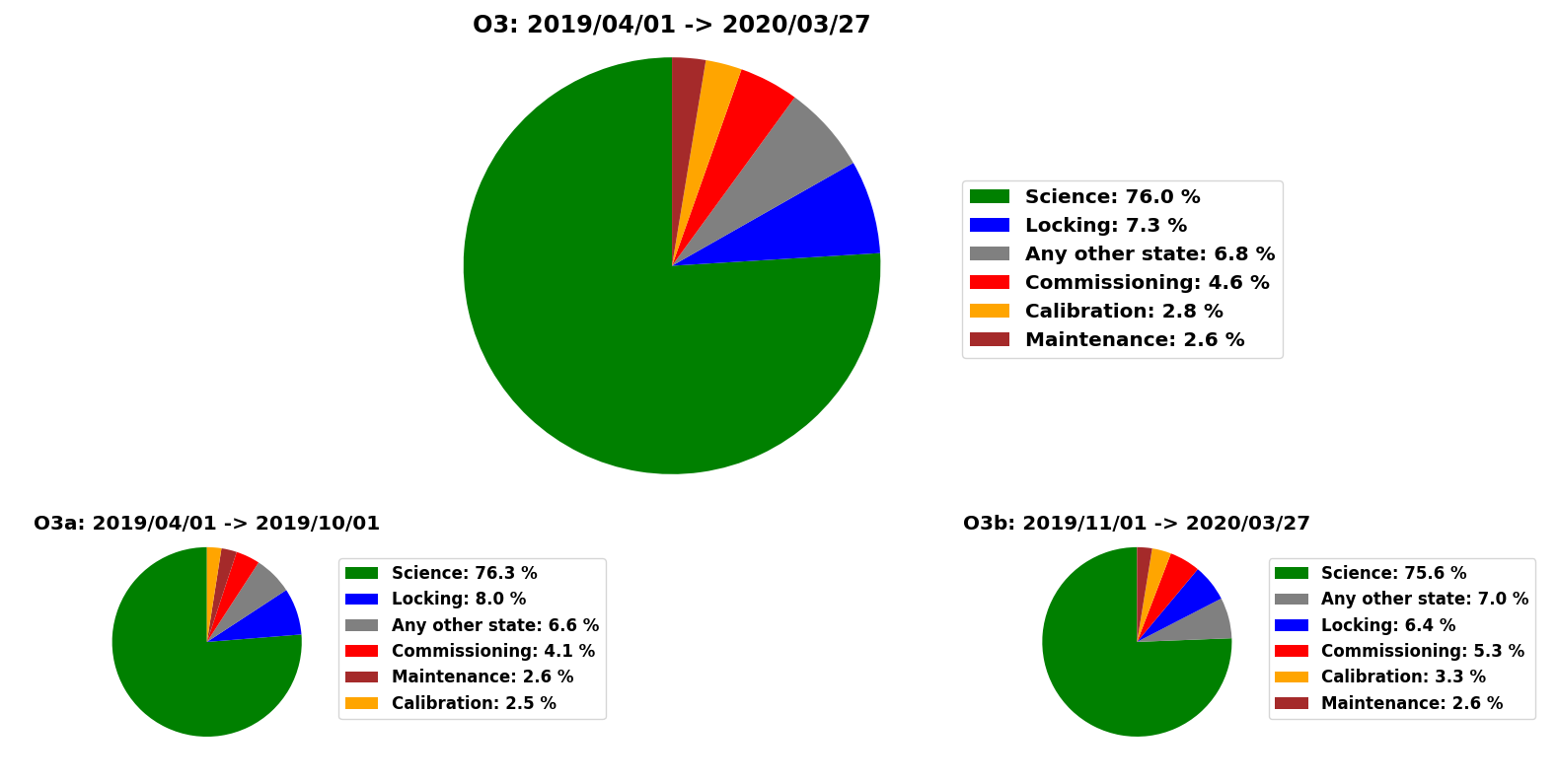}
  \caption{Breakdown of the time spent in different modes by the Virgo detector during the O3 run (larger, top-middle pie chart) and separately during the O3a and O3b sub-runs (pie charts at the bottom). The ``{\em Locking}'' mode corresponds to periods when the control of the detector is being acquired. The regular {\em maintenance} and {\em calibration} periods have been described in section~\protect\ref{subsection:data_taking}. Finally, the {\em Any other state} category includes all the other situations encountered during the whole run: troubleshooting periods, various kinds of tuning, etc. These results exclude the 1 month-long commissioning break that took place in October 2019, in between the O3a and O3b sub-runs. In each pie chart, the modes are sorted by decreasing percentage.}
  \label{fig:O3_Virgo_status}  
\end{figure}

Figure~\ref{fig:O3_Virgo_status} shows the breakdown of the time spent in different modes by Virgo during O3. Overall, the O3a and O3b distributions are quite consistent. Breaking these 11 month-averaged duty cycle figures down to a 24~hr period, Virgo took data during 18~hr, with the remaining 6~hr roughly divided into three blocks of the same duration: ${\sim}2$~hr for controlling the detector (Locking), ${\sim}2$~hr for recurring activities (Calibration, Commissioning and Maintenance) and ${\sim}2$~hr for solving issues (Any other state).

The analysis of these pie charts shows that increasing the duty cycle during future runs will not be straightforward. 
The room for improvement is limited in each area and so any significant duty cycle gain will likely stem from a combination of various small progresses, each made possible by the redesign or the optimization of a particular process.

\begin{figure}
  \center
  \includegraphics[width=\textwidth]{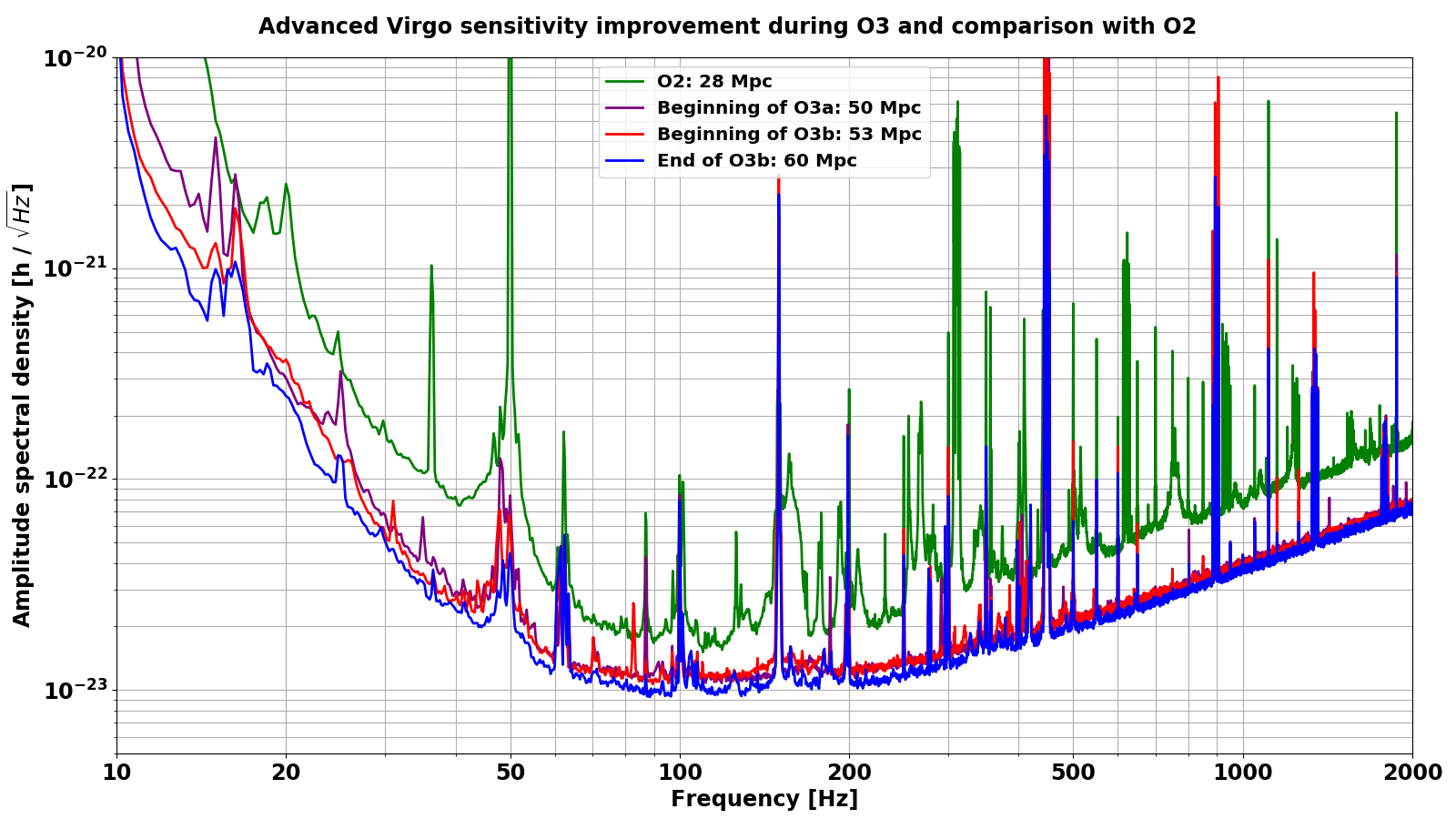}
  \caption{Comparison between four sensitivity curves of the \ac{adv} detector: during O2 (green trace), at the beginning of O3a (purple), at the beginning of O3b (red) and at the end of O3b (blue). The caption provides the corresponding estimated \ac{bns} ranges.}
  \label{fig:O3_sensitivities}  
\end{figure}

To conclude this overview, figure~\ref{fig:O3_sensitivities} summarizes the improvement of the sensitivity of the \ac{adv} detector. The \ac{bns} range associated to each curve is given in the legend. From O2 to O3b, the record \ac{bns} range has more than doubled from 28~Mpc to 60~Mpc, with a continuous improvement of the sensitivity in the whole bandwidth of the detector. Many spectral features of the residual noise structures have either been removed or significantly reduced over time.

%% file: O3_dataset.tex
The final Virgo O3 dataset consists of more than 250~days of data recorded during the O3a and O3b sub-runs and whose quality has been checked and validated (described in section~\ref{sec:dq_offline}). 
It is built upon and supersedes the online good-quality Science dataset that was used as input by the analysis pipelines that looked for \acp{gw} in real time (see section~\ref{sec:onlinedq}). Dedicated studies have been performed offline to refine the quality assessment of the data. In addition to running more in-depth analyses, new checks have been added during the run, as potential flaws got discovered in the existing analyses, or new problems identified at the detector level. Moreover, small sets of good data that had not been automatically included in the dataset (either because they were incorrectly labeled or because part of their data quality information was missing) were added by hand.

The main categories of checks applied to assess the quality of the Virgo data are the following:

\begin{itemize}
\item {\bf Are key components of the Virgo hardware (suspensions and photodiodes) having transient problems?} \\
These checks, described in Section~\ref{sec:onlinedq:cat1}, were fast enough to be performed online on live data.
\item {\bf Is the reconstruction of the \ac{gw} strain time series $h(t)$ nominal?} \\
This is a prerequisite for any further use of the Virgo data. The online reconstruction of the Virgo data was satisfactory during O3: only data from the very end of O3a (September 16-30, 2019) were reprocessed offline to increase the sensitivity by a few percents~\cite{VIRGO:2021umk}. Yet, during periods of high seismic activities (bad weather, high wind or the passing of seismic waves from strong and distant earthquakes), the nominal global control configuration could be replaced~\cite{o3virgoenv}) by a more robust one, the so-called ``earthquake (EQ)-mode''~\cite{VIRGO:2021umk}. Although that procedure saved some losses of the working point (whose recovery would have costed time), it could not be validated against the nominal reconstruction of the $h(t)$ strain stream until the final two months of O3b. Therefore, during most of the O3 run, data taken in these peculiar conditions had to be excluded from the final dataset. 
\item {\bf Do the data suffer from known problems?} \\
Tailored checks were run offline to identify and isolate periods during which the detector was not behaving nominally, although it was still controlled. One example of such studies is the fact that the North Input mirror suspension was randomly suffering from transient (a few second-long) losses of data. This was usually enough to lose the control of the entire detector, and hence to lose at least 20~min to 30~min of data: the time to reacquire the global working point and to restore Science data taking. Therefore, a patch was developed by experts to detect the data loss and switch in real time to a less robust --- but still available --- control until the missing data were back. This saved hours of running time for Virgo overall, but a dedicated scan of the data had to be performed offline to identify the occurrences of these control switches (potentially inducing transients and artifacts of instrumental origin in the data) and to remove them from the final dataset.
\item {\bf Are the data consistent?} \\
The last few seconds of a segment preceding a control loss of the detector have been removed offline, as the $h(t)$ data could be corrupted (see section~\ref{sec:dq_offline} for details). In addition, we have verified that the detector was nominally controlled during all segments flagged online as Science, and we also have looked for segments which could be included in the final offline dataset although they had not been categorized as Science online.
\item {\bf Is the dataset complete?} \\
There could be data segments with missing or corrupted $h(t)$ channel that would require a limited reprocessing. Or there could be segments with missing data segments due to problems in the \ac{daq}, etc. Dedicated checks were setup to target these problems specifically, before analysts would run into them when processing the data.
\end{itemize}

Data segments that fail one of the checks defined above are classified as ``Category 1'' (CAT1) vetoes and are excluded from all analyses. Overall, only 0.18\% of the Virgo O3 Science dataset have been CAT1-vetoed.

\begin{figure}
  \center
  \includegraphics[width=\textwidth]{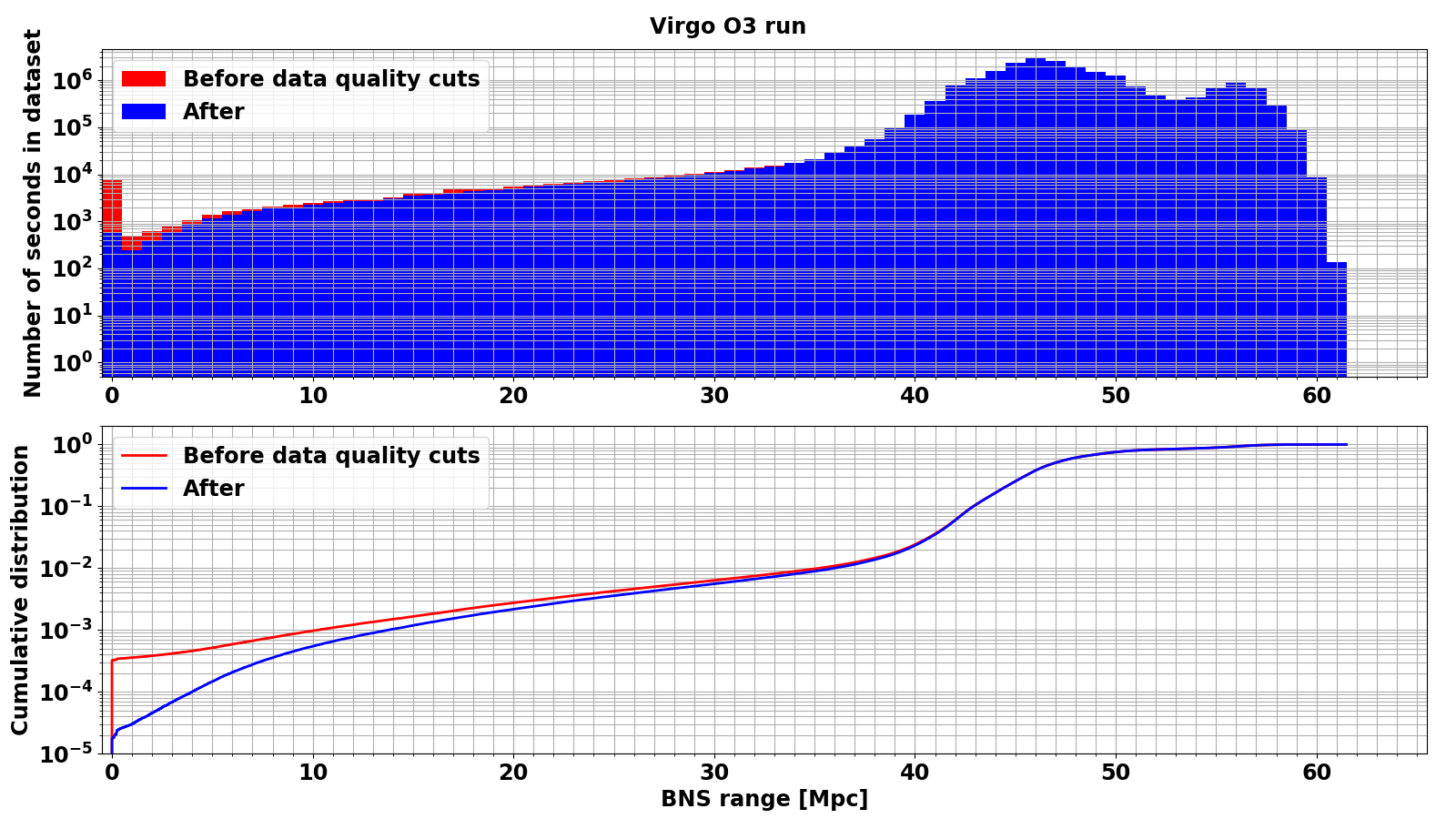}
  \caption{Distributions of the Virgo \ac{bns} range during O3 for the online (red histogram and trace) and offline (blue) dataset. The top (bottom) plot compares the histograms (cumulative distributions).}
  \label{fig:BNS_range_O3}
\end{figure}

To conclude this overview of the Virgo performance during the O3 run, figure~\ref{fig:BNS_range_O3} compares the Virgo \ac{bns} range distributions before (red) and after (blue) applying data quality cuts 
to determine the final O3 dataset. As expected, data quality requirements remove periods of low \ac{bns} range, i.e. when the sensitivity was poor. Yet, about 1\% of the data have a \ac{bns} range lower than 35~Mpc, that is significantly below the typical values achieved during O3 for that sensitivity estimator. While these data have not been flagged as bad by the various checks run on the dataset, they correspond to periods during which the detector was less accurately controlled, in particular due to bad weather. 

%% file: onlinedq_intro.tex
Online data quality was a key challenge to tackle for DetChar during the O3 run. The availability and the reliability of that information, supporting the data taking, had to be high in order to allow the real-time transient \ac{gw} searches to make the best use of the Virgo data. Significant candidates identified by those analyses ---  usually found in data from at least two of the three detectors of the global network, but sometimes identified in a single instrument ---  would then lead to public alerts, used by telescopes worldwide to search for counterparts of potential \ac{gw} signals.

In this section, we first describe the different blocks of the Virgo online data quality architecture, in use at EGO during the O3 run. This framework matches the dataflow shown in figure~\ref{fig:DataflowDetChar} and is complemented by the vetting of the most significant triggers identified in low latency, described in the following section~\ref{section:public_alerts}.

Real-time information about the detector status was combined with fast data quality estimators to produce a single integer channel sampled at 1~Hz, the {\em Virgo state vector}. That state vector was shipped alongside the \ac{gw} strain channel $h(t)$ to computing centers where data were analysed in real time. Its integer value was constructed by gathering several binary information (schematically: good vs. bad) encoded as bits; that bit pattern would later be decoded by the analysis frameworks to discard any bad data. Parallel to this data analysis stream, this information --- the detector status plus the real-time assessment of the data quality --- was automatically uploaded by a dedicated online process (called \texttt{SegOnline}) to the \ac{dqsegdb}~\cite{O3DetChar_tools}.

Finally, we present the experience gained during O3 with additional data-quality inputs, called {\em veto streams} whose aim is to help searches to reduce their false alarm rate by identifying triggers that are very unlikely to be of astrophysical origin.

%% file: onlinedq_architecture.tex
\begin{figure}
  \center
  \includegraphics[width=0.95\textwidth]{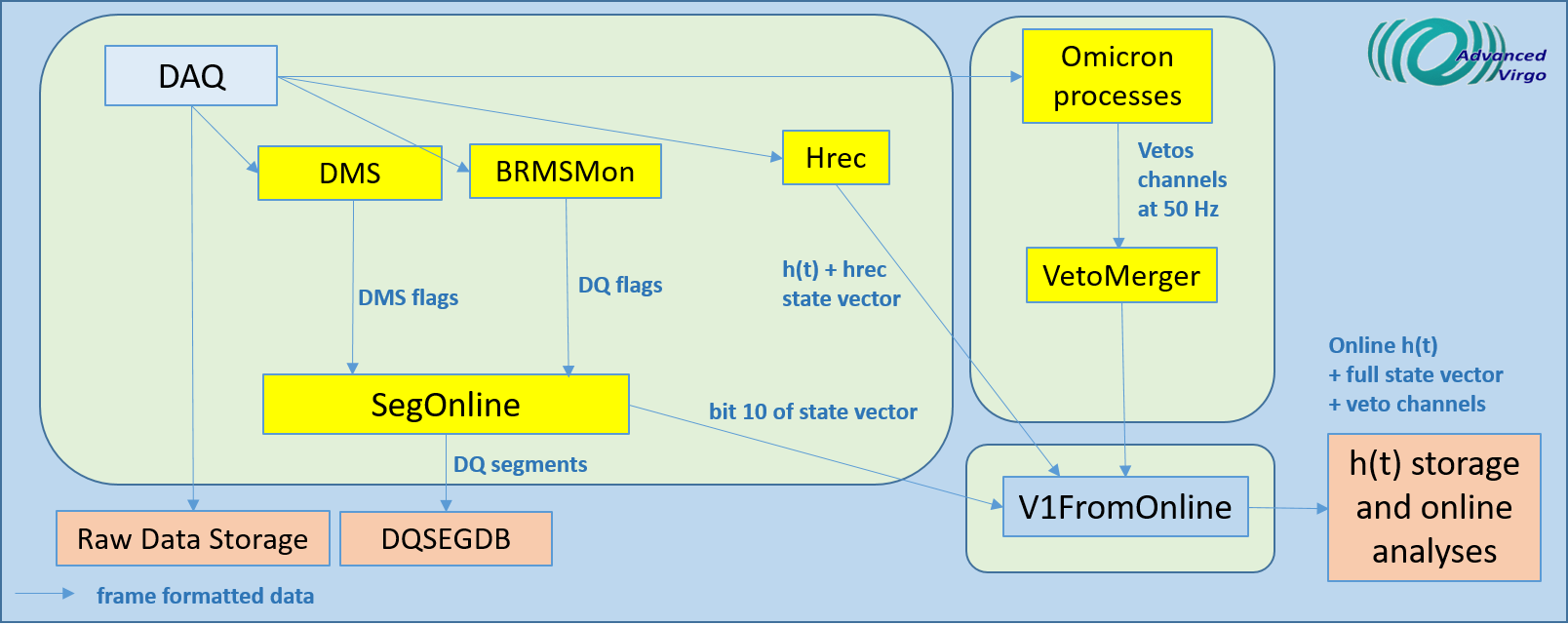}
  \caption{Online architecture to produce data quality products during the O3 run. The status of the interferometer is monitored by a dedicated \texttt{Metatron} server (see section~\ref{subsubsection:metatron}). Data quality flags are generated by dedicated servers described in~\cite{O3DetChar_tools}: the \ac{dms}, the \texttt{BRMSMon} process (environment), the \texttt{VetoMerger} process (large deviations in auxiliary signals), the \texttt{Hrec} process ($h(t)$ reconstruction), and the \texttt{Omicron} algorithm (glitches in $h(t)$ and auxiliary signals). Data quality segments are then generated by the \texttt{SegOnline} process and saved in the LIGO-Virgo segment database, while the online $h(t)$ stream, the state vector and the veto channels are sent to online data analysis pipelines through the \texttt{V1FromOnline} server.}
  \label{fig:onlinedq}  
\end{figure}

The online data quality architecture is designed to deliver data quality products to online transient searches.
It is based on a set of servers connected to the \ac{daq} and providing
relevant information about the quality of the data (the whole raw data, plus the reconstructed $h(t)$ stream).
In the following, the main elements of this architecture, summarized in figure~\ref{fig:onlinedq}, are presented.

%% file: onlinedq_statevector.tex
Table~\ref{table_state_vector} defines the 16 bits of the Virgo {\em state vector} integer channel
in use during the O3 run.
A bit is said to be {\em active} when its value is 1, meaning that the corresponding check is passed. A value at 0 means instead that a problem, or a non-nominal state, has been detected. The information provided by these bits is on purpose partially redundant, in the sense that several bits can be at 0 when proper data taking conditions are not met. During O3, the bits 0, 1 and 10 were required to be active to have the corresponding 1-second data frame processed by real-time analyses.

\begin{table}[htbp!]
\caption{\label{table_state_vector}Definition of the bits of the Virgo state vector during the O3 run (see text for details).}

\begin{tabular}{cc}
\toprule
Bit number & Active when \\
\midrule
0   & $h(t)$ successfully computed. \\
\hline
1-2 & Science mode enabled. \\
\hline
3   & $h(t)$ successfully produced by the calibration pipeline. \\
\hline
\multirow{2}{*}{4-7} & Bits irrelevant for the present discussion: \\ 
                     & either redundant with other bits or unused during O3. \\
\hline
\multirow{2}{*}{8}   & No DetChar-related hardware injection \\ 
                     & (see section~\ref{subsection:safety} for more details). \\
\hline
\multirow{3}{*}{9}   & No continuous wave hardware injection \\ 
                     & (the only type of non calibration-related injections \\ 
                     & performed for a short period during O3, while taking nominal data).\\
\hline
10  & Online data quality is good (no CAT1-type veto). \\
\hline
\multirow{2}{*}{11}  & Virgo interferometer fully controlled, \\ 
                     & with a nominal working point or close to it.\\
\hline
12-15 & Not used.\\
\bottomrule
\end{tabular}

\end{table}

%% file: onlinedq_CAT1.tex
During the O3 run, the problems detected online and leading to CAT1 vetoes are listed below. These are:

\begin{itemize}
\item No saturation of any of the 4 dark fringe photodiodes, using the `DC' (from 0 to a few~Hz) and `Audio' (from a few~Hz to 10-50~kHz) demodulated signals.
\item No saturation of the correction signal of any of the 16 suspension stages monitored.
\item No saturation of the rate of glitches reported by the online \texttt{Omicron} framework for the \ac{darm} correction channel\footnote{A more correct way to monitor the glitch rate would have been to scan $h(t)$, but the latency added by that \texttt{Omicron} check would have delayed too much the \ac{gw} strain channel for online processing. The offline equivalent version of that check did use $h(t)$, as latency was not an issue anymore in that case.}.
\end{itemize}

These saturation checks were combined using a logical OR to produce CAT1 vetoes with a 1~s granularity. Section~\ref{sec:dq_offline} describes the corresponding set of {\em offline} CAT1 vetoes, used by all analyses processing the final O3 Virgo dataset, including these {\em online} CAT1 vetoes.

%% file: onlinedq_segonline.tex
Any channel provided by the \ac{daq} or by the online processing (for instance \ac{dms} monitors or the \texttt{BRMSMon} process)
can be used by the \texttt{SegOnline} process to build segments of data quality flags which are sent online to
\ac{dqsegdb}~\cite{O3DetChar_tools}.

\texttt{SegOnline} writes down segments into XML files with a latency of about 10~s and those XML files are then read by a {\em rsync} process
to upload the segments into \ac{dqsegdb} every 5~min. Such data quality segments can then be used by any analysis,
or can be viewed and downloaded through a dedicated web interface~\cite{dqsegdb_web}.

%% file: onlinedq_vetostreams.tex
Low-latency transient searches are limited by glitches in the $h(t)$ data. Each search pipeline is sensitive to specific families of glitches. The online data quality architecture is designed to deliver a channel to flag glitches relevant to a given low-latency pipeline. These channels are called veto streams. A veto stream is a time series which can only take two values: 0 means good quality and 1 bad quality. A veto stream is generated by the \texttt{VetoMerger} process which combines information from many online data quality processes, carefully selected to target the glitches limiting the search of interest~\cite{O3DetChar_tools}.

Some \texttt{Omicron} processes~\cite{O3DetChar_tools} are configured to select triggers detected in auxiliary channels with a \ac{snr} above a threshold tuned with the \ac{upv} algorithm~\cite{O3DetChar_tools}. These triggers are known to witness glitches in the $h(t)$ channel. When this is the case, the veto channel is set to 1. \texttt{VetoMerger} also ingests the data quality flags generated by \texttt{BRMSMon}~\cite{O3DetChar_tools} to veto environmental disturbances.

In O3, the veto stream system was experimented as an input to one of the low-latency searches for compact binary mergers, \texttt{PyCBC} Live~\cite{PyCBCLiveO2,PyCBCLiveO3}. The veto stream, named \texttt{DQ\_VETO\_PYCBC}, combined two elements: a veto channel delivered by \texttt{Omicron} to target scattered-light glitches, and a data quality flag produced by \texttt{BRMSMon} to tag occasional glitches associated to lightning strikes. As explained in section~\ref{subsection:safety}, it is critical that a veto is constructed from channels which are insensitive to \acp{gw}: a channel (or veto) is then said to be \textit{safe}. The channel safety is tested with hardware injections that mimic the effects of \acp{gw} on the detector. The \texttt{DQ\_VETO\_PYCBC} veto stream is derived from safe channels: magnetometer signals are used to veto lightning strikes and \ac{daq} frequency modulation channels were found to witness scattered-light glitches. A conservative approach was adopted to tune the vetoes: their thresholds were set at high values to reliably flag really limiting glitches, while keeping the rejected time low. As a result, only 0.05\% of the O3 Science time was flagged by the \texttt{DQ\_VETO\_PYCBC} veto stream.
This is a conservative tuning as it means that there is only a 0.05\% probability that \texttt{DQ\_VETO\_PYCBC} would discard a true \ac{gw} signal, under the reasonable assumption that \acp{gw} are uncorrelated with scattered-light and lightning glitches.
\texttt{PyCBC} Live used the veto stream to simply prevent the generation of a candidate event from Virgo data, or remove Virgo's contribution from a LIGO-Virgo candidate, during periods of active veto.
In future runs, the veto streams may be integrated in a more general framework based on auxiliary channels to discard or down-weight transient noise events.

\begin{figure}
  \center
  \includegraphics[width=0.48\textwidth]{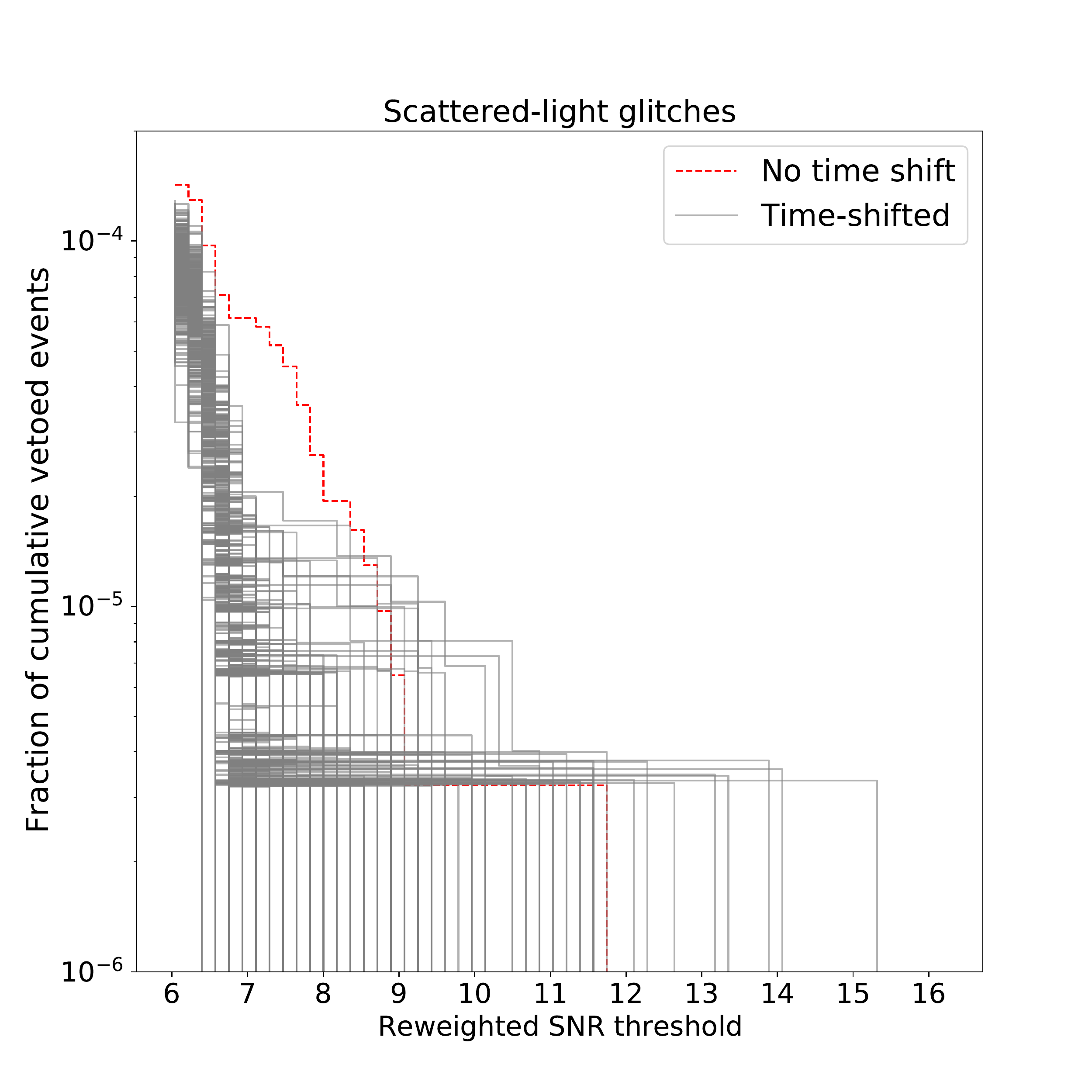}
  \includegraphics[width=0.48\textwidth]{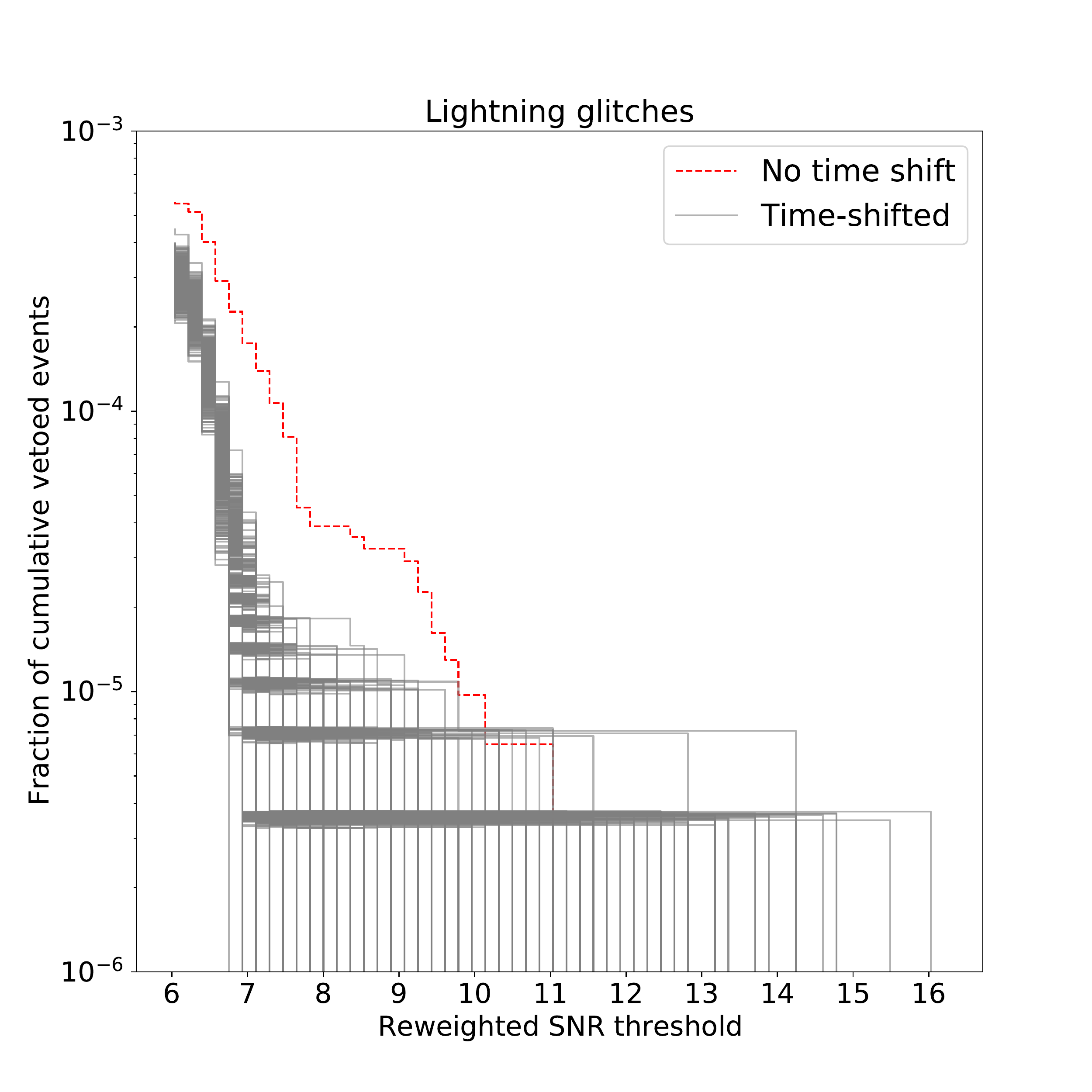}
  \caption{Effect of veto streams on triggers from the O3 archival search for compact binary mergers in Virgo data based on \texttt{PyCBC}. Each staircase shows the cumulative fraction of vetoed triggers with reweighted \ac{snr} higher than a given threshold, as a function of the threshold. The red curve shows what happens when removing triggers within segments flagged by the veto streams. The gray curves, instead, are samples from the null distribution, constructed by applying a time shift to the segments before vetoing the triggers. The fractions are relative to the overall number of triggers generated by the search. The left plot considers vetoes targeting scattered-light glitches, and the right plot considers vetoes associated with glitches from lightning. In both cases we can see that the red curve is significantly above the gray ones, at least for some reweighted \ac{snr} thresholds, indicating a significant correlation between the triggers and the vetoes.}
\label{fig:veto_streams_offline}
\end{figure}

Although the veto stream was only used for the online PyCBC analysis in O3, we can now evaluate its effect on the \texttt{PyCBC} offline analysis as well.
In particular, we consider here Virgo single-detector triggers generated by the broad-space \texttt{PyCBC} search~\cite{PyCBCOfflineO3} during the period from April 1$^{st}$ to May 11$^{th}$, 2019.
The study is performed under the assumption that such triggers are dominated by noise, and that scattered-light and lightning glitches are not correlated with \ac{gw} signals.
The search ranks the triggers by a quantity known as \emph{reweighted \ac{snr}} \cite{PyCBCOfflineO3}, i.e.\ the signal-to-noise ratio returned by the matched filtering technique~\cite{Davis1989,Abbott_2020}, weighted by the result of $\chi^2$ tests that quantify how well the time-frequency distribution of power observed in the data is consistent with the one expected from the matching template~\cite{Allen:2004gu, Nitz:2017lco}.
For practical reasons, only triggers with reweighted \ac{snr} higher than 6 are considered here.
After this selection, the sample is composed of roughly $2.5 \times 10^{5}$ triggers.
To evaluate the impact of the vetoes on the offline search, we remove triggers with a merger time belonging to a vetoed segment.
We carry out the study separately for vetoes targeting scattered-light glitches and glitches from lightning.
We show the fraction of vetoed triggers as red staircases in figure~\ref{fig:veto_streams_offline}, and in both cases it is found to be of the order of $10^{-4}$ or less, i.e.\ compatible with the expectation from the amount of vetoed time.
The majority of vetoed triggers have relatively small reweighted \acp{snr}.

As a next step, we would like to assess the statistical significance of the impact of the vetoes on the offline \texttt{PyCBC} triggers, i.e.\ calculate the probability that the vetoed triggers are simply explained by chance alignment with the veto segments.
To this end, we shift the veto segments rigidly by a constant time offset, so that the shifted segments maintain the general time structure of the original ones, but lose any correlation with glitches.
We then recompute the fraction of triggers vetoed by the shifted segments, obtaining a null sample that we can compare to the original fraction from the unshifted segments.
Note that the null fraction is rescaled to account for the overlap between the Science-mode segments and the time-shifted veto segments, which is a function of the time shift.
We construct 1000 such null samples by repeating the time-shifted analysis with time offsets covering the range of $[-50000, +50000]$~s in steps of 100~s.
The cumulative fraction of vetoed triggers for the null samples are shown in gray in figure~\ref{fig:veto_streams_offline}.
At a reweighted-\ac{snr} threshold of 6, the unshifted fraction is higher than any time-shifted fraction, for both scattered-light (left plot) and lightning (right plot) vetoes.
We conclude that the probability for the observed effect of the vetoes on the \texttt{PyCBC} offline triggers to be a statistical fluctuation is less than the inverse size of our null sample, i.e.\ $10^{-3}$.
For higher reweighted-\ac{snr} thresholds of $8.5$ ($10.5$), this probability is $2 \times 10^{-3}$ ($2 \times 10^{-2}$) for vetoes targeting scattered-light glitches, and less than $10^{-3}$ ($6 \times 10^{-3}$) for vetoes associated with lightning glitches.
It does therefore appear that scattered light and lightning strikes are correlated with a small population of (relatively weak) \texttt{PyCBC} triggers, and that the veto streams can in principle be used to remove or down-weight these triggers.
Whether this is beneficial or not should be established by carrying out a complete simulation with injected signals, in order to measure the effect of the veto streams on the sensitive time-volume of the search, as routinely done when optimizing searches for compact binary mergers.
We reserve this to a more detailed future investigation, however our results indicate that this effect would not be very large (assuming O3-like data) as the fraction of vetoed signals is of order $10^{-4}$ and the reduction of the background would only impact relatively quiet Virgo triggers anyway.

For the O3 run we demonstrated the possibility of delivering search-specific veto streams to online pipelines to reject transient noise events. This first experiment with \texttt{PyCBC}, although with limited performance, has validated the online architecture. For the next run, O4, we plan to generalize this framework to other pipelines and plug in more veto streams to target other families of glitches.

%% file: alerts_intro.tex
As demonstrated with the extraordinary GW170817~\cite{TheLIGOScientific:2017qsa} event from the O2 run, public alerts sent by the LIGO-Virgo network are key deliverables targeting the astronomy community. Yet, how successful these are depends on the accuracy of the information provided, and of the latency at which they are delivered. For O3, the main contribution of the DetChar group to this effort has been the design and the implementation of the \ac{dqr} framework. A \ac{dqr} is a set of data quality checks, automatically triggered by the finding of a new \ac{gw} candidate. Its output allowed the \ac{rrt} team to vet the associated data in a timely way. Moreover, its usage extended way beyond the data taking period, as it was the main tool used to assess the data quality of all \ac{gw} candidates identified by analyses, in some cases with a latency longer than a year (compared to when the corresponding data were acquired).

The performance of the Virgo O3 \ac{dqr} is described below, before summarizing how Virgo contributed to the LIGO-Virgo public alerts during the O3 run. The Virgo \ac{dqr} implementation can be found in~\cite{O3DetChar_tools}.

%% file: dqr.tex
This section briefly summarizes the performance of the Virgo \ac{dqr}, via statistical analyses using data from O3b that correspond to the final, most complete, version of that framework during the O3 run. Emphasis is put on latencies and running times, as these are key quantities to vet public alerts in a timely way. As those time distributions can include tails due to occasional technical problems impacting the \ac{dqr} dataflow somewhere along its way (from the \ac{gracedb}~\cite{O3DetChar_tools} to the EGO HTCondor farm and back) while being external to it, the results presented in the following two tables include the $50^{\rm th}$ and $95^{\rm th}$ percentiles in addition to the mean values.

Table~\ref{table:DQR_perf_1} provides the measured latencies for the processing steps that occur upstream of the \ac{dqr} with the following meaning:

\begin{itemize}
\item The first line is the difference between the time when the trigger is recorded in \ac{gracedb} and the time when the corresponding data were acquired.
\item The second measures the time needed for \ac{gracedb} to notify the \ac{lvalert} and to have this message trigger the Virgo \ac{dqr} framework upon reception.
\item The third reports the time needed to create and configure a new \ac{dqr} instance, until it is ready for processing. One should note that this duration includes a 300~s wait time, imposed in order to allow \ac{gracedb} to receive, process and gather all triggers found by the different online searches that analyse strain data in parallel and independently. The assumption is that, after these 5~min, the low-latency information available in \ac{gracedb} should be optimal and stable in the vast majority of cases. Therefore, the actual \ac{dqr} configuration phase only takes a few tens of seconds: the needed data are located in the low-latency streams just made available by the \ac{daq} and more than 30 scripts (each corresponding to a data quality check) are generated one after the other.
\item Finally, the last reported duration accounts for the time needed to start processing the \ac{dqr} on the EGO HTCondor farm. This depends on the occupancy of the farm and of the EGO internal network performance.
\end{itemize}

\begin{table}[htbp!]
\caption{\label{table:DQR_perf_1}Summary of the performance of the low-latency with Virgo \ac{dqr} dataflow during O3b, from the GPS time of a trigger to the start of the Virgo \ac{dqr} on the EGO HTCondor farm: see text for details.}
\footnotesize
\centering
    \begin{tabular}{r|ccc}
    \toprule
    \multirow{2}{*}{Operation} & \multicolumn{3}{c}{Time taken [s]} \\
    \cline{2-4}
    & Median & Mean & $95^{\rm th}$ percentile \\
    \hline
    Data acquired $\to$ Candidate on \ac{gracedb}          & 52  & 166 & 331 \\
    Candidate on \ac{gracedb} $\to$ \ac{lvalert} trigger   & 4   & 4   & 11 \\
    \ac{lvalert} trigger $\to$ Virgo \ac{dqr} configured   & 331 & 339 & 383 \\
    Virgo \ac{dqr} configured $\to$ Virgo \ac{dqr} started & 8   & 10  & 21\\
    \bottomrule
    \end{tabular}

\end{table}

We can see that the mean time elapsed between the recording of the data by the different detectors and the creation of a new record in \ac{gracedb} is under 3~min. The median time is even under 1~min while the tail of the time distribution extends beyond 5~min. This includes the reconstruction of the \ac{gw} strain channels, the transfer of these data alongside the associated online data quality information to computing centers, the processing of these data by real-time \ac{gw} searches, the automated analysis of the results and the final transfer of trigger information to \ac{gracedb} where it is made centrally available. Then, the new alert is received at EGO a few seconds later, triggering the creation and the configuration of a new \ac{dqr} instance. Removing the compulsory wait time of 300~s, the \ac{dqr} configuration takes a few tens of seconds only. Finally, 10 additional seconds are needed on average to have the first \ac{dqr} jobs processed on the EGO HTCondor farm.

Table~\ref{table:DQR_perf_2} summarizes the performance of the Virgo O3 \ac{dqr} framework in terms of running time. Each row corresponds to a category of checks. The quoted durations increase from one row to the next, as each new set of checks includes the previous ones.

The quick checks whose outputs are mandatory to vet a trigger take about 6~min to be all available, with a few minutes spread. Adding information about the \texttt{Omicron} triggers around the candidate takes about 10 more minutes. During O3, this latency was dominated by the fact that \texttt{Omicron} triggers were computed in real time and stored internally by the online server: they were only written to disk every 600~s, in order to allow the framework to cope with the incoming data flow. Work will be done prior to O4 to optimize this latency and to make the \ac{dqr} aware of when the needed data have been written to disk, so that their processing can start immediately after. \texttt{Omicron}-scanning all the available channels (more than 2000 in total, with the vast majority of them sampled at 10~kHz) around the trigger time requires 15-20 additional minutes. Finally, the full \ac{dqr} took from 1.5 to 2~h to complete. The longest checks were \ac{bruco}~\cite{O3DetChar_tools} and \ac{upv}, plus the scan of all online logfiles described above. 

\begin{table}[htbp!]
\caption{\label{table:DQR_perf_2}Summary performance of the Virgo \ac{dqr} processing during the last $\sim$100 days of the O3b run. The quoted durations include the time to upload \ac{dqr} check results back to \ac{gracedb} that usually takes from $\sim$5 to $\sim$20~s.}
    \begin{center}
        \begin{tabular}{r|ccc}
        \toprule
        \multirow{2}{*}{Operation} & \multicolumn{3}{c}{Time from \ac{dqr} start [s]} \\
        \cline{2-4}
        & Median & Mean & $95^{th}$ percentile \\
        \hline
        Quick key checks & 374 & 383 & 619 \\
        Adding \texttt{Omicron} trigger distributions & 868 & 816 & 935 \\
        Adding full \texttt{Omicron} scans & 1740 & 2159 & 4690 \\
        End & 5185 & 4954 & 6330\\
        \bottomrule
        \end{tabular}
    \end{center}

\end{table}

The Virgo \ac{dqr} reliability --- that is how efficient a \ac{dqr} is in providing accurate information on a \ac{gw} trigger --- is another key figure of merit of that framework. A good indicator of this is the number of (software) check failures per \ac{dqr} instance, as any check not properly completing would prevent analysts from accessing part of the available data quality information. During the whole O3 run, there was no case of a public alert for which the rapid vetting of the Virgo data was delayed due to Virgo \ac{dqr} issues. In addition, we used the large sample of \acp{dqr} automatically processed during O3b --- by design, all \ac{gw} candidates with a low-latency false-alarm rate below 1/day triggered a \ac{dqr}. The results are in table~\ref{table:DQR_perf_3}: only 13\% (2\%) of the \ac{dqr} had 1 (2) failed checks --- and none had more than 2 failures, while an O3 \ac{dqr} included more than 30 checks in total. No exhaustive analysis of these failures has been performed, as most of these \acp{dqr} were never checked by hand because the associated trigger was not significant enough. The two main causes of problems were, however, incomplete handling of edge-cases with the input data and actual bugs in \ac{dqr} check algorithms. These issues are being addressed as part of the upgrade of the Virgo \ac{dqr} framework for the O4 run.

\begin{table}[htbp!]
\caption{\label{table:DQR_perf_3}Percentages of the O3b Virgo \acp{dqr} with 0, 1 and 2 unsuccessful checks respectively.}
\centering
    \begin{tabular}{c|ccc}
    \toprule
    Number of unsuccessful checks & 0 & 1 & 2 \\
    Percentage of O3b automatically processed \acp{dqr} & 85\% & 13\% & 2\%\\
    \bottomrule
    \end{tabular}

\end{table}

%% file: alerts_public.tex
\subsubsection{Public alerts retracted because of an issue with Virgo data}

\begin{figure}[htb!]
  \center
  \includegraphics[width=\textwidth]{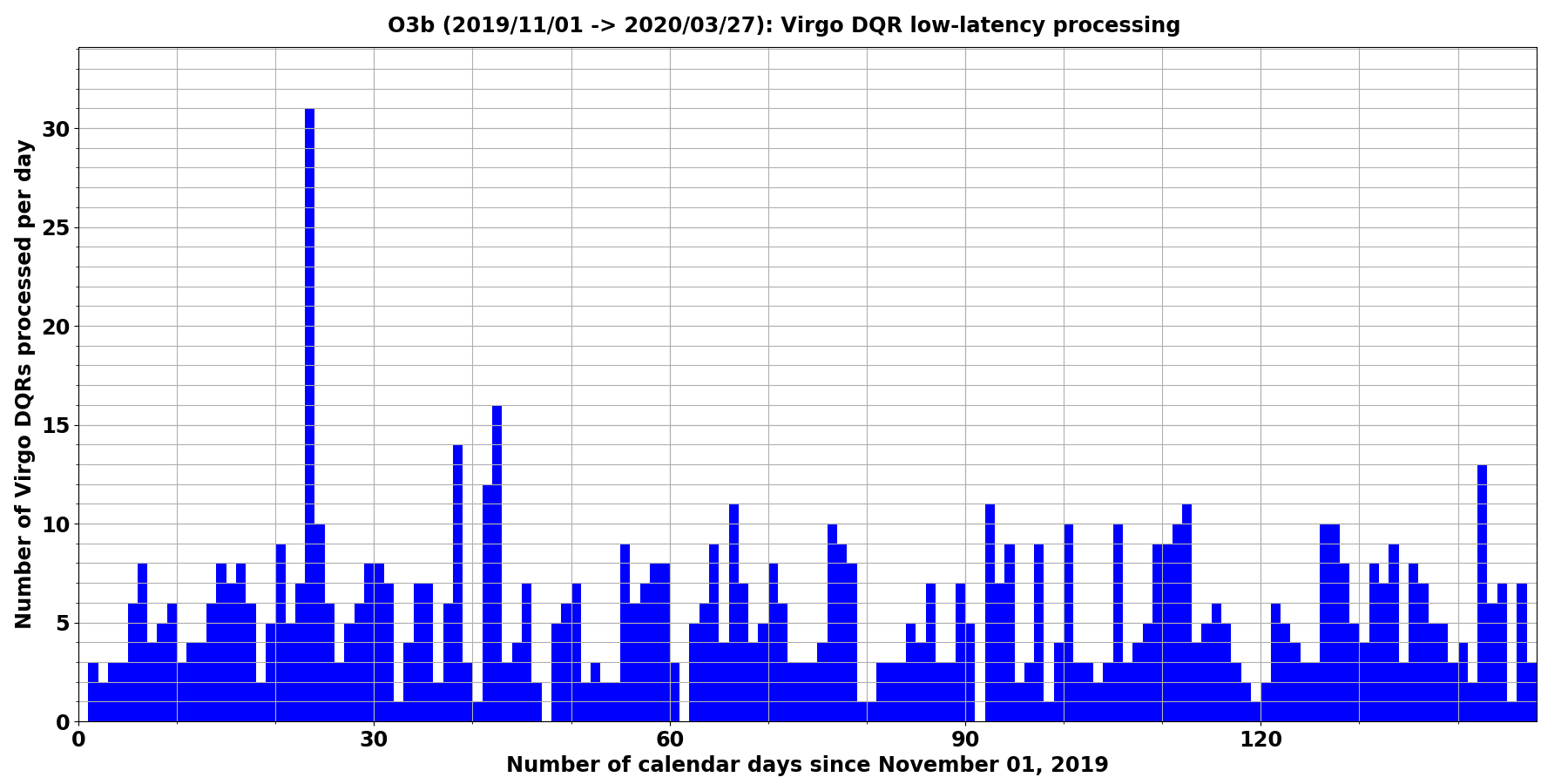}
  \caption{Number of Virgo \acp{dqr} automatically processed per day during the O3b run. The peak of 31 entries corresponds to November 24$^{th}$, 2019 when there was a transient problem with the Virgo $h(t)$ reconstruction: that generated several online triggers, finally including S191124be that passed the public alert threshold and was promptly retracted.}
  \label{fig:nDQRsPerDay_O3b}  
\end{figure}

During O3, 24 public alerts out of 80 have been retracted: 8 during O3a and 16 during O3b. Out of these retractions, only two were due to Virgo data:
\begin{itemize}
\item S191124be~\cite{S191124be} was due to a problem in the noise removal procedure included in the online reconstruction of the $h(t)$ \ac{gw} stream~\cite{VIRGO:2021umk}. Two such cleaning algorithms running in sequence started interfering, leading to a noise increase over time. An online pipeline started triggering on that excess noise, creating several non-astrophysical \ac{gw} candidates in rapid succession (figure~\ref{fig:nDQRsPerDay_O3b}), until one of them had a false alarm rate lower than the public alert threshold. That led to the generation of an automated alert that was then quickly retracted, the problem of the noise cleaning procedure was fixed as well.
\\ A similar problem should not happen again in future runs for three reasons: i) improved noise cleaning procedures are being developed within the Virgo $h(t)$ reconstruction; ii) an online monitoring dedicated to such noise removal interferences will be in place during O4; iii) a monitoring of the pipeline trigger rates in \ac{gracedb} will be running as well during future data taking periods, in order to spot quickly any misbehavior, like an excess trigger rate (in the case of S191124be) or the opposite: a too long data-taking time period without any trigger, even of low significance. 
\item S200303ba~\cite{S200303ba} was a single-pipeline trigger with most of its \ac{snr} concentrated in Virgo. At that time, Virgo data were quite noisy due to bad weather. An \texttt{Omicron}-scan around the trigger time (figure~\ref{fig:S200303baOmicron}) showed evidence of scattering light noise at low frequency. The unusually-long delay to send out the retraction circular (80~min) was partly due to an issue with the Gamma-ray Coordinates Network (GCN) broker connection.
\begin{figure}[htb!]
  \center
  \includegraphics[width=0.9\textwidth]{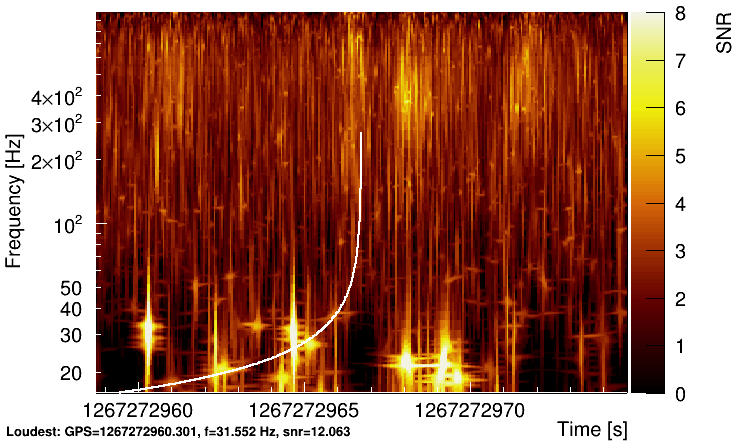}
  \caption{\texttt{Omicron} spectrogram around the time of the S200303ba trigger. The search template time-frequency track (solid line) overlaps with low-frequency scattering-light glitches (yellow color) caused by bad weather.}
  \label{fig:S200303baOmicron}  
\end{figure}

\end{itemize}

\subsubsection{Virgo contribution to O3 public alerts}

Out of the 56 non-retracted O3 public alerts, 42 involved the Virgo detector.
For 10 out of the 14 LIGO-only alerts, Virgo was not controlled in its nominal configuration at the GPS time of the trigger.
This fraction is consistent with the average duty cycle of Virgo during O3 (see section~\ref{subsection:O3_perf}).
For the four remaining alerts, described in detail next, Virgo was fully controlled at the time of the trigger and had a \ac{bns} range consistent with its typical performance at that moment.

S190720a occurred during a ${\sim} 1$ min segment in between the end of the acquisition of the detector global working point and the beginning of the nominal Science mode. Therefore, Virgo data were not used for low-latency analyses.
Offline analyses later confirmed S190720a as a significant detection and were able to use the low-noise Virgo data, finding a non-negligible amount of signal power in them.
S190720a was published as GW190720\_000836 in GWTC-2~\cite{Abbott:2020niy}.

S190910d occurred during nominal Science mode data-taking in Virgo.
It was a marginal candidate, only reported by a subset of the low-latency searches.
These searches did not find a significant amount of signal power in Virgo data, and did not report Virgo as being used for the candidate.
S190910d was not confirmed by offline analyses.

S190923y occurred while Virgo was undergoing commissioning activity.
It was not confirmed by offline analyses.

S200225q occurred while Virgo was undergoing a calibration run.
Offline analyses confirmed S200225q as a significant detection and were able to include the low-noise Virgo data, although no significant signal power was found there.
S200225q was published as GW200225\_060421 in GWTC-3~\cite{GWTC3}.

%% file: dq_intro.tex
This final section presents examples of global data quality studies made during or after the O3 run: noise transients, spectral analyses, 
classification of auxiliary channels based on their potential sensitivity to \ac{gw} signals
and offline data quality studies leading to the final Virgo O3 dataset.

%% file: dq_pipelines.tex
\subsubsection{Glitch rates during the O3 run}

\begin{figure}
    \begin{center}
      \includegraphics[width=0.98\columnwidth]{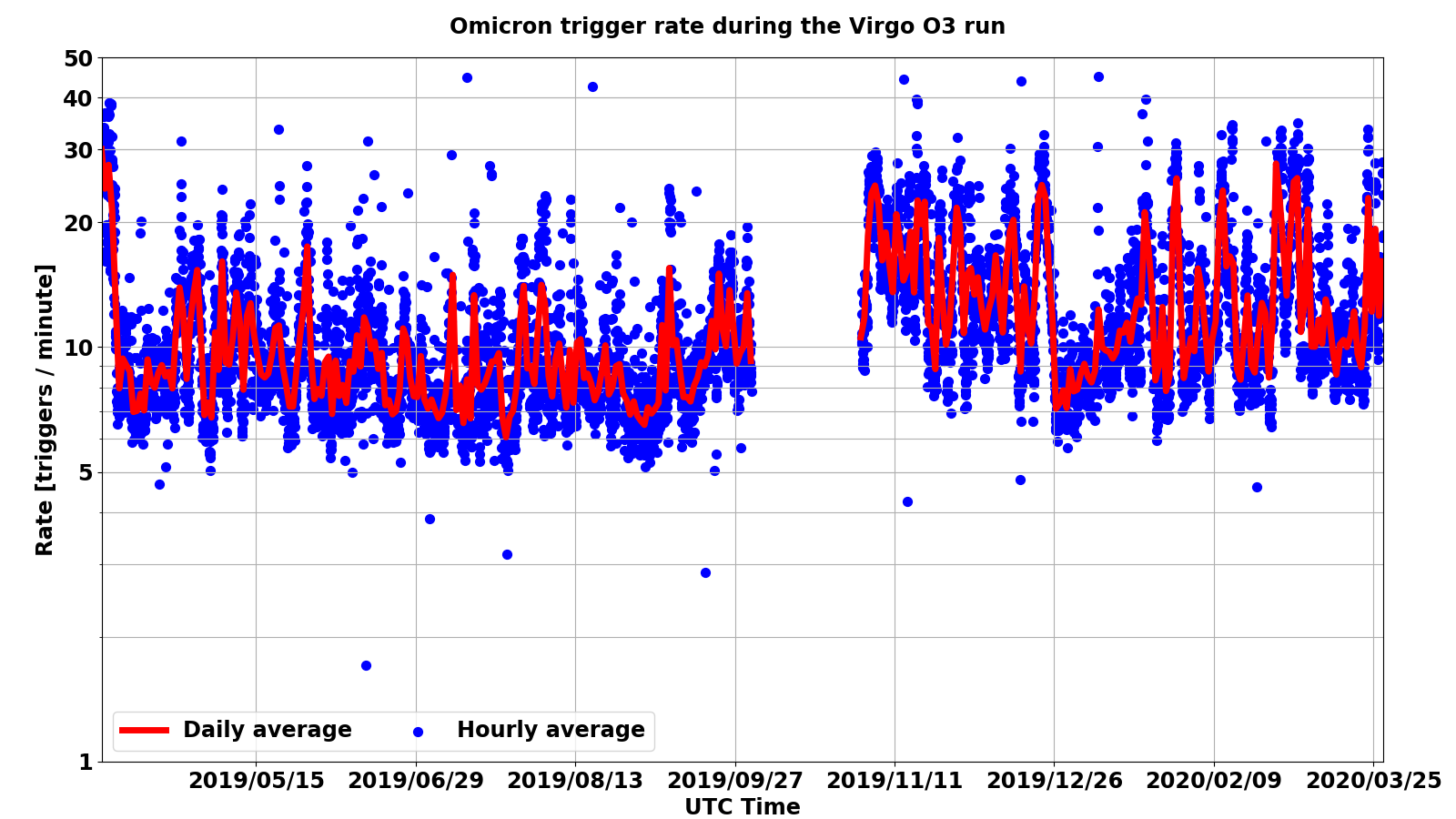}
    \end{center}
    \caption{Virgo glitch rate, using \texttt{Omicron} triggers, for the final O3 dataset (Science segments that have not been CAT1-vetoed). The blue dots are averages over one hour while the red curve shows the corresponding daily moving average. The gap in between O3a and O3b corresponds to the 1-month commissioning break.}
    \label{fig:omicron_trigger_rates:global}
\end{figure}

\begin{figure}
    \begin{center}
      \includegraphics[width=0.98\columnwidth]{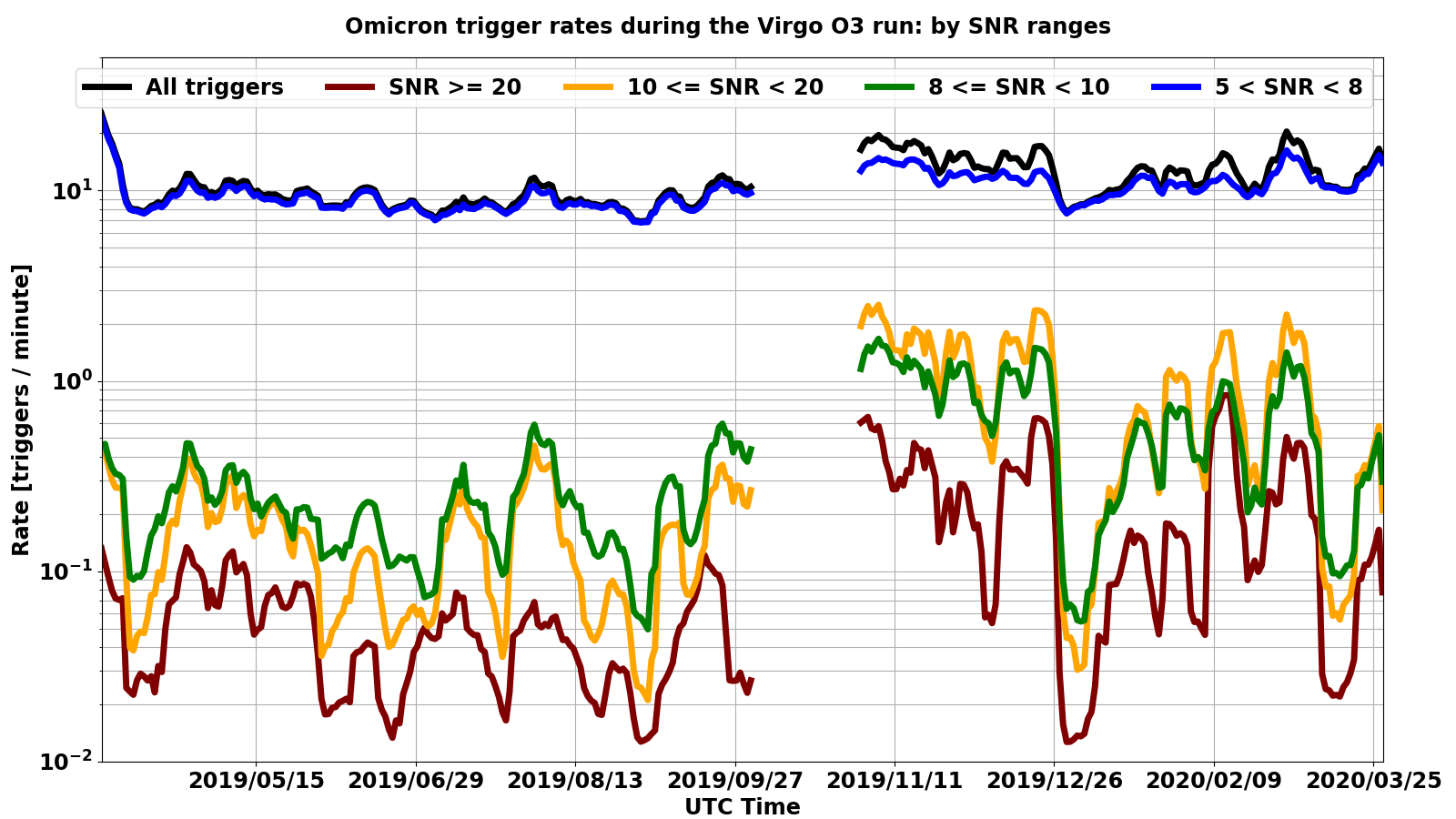}
      \includegraphics[width=0.98\columnwidth]{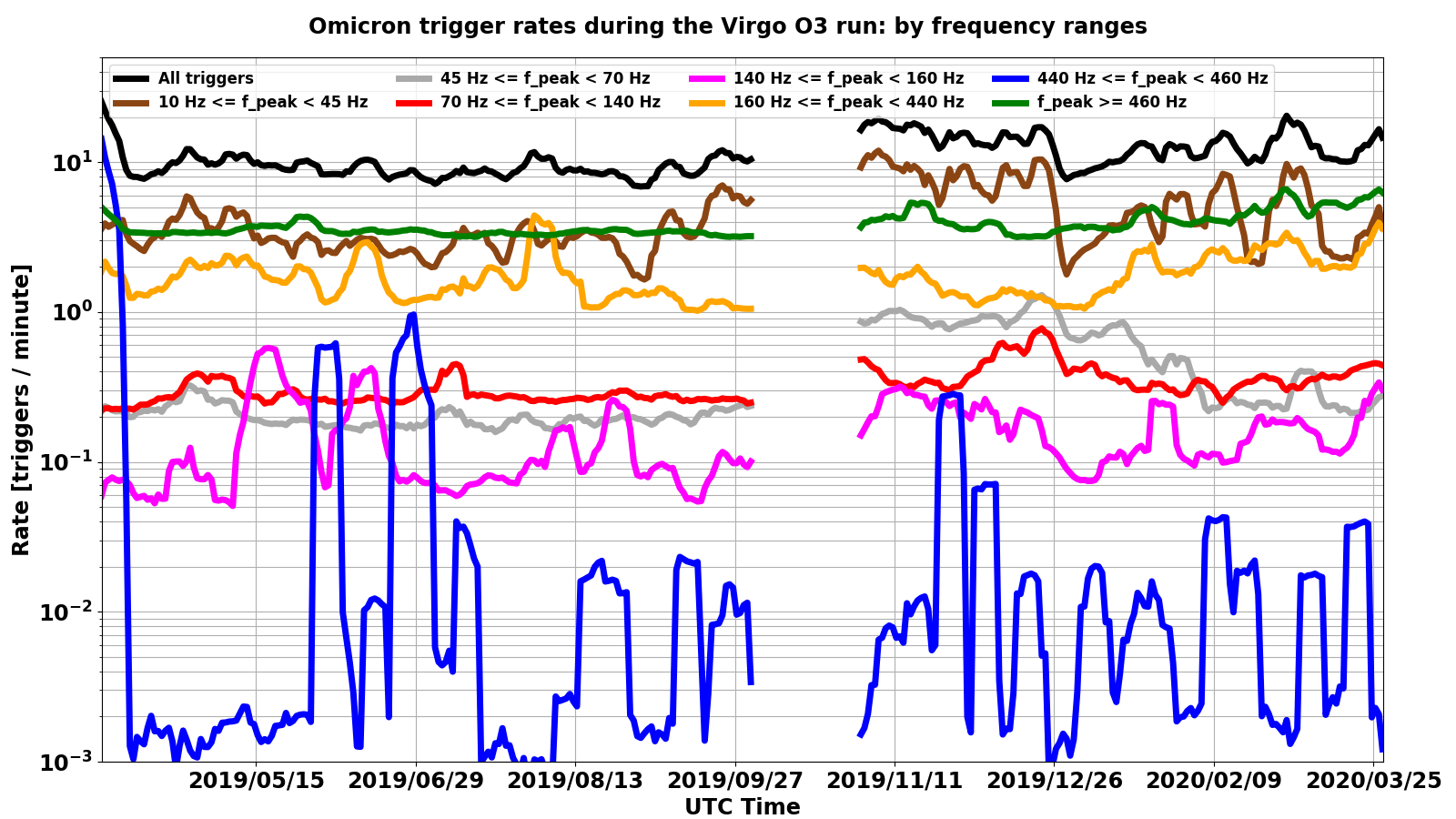}
    \end{center}
    \caption{Glitch rates (daily moving average) using \texttt{Omicron} triggers during the O3 run for Virgo. The gap in between O3a and O3b corresponds to the 1-month commissioning break. The top plot breaks the glitch rate into \ac{snr} ranges, while the bottom one categorizes it in terms of frequency ranges for the glitch peak frequency.}
    \label{fig:omicron_trigger_rates:details}
\end{figure}

During data taking, \texttt{Omicron} runs online on a few hundred channels, including the \ac{gw} strain $h(t)$, and monitors glitches in real time: these triggers are stored on disk with a few minutes latency.
Figure~\ref{fig:omicron_trigger_rates:global} displays the evolution of the glitch rate during the O3 run, and constitutes a reference for the detailed view of figure~\ref{fig:omicron_trigger_rates:details}, where the global \texttt{Omicron} glitch rate has been broken down into \ac{snr} (top plot) and peak frequency (bottom plot) bands.
In these plots, the rates have been averaged by computing their daily moving median to ease the reading.
From the top plot in figure~\ref{fig:omicron_trigger_rates:details}, we can notice that the ratio of glitches in the various \ac{snr} bands is approximately constant.
On the contrary, the bottom plots highlight several temporary increases in glitch rates, different for the various frequency bands.
The choice of these frequency bands has been made to try to isolate some possible classes of glitches, and consequently of their sources: the region below $45~\mathrm{Hz}$ is characteristic of scattered-light glitches, enhanced during bad weather conditions; the regions around $50$ and $150~\mathrm{Hz}$ contain the mains fundamental frequency in Europe and its second harmonics, hence is likely related to glitches of electrical origin; the region around $450~\mathrm{Hz}$ contains the frequencies of the suspension wire violin modes for the test masses, and of another harmonic of the mains frequency. 

The large majority of glitches identified by \texttt{Omicron} have a moderate \ac{snr}: between 5 (the minimum value from which the \texttt{Omicron} trigger is kept) and 8. The highest trigger rate at the very beginning of O3a corresponding to glitches with a peak frequency between 440~Hz and 460~Hz is an artefact due to a mis-configuration of the \texttt{Omicron} online server that was quickly fixed. The significant increase of the trigger rate in O3b with respect to O3a is mainly due to the bad weather conditions during the fall and winter seasons (see~\cite{o3virgoenv} for more details). The weather was actually very quiet in January 2020 and the associated drop in glitch rate is quite strong.

\subsubsection{Offline searches}

Non-stationary instrumental noise can potentially impact searches for transient
\acp{gw}, which must include methods to robustly separate
astrophysical candidates from noise fluctuations. Despite the power of such
methods, inspecting the candidates produced by a search remains an important way
to identify problematic operating conditions of \ac{gw} detectors, and to
understand if the search needs to be tuned in different ways for different
detectors.

In this section, we perform a detailed inspection of the candidates produced
from Virgo O3 final data by one of the analyses used for compiling the GWTC
catalog, namely the \pycbc offline analysis \cite{PyCBCOfflineO3} which we
already considered in Section \ref{sec:onlinedq:vetostreams}.
As a reminder, this analysis performs a broad-space search for compact binary
mergers involving neutron stars, black holes, or both.
It uses a bank of model waveforms and matched filtering to generate candidates
from LIGO and Virgo data.
Each single-detector candidate is ranked by a reweighted \ac{snr}, i.e.\ a
combination of its matched-filter \ac{snr} and two $\chi^2$ signal-based
discriminators \cite{Allen:2004gu, Nitz:2017lco} designed to reject candidates
produced by non-stationary noise.
Such discriminators have been mainly tuned on noise from Advanced LIGO so far,
and to the best of our knowledge, their behavior on Virgo data has not been
published before.

\begin{figure}
    \begin{center}
        \includegraphics[width=0.48\columnwidth]{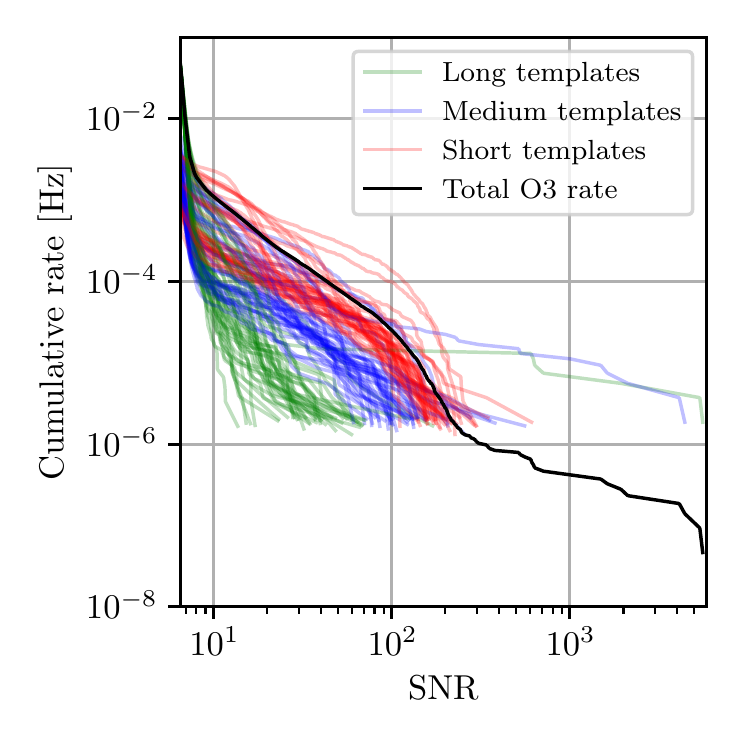}
        \includegraphics[width=0.48\columnwidth]{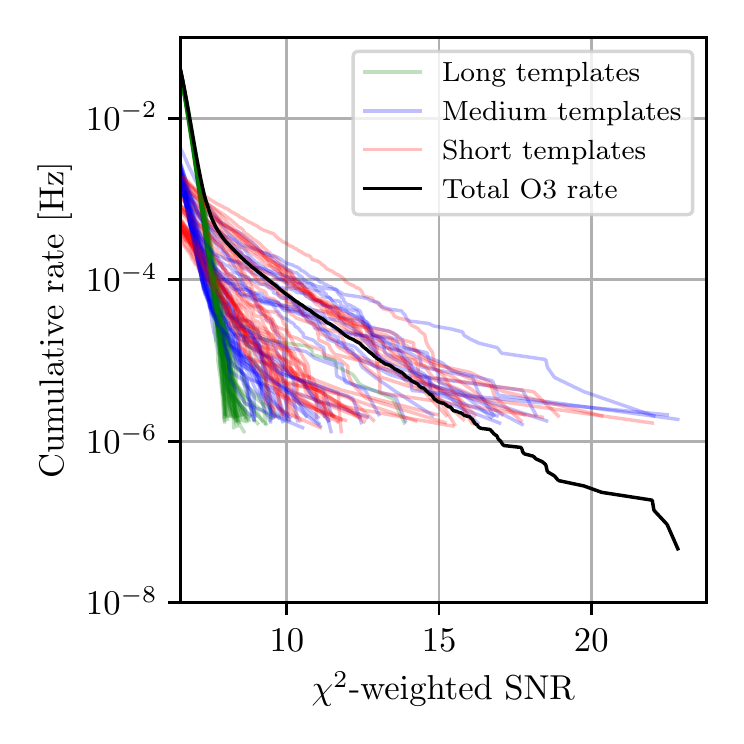}
    \end{center}
    \caption{Rate of compact binary merger candidates produced from Virgo
             O3 data using a broad-space search based on \pycbc.
             Left: rate as a function of the matched-filter \ac{snr}.
             Right: rate as a function of the reweighted \ac{snr},
             which combines the \ac{snr} and two $\chi^2$ discriminators.
             Red, blue and green curves correspond respectively to templates
             lasting less than 0.6~s, 0.6---4~s, and more than 4~s.
             Curves of the same color represent different chunks of Virgo data,
             each lasting $\sim $5~days.
             The black curves show the rate over the entire run and search space.
             They extend to lower rates than the individual chunks due to the
             much longer duration of the entire run.}
    \label{fig:pycbc_trigger_rate}
\end{figure}

Figure~\ref{fig:pycbc_trigger_rate} shows the rate of candidate events recorded
by \pycbc from Virgo data.
The horizontal axis shows either the matched-filter \ac{snr} of the candidate
(left plot) or its reweighted \ac{snr} (right plot).
The vertical axis shows the rate of candidates that are ranked higher than the
value in the horizontal axis\footnote{At this early stage of the \pycbc analysis, candidates with
merger times within fractions of a second from each other can be highly
correlated, because a given transient in the data typically ``rings off''
several templates with high overlaps between each other.
The estimated rate of candidates is biased if this correlation is not accounted for.
We do so by means of a clustering procedure: a given candidate is ignored if a
higher-ranked one exists within a time window of $\pm$5~s.}.
If the Virgo noise had been perfectly Gaussian and stationary throughout O3, we would expect the
rate to decrease exponentially for larger and larger values of the matched-filter \ac{snr}, and
be independent on the template parameters and particular chunks of data (subsets of
the full O3 dataset, each lasting about five days and analysed as a single block of data).
Instead, the rate-vs-\ac{snr} curves show a more complicated behavior, with a large
variation across the search space and particular data chunks. We observe a
non-negligible rate at \acp{snr} as high as 100, while astrophysical signals are
typically expected to have \acp{snr} between ${\sim}$5 and ${\sim}$10.
After the application of the $\chi^2$ discriminators, the behavior changes
drastically, and the exponential behavior of the rate is recovered, at least as
long as we restrict to a subset of the search space.
We still observe a large variation of the exponential slope and amplitude across
the search space and data chunks, except for the longest templates (green curves
of figure~\ref{fig:pycbc_trigger_rate}), which are more robust to instrumental
artifacts due to their particular time-frequency signature.
The same variation is also seen with candidates from the LIGO detectors, and it
is taken into account by the analysis when ranking the multi-detector
candidates~\cite{Nitz:2017svb}.

Even if properly accounted for, however, a higher rate of noise triggers still
reduces our ability to discover compact binaries.
Hence, it is informative to inspect the data quality around the triggers in the
tail of the curves, in order to understand if a particularly problematic
behavior of the detector or analysis can be improved in the future, or if a more
robust ranking of the triggers can be found.
To this end, we find that the highest \acp{snr} can be attributed to a single
segment of $\sim 15$~min of data on November 11$^{th}$, 2019.
These data contain narrow-band, loud and rapidly-varying excesses of power which
temporarily affected the data conditioning algorithm used by \pycbc.
We determined these excesses to be coming from transient problems with the noise
subtraction algorithms used to reconstruct the \ac{gw} strain channel $h(t)$.
Most of the associated high-\ac{snr} triggers were automatically removed by the
$\chi^2$ discriminators, effectively vetoing the entire problematic segment,
however it would have been unlikely for a signal to be detected in Virgo data
during this segment.
When inspecting the top candidates by $\chi^2$-weighted \ac{snr}, instead, we
find that most of the tail is clearly associated with scattered-light glitches.

We conclude from this section that the data conditioning procedure and $\chi^2$
discriminators used by \pycbc, albeit designed for and tuned on LIGO data, are
also reasonably effective in Virgo noise.
Nevertheless, there seems to be room for further sensitivity gain via a more
effective removal of scattered light.
One possible way is through better tuning of the Virgo veto stream mechanism
introduced in Section \ref{sec:onlinedq:vetostreams}, provided that the fraction
of affected astrophysical signals can be kept to negligible levels.
Another avenue to investigate in the future is a signal-based discriminator or
conditioning procedure specifically targeting scattered-light transients, as
proposed for example in \cite{ArchEnemy}.
Improved tuning and vetoes would have to be evaluated with injection studies and
a corresponding calculation of the relative sensitive time-volume of the search.

%% file: dq_channel_safety.tex
Many Virgo data quality analyses aim at ensuring that \ac{gw} candidates are of astrophysical origin and not caused by terrestrial noises. Typically, searches for correlations between auxiliary channels (monitoring the environment, the detector status, the accuracy of its control, etc.) and the $h(t)$ strain channel are run to produce vetoes, that reject times when such correlations are identified. This strategy can lead to a loss of interesting signals if any of the auxiliary channels is sensitive to \acp{gw}, which means that it picks up disturbances induced in the detector by these. Hence, a good knowledge of the couplings of auxiliary channels to $h(t)$ is essential. To gather such information, a statistical analysis of all auxiliary channels is performed, using the approach proposed in~\cite{Essick2021}.

This method relies on hardware injections that mimic the effects of \acp{gw} on the detector, by moving in a deterministic way one of its test masses. They are used to work around the fact that the transfer functions between $h(t)$ and most auxiliary channels are not well-known, nor understood. The injected signals are 0.6~s long sinusoidal Gaussian functions of various frequencies (between 19~Hz and 811~Hz) and amplitudes (\ac{snr} between ${\sim}$20 and ${\sim}$500). The frequencies injected are chosen to scan the entire detection band while avoiding any known resonant frequency (like violin modes). Each waveform is injected three times, spaced by 15~s.

This {\em safety} analysis assumes that glitches in a given auxiliary channel are distributed according to a stationary Poisson process, whose rate and $p$-value time series are measured using stretches of data during which no hardware injection is performed. These $p$-value time series are used to define a classification threshold. Then, a null test is applied to see whether the $p$-value distribution changes significantly in the presence of hardware injections. Auxiliary channels that exhibit anomalously small $p$-values (i.e. lower than the defined threshold) are classified as {\em unsafe}, meaning that they are likely to mirror excess power coming from the strain channel. The other channels, called {\em safe}, are the only ones used to produce vetoes.

Virgo DetChar hardware injections were organized at short notice, in the few days between the anticipated end of the O3b run and the moment when the detector had to be switched off because of the pandemic. Among the ${\sim}$2500 auxiliary channels analysed, 69 were found to be unsafe. 
That analysis confirmed the existing sets of safe and unsafe Virgo channels, as determined by a previous study. Moreover, its results were in agreement with the safe/unsafe status one could assign a priori to auxiliary channels, based on its definitions --- that is what quantity they measure, and how these measurements are done. The results validate the Virgo O3 data quality vetoes, both online and offline, which must be based on safe channels to avoid the risk of rejecting real \ac{gw} signals. The channels identified as unsafe belong to a few well-defined categories: error or correction signals from the \ac{darm} control feedback system; correction signals from test mass suspensions; readout channels from the B1 and B1p photodiodes located at the interferometer output (see figure~\ref{fig:AdVsetup}); finally, signals monitoring the quality of the detector working point, or used to reconstruct the \ac{gw} strain channel h(t).

%% file: dq_spectral_noise.tex
The term spectral noise, introduced in section~\ref{section:data_detchar}, identifies the class of detector disturbances appearing as a persistent excess in the noise power spectrum estimation of the data.

Spectral noise has a negative impact especially on searches for persistent \acp{gw}, which aim at detecting astrophysical or cosmological signals mainly through the identification of their spectral features.
Two typical signal categories of persistent waves are \ac{cw}~\cite{cw_review_keith} and a \ac{sgwb}~\cite{stoch_review_nelson}. The signals are very weak with respect to the already detected coalescing binary emission. Due to their persistent nature, they can be looked for in the frequency domain where the accumulated power over long observation times can show up at a detectable level, after applying effective signal processing techniques.
Moreover, some spectral features of the signals can help in discriminating them from detector noise. On the other hand, spectral noise can mask signals, or produce false candidates, in both cases reducing the search sensitivity.

Searches for persistent signals are typically run off-line, once long stretches of data have been collected.
An early identification of spectral disturbances and of their instrumental source would allow to remove, or at least reduce, the source of noise, thus improving the quality of the data. 

Different actions can be accomplished at the detector characterization level in support of data analysis.
A first action is to identify, and possibly remove, the instrumental source of spectral noises as soon as possible during a data taking period. This is a non trivial task that typically requires a significant amount of work to nail down which detector component is responsible for a given disturbance and to eliminate the noise source, which may imply to replace the noisy component (for instance a cooling fan, an electric motor, etc.)~\cite{EnvHuntVirgoO3}, to shut it down (if not needed) or to modify it properly.
This could consist, for instance, in shifting the frequency of a calibration line which non-linearly couples to another noise source, in order to move the noise line frequency into a band less relevant for \ac{gw} searches~\cite{Aasi:2012wd,cw_vsr4_nb}.

A second action is the use of additional techniques to
 differentiate between possible signals and other spectral features. These methods strongly depend on the analysis and 
on the type of \ac{gw} signals searched.
An example of such techniques
relies on the Doppler effect. An astrophysical \ac{cw} signal is expected to be modulated in frequency by the Doppler effect, due to the Earth rotation, which induces a shift 

\begin{equation}
\Delta f (t)\simeq f_0\frac{\vec{v}(t)\cdot \hat{n}}{c},
\nonumber
\end{equation}

 where $f_0$ is the source frequency, $\vec{v}$ the detector velocity, $\hat{n}$ the unit vector identifying the sky direction of the \ac{gw} source, and $c$ the speed of light. 
\ac{cw} searches correct this Doppler effect, thus any monochromatic line present in the $h(t)$ signal is spread
by a maximum amount of $\Delta f_{max} \simeq 10^{-4}f_0\cdot \cos{\beta}$, where $\beta$ is the ecliptic declination. This shift corresponds to up of hundreds or even thousands of frequency bins for typical \ac{cw} searches. 

Potential candidates found in the analysis lead to follow-up 
investigations to identify a possible instrumental source. This follow-up is also based on a combination of DetChar activity, to spot the source of the disturbances, and application of \ac{cw} or \ac{sgwb} algorithms to build confidence in the astrophysical nature of the candidate, see for example~\cite{2022arXiv220100697T}.

Although spectral noises cannot always be removed, it is still useful to characterize them by constructing a list of noisy lines.
This list can be used to exclude those disturbing frequency bands from the analyses, or to veto candidates with frequency 
too close to those noisy lines

The identification of lines is typically done by automated pipelines~\cite{O3DetChar_tools}, based on:
\begin{itemize}
\item User-defined thresholds set on data power spectrum or on line {\it persistency}, defined as the fraction of \acp{fft}, compared to the total number covering the full observation time, in which the ``normalized'' power content of a given frequency bin was above such a threshold (typically set to six times the average value);
\item Highlighting coincidences or significant coherence among different channels;
\item Highlighting a pattern in time-frequency maps of the data. 
\end{itemize}

Candidates found in \ac{gw} searches are subject to verification steps, in which the identification of possible noise counterparts is done by processing the data in the relevant frequency band and period of time and/or running manually one or more of the line identification pipelines.
In the following, we report and discuss a few examples of lines identified in Virgo O3 data. Readers can refer to the LIGO-Virgo \ac{gwosc}~\cite{GWOSC_O3_lines} for the official full list of lines.

\subsubsection{Combs}

\begin{figure}
    \begin{center}
        \includegraphics[width=\columnwidth]{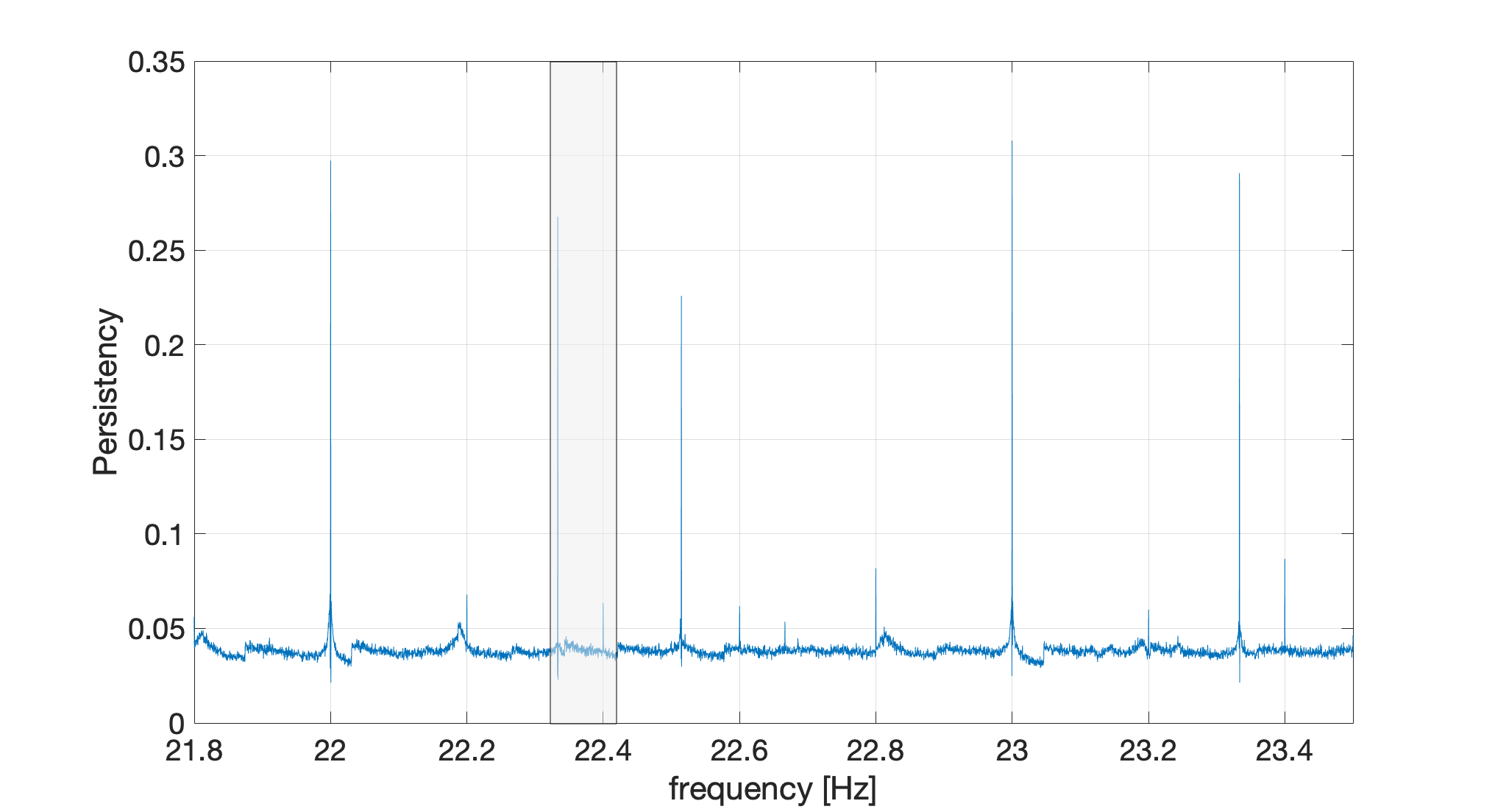}
    \end{center}
    \caption{Plot of line {\it persistency} over the frequency band 20~Hz to 30~Hz. The grey box identifies the frequency range covered by a narrow-band search of \ac{cw} signals from the Vela pulsar in O3a data. The line at 23.333~Hz, which contributed to produce a candidate in the search is clearly visible. In fact, several other lines belonging to the 1~Hz {\it comb}, both at integer frequencies and shifted by 0.333~Hz, and to a weaker 0.2~Hz-spaced comb, are present.}
    \label{fig:line23}
\end{figure}

Combs are families of lines separated by a constant frequency interval. 
Typically, noise combs are electromagnetic disturbances 
generated by digital devices (e.g. microprocessors, programmable communication devices like logical controllers, Ethernet cables, wireless repeaters) that leak into the \ac{gw} strain channel. 
Comb lines can have 
an impact on searches for persistent \acp{gw} due to their large number and usually high strength. This makes the identification of combs an important task. There are several combs present in Virgo O3 data, which we describe in the following. They are typically made of many (up to O(10)) lines, covering from a few to several Hertz in frequency.. 

A 1~Hz-spaced comb with 0~Hz offset was already present during previous runs. 
A new 1~Hz-spaced comb discovered during O3 has a 0.333~Hz offset with respect to integer frequencies. This comb was found
during investigations of a line at 22.333~Hz that falls 
within a region of interest for the Vela pulsar \ac{cw} search.
The instrumental origin of the comb has been confirmed by finding lines at the same frequency in the magnetometers deployed at EGO.

Figure~\ref{fig:line23} shows the line {\it persistency} computed over the frequency range 21.8-23.5~Hz on O3 Virgo data. Both 1~Hz combs
are clearly visible. Furthermore, there is a comb with 0.2~Hz spacing, whose origin is unknown. The grey shaded area indicates the frequency region explored by a narrow-band \ac{cw} search targeting the Vela pulsar. The strong line at 22.333~Hz produced an outlier in the search, which was discarded after its instrumental origin was understood.

Finally, two more combs which have been identified by DetChar studies, have both $\sim$9.99~Hz spacing, one with 0~Hz offset and the other with 0.5~Hz offset.

\subsubsection{Wandering line around 83~Hz -- 84~Hz}
\label{sec:wanderingline83}

A wandering line is a peculiar kind of spectral noise where the frequency of a spectral line changes with time, with no apparent reason.
This is also called a \emph{drifting line} once the mechanism driving the frequency change is at least partially identified, making its variations not entirely random anymore.

\begin{figure}
  \center
  \includegraphics[width=0.9\textwidth]{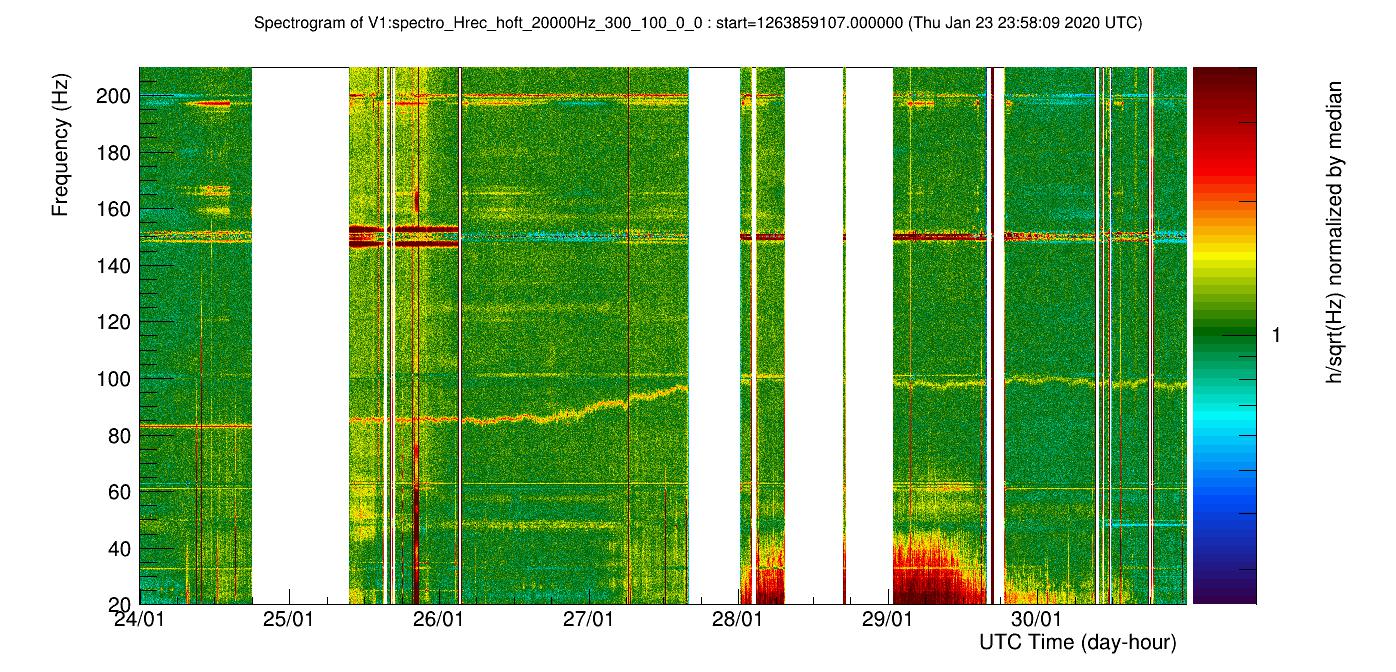}
  \caption{A typical 7-day spectrogram of the $h(t)$ channel, generated by the \texttt{Spectro} tool~\cite{O3DetChar_tools} and allowing the monitoring of a wandering spectral line whose main frequency was first stable around 83~Hz before increasing up to around 100~Hz in about a day.}
  \label{fig:o3_spectro} 
\end{figure}

An example that triggered lots of DetChar investigations during O3 is the line, normally located between 83 and 84~Hz, as shown in figure~\ref{fig:o3_spectro},
that reached about 110~Hz at the maximum of its excursion and had variations of a few Hz over about one hour~\cite{DiRenzo2019:elog,Fiori2018:elog}.
Its origin dates back to the Virgo commissioning run 10 (C10) of August 2018~\cite{DiRenzo:2020}, and possibly even earlier, in the preparatory phase preceding O2~\cite{Mantovani2017:elog}.
Neither of the mechanisms that make the line to depart from its typical frequency of about 83~Hz or what produces its variations with time have ever been understood, although several data analysis techniques have been applied and newer ones developed 
for \emph{line tracking}~\cite{DiRenzo:2020}.
An analysis with Bruco~\cite{O3DetChar_tools} revealed no witness channel coherent with $h(t)$ around that line.
Moreover, we tracked the frequency evolution of this line, and we correlated the corresponding time series with the auxiliary channels monitoring Virgo~\cite{DiRenzo2019:elog}.
This technique has proven successful in the past, in the case of drifting lines driven by the temperature of some optical components~\cite{Fiori2017:elog}, but has produced no convincing correlation in the case of this line, whose origin remains unknown. 

\subsubsection{Spectral noise bump around 55~Hz}

Figure~\ref{fig:spectral_noise}a) shows the power spectrum of the Virgo \ac{gw} strain channel $h(t)$ computed at two different dates, February 26 and March 2, 2019 (before the start of the O3 run). Comparing the two plots, one can see that a wide bump around 55~Hz was cured in the meantime. Indeed, a detailed study showed that this disturbance was present most of the time and was observed also in the \ac{prcl} channel. This allowed to remove most of this noise excess when producing the reconstructed strain $h(t)$, by accurately subtracting the remaining \ac{prcl} contribution~\cite{VIRGO:2021umk}. Note that this 55~Hz bump affected the frequencies around 55.6~Hz, where the \ac{cw} signal possibly emitted by pulsar PSR J1913+1011 is expected. Furthermore, this bump was located within the most sensitive region of the Virgo spectrum for an (isotropic) \ac{sgwb} search.

\subsubsection{Spectral noise around the 50~Hz power line frequency}

The \ac{gw} strain channel $h(t)$ in the frequency region between 45~Hz and 55~Hz was significantly affected by ambient electromagnetic fields originating from the interferometer infrastructure. This noise was studied and mitigated in subsequent steps during the run~\cite{EnvHuntVirgoO3}.

\begin{figure}
    \begin{center}
        \includegraphics[width=\columnwidth]{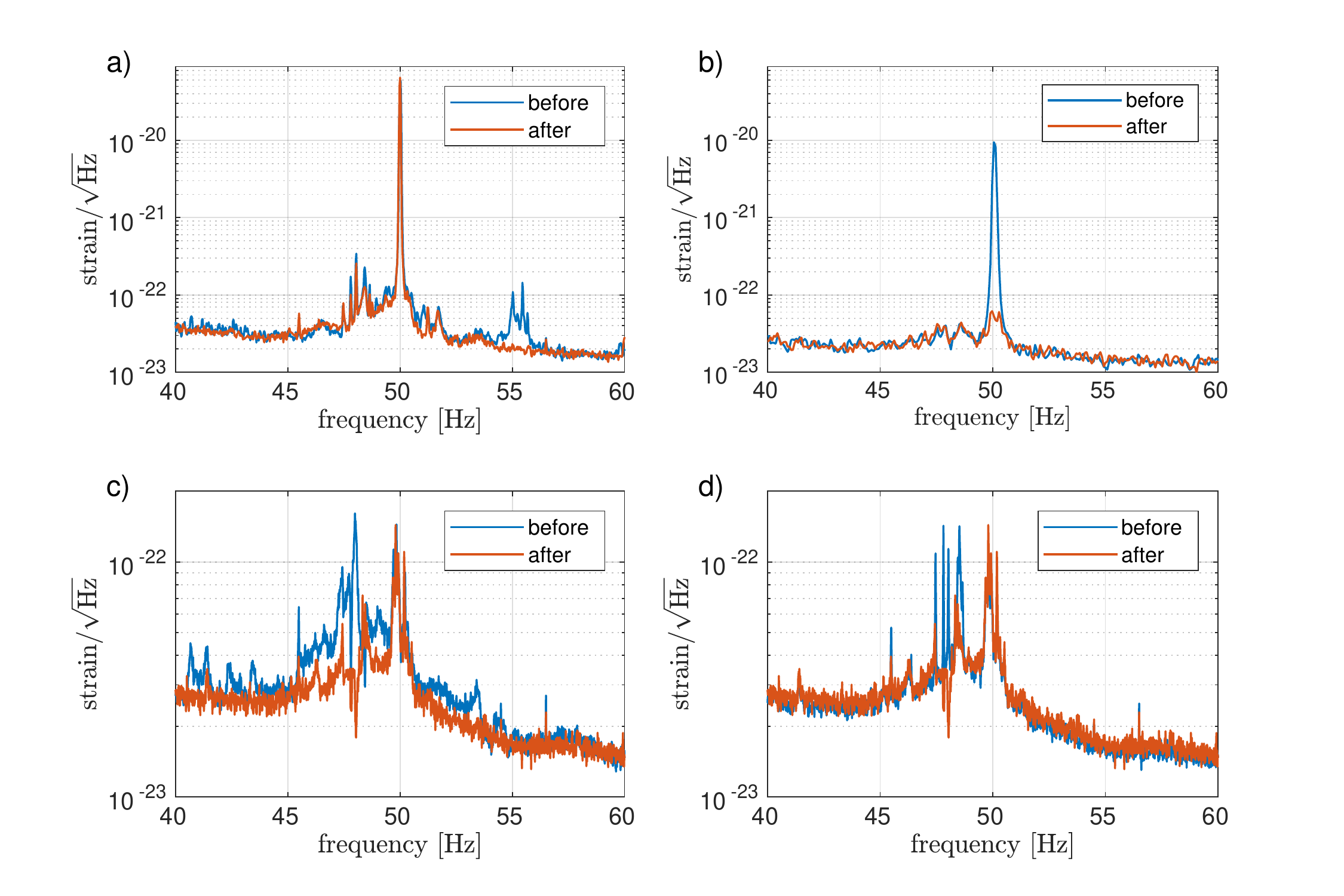}
   \end{center}
    \caption{Four steps of the reduction process of the strain spectral noise between 40~Hz and 60~Hz during O3. 
In all sub-figures the blue curve is before the mitigation, and the red curve is after the mitigation. a) Cancellation of the spectral noise structure around 55~Hz which was common to the \ac{prcl} signal.
b) Subtraction of the 50~Hz line associated to the power grid.
c) Mitigation of a wide-band noise associated to the motor driver crate of the WE suspension.
d) Suppression of a few noise lines associated to mechanical modes of the last stage of the test mass suspension (the payload~\cite{art2-PayloadProceeding}).
}
    \label{fig:spectral_noise}
\end{figure}

The intense 50~Hz line, corresponding to the frequency of the electricity mains, was mitigated and substantially eliminated from $h(t)$ (see figure~\ref{fig:spectral_noise}b), by implementing a feed-forward noise cancellation scheme using as sensor a voltage monitor of the detector uninterruptible power supply system~\cite{EnvHuntVirgoO3}.
This operation did not reduce the 50~Hz harmonics also present in the $h(t)$ spectrum
(see figure~\ref{fig:O3_sensitivities}) because they are not due to a non-linear response of the interferometer. 
They are present in the global environmental disturbances and enter the \ac{gw} strain channel through different 
coupling paths.

Sidebands of the mains frequency, at approximately 49.5~Hz and 50.5~Hz, were generated by the pulse width modulation of the electric heater controller of the \ac{imc} building. The noise was mitigated by decoupling the electric ground of the building from the central experimental area with an isolation transformer.

Figure~\ref{fig:spectral_noise}c) illustrates a wide-band noise affecting the same region. 
The origin of this noise was eventually found to be a noisy static voltage accidentally 
applied to the signal wires of the motors used for positioning and balancing the \ac{we} mirror suspension, then coupling capacitively to the mirror coil actuator wires. 
The noise was mitigated by un-plugging the drivers of the motors, which are not used in Science mode.

Finally, figure~\ref{fig:spectral_noise}d) illustrates a family of lines between 47~Hz and 49~Hz which
have been identified as vertical mechanical modes of the last stage of the test mass suspension
system. These modes are excited by ambient magnetic fields coupling to the magnetic actuators along the suspension chain. This noise was suppressed by an active mechanical damping of the modes.

%% file: dq_offline.tex
While section~\ref{sec:onlinedq:cat1} described the {\em online} CAT1 vetoes, we focus here on the final set of {\em offline} CAT1 vetoes. They supersede online vetoes and have been used by all analyses processing the final O3 Virgo dataset. These include analyses using the O3 LIGO-Virgo public dataset: that is why the \ac{gwosc} website~\cite{RICHABBOTT2021100658} includes detailed public information about these vetoes~\cite{GWOSC_O3_vetoes}.

Like the online CAT1 vetoes, all these veto flag segments of bad data that are unusable. They are defined with a 1~s granularity and the figure-of-merit used to quantify their impact is their dead time, that is the fraction of Science time that is removed by applying them individually. Yet, the vetoes are not independent and they may overlap. Therefore, they are meant to be applied globally on the dataset, by taking the logical OR of all of them.

Offline vetoes are computed by reanalyzing the whole dataset, using more complex algorithms and additional input information not available online. All the online data quality assessments are reviewed and updated where needed. GPS segments are added to (or removed from) the final dataset by this, possibly iterative, procedure.
The offline vetoes defined during the O3 run can be classified into three main categories.

\begin{itemize}
\item The duplication --- after crosscheck and potential additions or fixes --- of online CAT1 vetoes: this includes the saturations of dark fringe photodiodes or mirror suspensions, and the monitoring of the reconstructed \ac{gw} strain $h(t)$.
\item The upgrade of existing online vetoes: the excess rate of glitches is monitored offline using $h(t)$, whereas only the \ac{darm} channel could be used online due to latency constraints.
\item The addition of new vetoes, based on information that was not available in low latency, or that was not known at the time online flags were generated. These categories are:
    \begin{itemize}
    \item Checks of the consistency and of the completeness of the files storing the $h(t)$ \ac{gw} stream: these vetoes flag segments in which $h(t)$ is missing or contains missing samples.
    \item The $h(t)$ stream is reconstructed by blocks of eight consecutive seconds of data. Therefore, a control loss can possibly impact up to the eight seconds of data that predate it. As the exact time of a control loss is not easy to define, the last ten seconds preceding each recorded control loss have been removed.
    \item The Science dataset has been scanned accurately to identify segments during which the detector was not taking good quality data, contrary to what its status was indicating. These segments were removed from the final dataset.
    \item Finally, a workaround was applied to the detector control system during some weeks in O3b in order to mitigate transient data losses due the failure of an hardware component. That patch allowed to maintain the working point of the instrument, thus sparing a ${\sim}$20~min control acquisition procedure each time it prevented a global control loss. Yet, the application of that workaround could degrade the quality of the data. Thus, the impacted segments were removed from the final dataset, with some safety margin on both ends (the last 10~s before having the control patch be applied automatically, and the first 110~s following the end of the transition back to the nominal control system).
    \end{itemize}
\end{itemize}

Table~\ref{table:offline} summarizes the impact of CAT1 vetoes on the final O3 Science dataset: overall, only 0.2\% of the Science data have had to be removed due to various problems.

\begin{table}[htbp!]
\caption{\label{table:offline}Virgo O3 offline Science dataset and CAT1 vetoes.}
\centering
    \begin{tabular}{c|ccc}
        \toprule
            & O3a              & O3b               & O3a + O3b\\
        \hline
        Science dataset & 12,057,731~s & 9,611,843~s & 21,669,574~s \\
        Logical OR & 18,802~s & 20,636~s & 39,438~s \\
        of all offline CAT1 vetoes & (0.16\%) & (0.22\%) & (0.18\%) \\
        \bottomrule
    \end{tabular}
\end{table}

Conversely, a few minutes of good quality data that had not been included in the online Science dataset for various and clearly understood reasons (software issue, human error, etc.) were added to the offline, final, dataset.

%% file: validation.tex
To assess whether the detection alerts produced by transient searches~\cite{PyCBCLiveO3,PyCBCLiveO2,Adams:2015ulm,Klimenko:2016} should be considered as ``candidate events'', a procedure of \emph{validation} is implemented for each generated trigger~\cite{GWTC3,LIGO:2021ppb}. 
This task has the role to verify if data quality issues, such as 
instrumental artifacts, environmental disturbances, etc., can impact the analysis results and decrease the confidence of a detection,
or even foster a rejection~\cite{2019ApJ_low-lat}.

The validation of the online triggers found by \ac{gw} transient searches includes two separate stages.
A prompt evaluation is typically completed within few tens of minutes after an event trigger has been generated, as represented by the data flow in figure~\ref{fig:DataflowDetChar}.
Its goal is to determine a preliminary detection confidence and sky localization, in order to deliver public alerts to the astronomy community and support for multi-messenger follow up observations~\cite{2019ApJ_low-lat}, as described in section~\ref{section:public_alerts}, or to vet that trigger if evidence of severe contamination from non-astrophysical artifacts is present.
A team of Virgo DetChar shifters is in charge of this task as part of the \ac{rrt} (see section~\ref{section:oncall_RRT}). 
The decision about the event is primarily based on the quick results provided by the \ac{dqr}, within a few minutes from the trigger.
This decision takes into account the evaluation of the operational status of the detector and its subsystems, the environmental conditions, as well as preliminary checks on the strain data.
In particular, the shifters are asked to verify the presence of excess noise, namely glitches, around the time of the trigger and the validity of the hypotheses of stationarity and Gaussianity of the data, as discussed in~\cite{O3DetChar_tools}.
Moreover, it is examined the possible presence of correlations between the strain data and the auxiliary sensors, which may advise a non-astrophysical origin of the trigger.  

With higher latency, a second stage of validation is performed by a \emph{validation team} to finally check candidate events before publications, including those found by offline analyses~\cite{PyCBCOfflineO3,Aubin:2020goo}.
Besides of (double-)checking the astrophysical origin of the event trigger, the main purpose of this process is to carefully assess whether the parameter estimation of the source properties can be affected by noise artifacts.   
This procedure takes advantage of dedicated reruns of the \ac{dqr}, as well as from additional tools and metrics, including, for example, signal consistency checks~\cite{LIGO:2021ppb,Mozzon:2020gwa}.

For those events where non-stationary noise, such as glitches, are found in the vicinity of the putative \ac{gw} signal, or even overlapping with it, a procedure of noise \emph{mitigation} is implemented~\cite{Davis:2022dnd,Cornish:2014kda}.
During O3b, such process has involved 12 events, including one with Virgo data, GW191105e~\cite{GWTC3,GCN26182}, where the process of mitigation and validation of the data quality has improved the parameter estimation results and credibility. Various O3a events have undertaken a preliminary version of this procedure~\cite{Abbott:2020niy}.

%% file: outlook.tex
The LIGO-Virgo O3 run has lead to the discovery of dozens of new \ac{gw} signals from compact binary mergers, which have boosted our knowledge of these populations in our local Universe and allowed further, more stringent, tests of general relativity. The O3 run has also been the first long data-taking period for the \ac{adv} detector. Thus, it represents a full-scale, extended and non-stop stress test of the organization and work methods of the Virgo DetChar group. The experience accumulated during these 11 months form the base of the DetChar activities, both to prepare and operate for O4 and the following runs.

Although the Virgo DetChar group has fulfilled all its main requirements during the O3 run, work has been going on since then to improve its performance and extend its activities. In particular, the anticipated differences between the O3 and O4 runs lead to new challenges that the group should tackle. The \ac{adv} detector will have evolved significantly, with the completion of the Phase I of the \ac{adv}+ project~\cite{AdVPlus}. The main changes on the instrument side are the addition of the signal-recycling mirror in between the beam splitter and the output port of the Virgo interferometer, a higher input laser power and the implementation of frequency-dependent squeezing. This new configuration will require dedicated instrument characterization activities, while many new data quality features will have to be discovered, understood and later mitigated or solved. On the data analysis side, progress in terms of sensitivity while keeping the network duty cycle high will lead to more \ac{gw} detections. On the one hand, more work will be required to validate this excess of signal candidates compared to O3. On the other hand, the triggers passing a given false alarm rate threshold will remain dominated by noises, meaning that the bulk of computing resources used by the Virgo DetChar group will not change significantly.

Gathering experience from the past and predictions for the future, a few top priorities have emerged for the DetChar group. A first and obvious one is to broaden the scope of the DetChar monitoring, to make sure that no relevant area remains uncovered, from raw data to the final analyses. Then, the latency of the various DetChar products should be decreased when it is relevant and possible: either by making the corresponding software framework more efficient, or by processing new data more regularly. Finally, some emphasis should be put on increasing the automation of the DetChar analyses and the reporting of their results. In that respect, the \ac{dqr} is a good example of the realization of these plans. Parallel to common LIGO-Virgo-KAGRA developments on the framework architecture to make \acp{dqr} more uniform among the three collaborations and to improve its performance, additional data quality checks will be implemented. They will provide combined results that should give a partial digest of the global vetting of a given \ac{gw} signal candidate. 

The increase of the information available and the help to identify quickly its most relevant points should allow maintaining, if not improving, the high and steady level of Virgo performances observed during O3.

%% file: acknowledgments.tex
The authors gratefully acknowledge the Italian Istituto Nazionale di Fisica
Nucleare (INFN), the French Centre National de la Recherche Scientifique (CNRS)
and the Netherlands Organization for Scientific Research (NWO), for the construction
and operation of the Virgo detector and the creation and support of
the EGO consortium. The authors also gratefully acknowledge research support
from these agencies as well as by the Spanish Agencia Estatal de Investigaci\'on,
the Consellera d'Innovaci\'o, Universitats, Ci\`encia i Societat Digital de la
Generalitat Valenciana and the CERCA Programme Generalitat de Catalunya,
Spain, the National Science Centre of Poland and the European Union --- European
Regional Development Fund; Foundation for Polish Science (FNP), the
Hungarian Scientific Research Fund (OTKA), the French Lyon Institute of Origins
(LIO), the Belgian Fonds de la Recherche Scientifique (FRS-FNRS), Actions
de Recherche Concert\'ees (ARC) and Fonds Wetenschappelijk Onderzoek ---
 Vlaanderen (FWO), Belgium, the European Commission. The authors gratefully
acknowledge the support of the NSF, STFC, INFN, CNRS and Nikhef for
provision of computational resources.

{\it We would like to thank all of the essential workers who put their health at risk
during the Covid-19 pandemic, without whom we would not have been able to
complete this work.}

{\it The authors would also like to thank Samuel Salvador for his extensive and careful proofreading of the manuscript.}